\documentclass[a4paper,11pt]{article}
\usepackage[
  left=2.2cm,
  right=2.2cm,
  top=2cm,
  bottom=2.5cm,
  footskip=35pt 
]{geometry}

\renewcommand{\arraystretch}{1.2}

\usepackage{caption}
\DeclareCaptionFormat{myformat}{%
  #1#2#3\par\vspace{4pt}\centerline{\rule{0.77\linewidth}{0.4pt}}
}
\usepackage[utf8]{inputenc}
\usepackage{multirow}
\usepackage[pdftex]{graphicx}
\usepackage{amssymb}
\usepackage{amsmath}
\usepackage{cite}
\usepackage{braket}
\usepackage{slashed}
\usepackage{longtable}
\usepackage{booktabs}
\usepackage{array}
\usepackage[normalem]{ulem}
\usepackage{makecell}
\usepackage{bbm}
\usepackage{subcaption}
\usepackage{enumitem}
\usepackage{calc}
\usepackage{multicol}

\usepackage[dvipsnames]{xcolor}
\usepackage{xspace}

\usepackage[hidelinks]{hyperref}
\allowdisplaybreaks
\numberwithin{equation}{section}

\usepackage{tcolorbox} 

\usepackage{soul} 
\sethlcolor{myliblue} 

\usepackage{tocloft} 
\newcommand{\appsection}[1]{%
  \refstepcounter{APP}%
  \section*{Appendix \theAPP: #1}%
  \addcontentsline{toc}{section}{\theAPP~#1}%
  \setcounter{equation}{0}%
}

\newcommand{\refapp}[1]{\hyperref[app:#1]{Appendix~\ref*{app:#1}}}
\newcommand{\reffig}[1]{\hyperref[fig:#1]{Figure~\ref*{fig:#1}}}
\newcommand{\refeq}[1]{\hyperref[eq:#1]{Eq.~(\ref*{eq:#1})}}
\newcommand{\refeqs}[2]{\hyperref[eq:#1]{Eqs.~(\ref*{eq:#1})-(\ref*{eq:#2})}}
\newcommand{\refeqa}[2]{\hyperref[eq:#1]{Eqs.~(\ref*{eq:#1})}~and~\hyperref[eq:#2]{(\ref*{eq:#2})}}
\newcommand{\refsec}[1]{\hyperref[sec:#1]{Section~\ref*{sec:#1}}}
\newcommand{\refsubsec}[1]{\hyperref[sec:#1]{Subsection~\ref*{sec:#1}}}
\newcommand{\reftab}[1]{\hyperref[tab:#1]{Table~\ref*{tab:#1}}}

\newcommand{\B}{\mathcal{B}}
\newcommand{\C}{\mathcal{C}}
\newcommand{\Chi}{\mathcal{X}}
\newcommand{\E}{\mathcal{E}}
\newcommand{\F}{\mathcal{F}}
\newcommand{\G}{\mathcal{G}}

\renewcommand{\O}{\mathcal{O}}
\renewcommand{\P}{\mathcal{P}}
\renewcommand{\S}{\mathcal{S}}

\newcommand{\EOS}{\texttt{EOS}\xspace}
\newcommand{\OPE}{\text{OPE}}
\newcommand{\QCD}{\text{QCD}}

\renewcommand{\th}{\text{th}}
\newcommand{\eps}{\varepsilon}

\renewcommand{\Im}{\text{Im}}
\newcommand{\lamkin}{\lambda_{\text{kin}}}
\newcommand{\MeV}{\text{MeV}}
\newcommand{\GeV}{\text{GeV}}
\newcommand{\sopt}{s_0^{\text{opt}}}
\newcommand{\para}{{\parallel}}

\definecolor{mygreen}{RGB}{0, 200, 120}
\definecolor{myblue}{RGB}{70, 45, 255}
\colorlet{myinblue}{myblue!40!white}
\colorlet{myliblue}{myblue!13!white}
\definecolor{mymagenta}{RGB}{220, 0, 157}
\definecolor{myorange}{RGB}{255, 50, 0}

\newtcolorbox{mybox}[1][]{
    colback=myliblue,
    colframe=myblue,
    left=3mm,
    right=3mm,
    boxrule=0.3mm,
    #1
}

\newcommand{\bs}[1]{\boldsymbol{#1}}
\renewcommand{\bf}[1]{\textbf{#1}}

\newcommand{\hi}[1]{{\setulcolor{myinblue}\ul{#1}}} 

\newcounter{TODO}

\newcounter{TODOd}

\begin{document}

\vspace*{2mm}
\begin{center}
\fontsize{16}{20}\selectfont
\bf{Unitarity bounds and form-factor predictions\\
for \textit{B}-meson decays}
\\[0.2cm]
\fontsize{12}{14}\selectfont
\textcolor{white}{\it{Everything you always wanted to know about form factors$^*$ ($^*$But were afraid to ask) }}
\end{center}

\vspace{-7mm}

\renewcommand{\thefootnote}{\fnsymbol{footnote}}
\setcounter{footnote}{1}

\begin{center}
{\large Nico Gubernari}\\[10mm]
{\it
    DAMTP, University of Cambridge, Wilberforce Road,
    Cambridge, CB3 0WA, United Kingdom\footnote{Previous affiliation}
    \\[2mm]
    Helmholtz-Institut f\"ur Strahlen- und Kernphysik, Universit\"at Bonn, 53115 Bonn, Germany\hspace{0.3mm}\footnote{Current affiliation}
    \\[6mm]
    E-mail: \textnormal{\texttt{nicogubernari@gmail.com}}
    \\[2mm]
    Preprint number: \textnormal{\texttt{EOS-2026-03}}
}
\end{center}

\renewcommand{\thefootnote}{\arabic{footnote}}
\setcounter{footnote}{0}

\vspace*{1.5cm}

\begin{abstract}

\noindent
This paper has three main aims.
First, I review in a pedagogical way the unitarity bounds for form factors in $B$-meson decays, together with the parametrisations most commonly used in phenomenological analyses.
These include BGL, BCL, CLN, and the Dispersive Matrix (DM) method.
I also clarify the relation between BGL and DM, showing that they are two equivalent implementations of the same unitarity information.
Second, I demonstrate that the standard BGL and DM constructions are strictly rigorous only when no subthreshold cuts are present.
For $B$-meson decays, this requirement is fulfilled exclusively by the $B\to\pi$ FFs.
To treat the generic case, I fully develop the GG parametrisation introduced in a previous paper and show how the same logic extends to a DM-like construction.
Third, I perform three novel combined analyses of form factors and obtain predictions over the full semileptonic region.
The first analysis concerns the form factors for $B\to\pi$ and $B_s\to K^{(*)}$, the second those for $B\to K^{(*)}$ and $B_s\to \phi$, and the third those for $B\to D^{(*)}$ and $B_s\to D_s^{(*)}$.
All numerical results, posterior samples, analysis files, and plots are provided in the supplementary material~\cite{suppl_unitb}.
\end{abstract}

\thispagestyle{empty}

\newpage

\section*{Abbreviations}

\vspace*{-0.3cm}
\begin{multicols}{2}

\noindent
BCL  = Bourrely--Caprini--Lellouch \\
BGL  = Boyd--Grinstein--Lebed \\
BSZ  = Bharucha--Straub--Zwicky \\
CLN  = Caprini--Lellouch--Neubert \\
DA   = distribution amplitude \\
DM   = Dispersive Matrix \\
FF   = form factor \\
FP   = fictitious pole \\
GG   = Gopal--Gubernari \\
HQET = heavy-quark effective theory \\
LCSR = light-cone sum rule \\
LQCD = lattice QCD \\
NLO  = next-to-leading order \\
NNLO = next-to-next-to-leading order \\
OPE  = operator product expansion \\
SPP  = simplified power-series parametrisation 

\end{multicols}

\tableofcontents

\thispagestyle{empty}

\newpage

\section{Introduction}

Exclusive $B$-meson decays are among the main tools for testing the Standard Model and for determining Cabibbo--Kobayashi--Maskawa matrix elements.
Among the phenomenologically most relevant channels are the semileptonic decays $B\to \pi \ell \nu$ and $B\to D^{(*)}\ell\nu$, together with the rare semileptonic transitions $B\to K^{(*)}\ell^+\ell^-$.
Achieving precise phenomenological predictions, however, requires reliable control over the corresponding hadronic matrix elements.

For local quark currents, these matrix elements can be decomposed into a finite set of Lorentz structures, multiplied by scalar functions of the squared momentum transfer between the mesons $q^2$.
These scalar functions are the local hadronic \emph{form factors} (FFs).
This decomposition follows solely from Lorentz covariance.
In rare decays additional non-local matrix elements also appear, but the local FFs remain indispensable ingredients and often provide the basic hadronic input on top of which the more complicated non-local effects are built.

The main obstacle is that FFs are governed by low-energy QCD and therefore cannot be calculated in perturbation theory.
At present, the two main QCD-based approaches to calculate FFs are lattice QCD and light-cone sum rules (LCSRs).
These methods are often complementary: lattice QCD is particularly powerful at large $q^2$, whereas LCSRs are applicable at low $q^2$.
Here, the important fact is that both approaches provide results only at a finite set of kinematic points and with non-trivial correlations.
A FF parametrisation is therefore required to interpolate and extrapolate these theoretical inputs across the full semileptonic region and to combine different calculations in a statistically consistent manner.
\\

A good parametrisation should do more than reproduce the available points.
It should respect the FFs analytic structure implied by QCD, treat poles and thresholds correctly, and remain flexible enough not to bias the result.
This matters because in practical analyses only a finite number of parameters can be determined, and hence every series expansion has to be truncated.
For a generic truncated expansion, the size of the neglected higher-order terms is not known a priori.
In this regard, unitarity bounds are extremely valuable because they turn analyticity and dispersion relations into quantitative constraints on the FFs.
In practice, they provide a model-independent way to control truncation effects and to test whether a fitted parametrisation is compatible with unitarity.

This is the basic logic behind the $z$-expansion literature.
Simplified power-series parametrisations are often sufficient when one only needs a flexible description of the $q^2$ dependence.
If one also wants a rigorous estimate of the truncation error, however, the problem becomes more restrictive.
The most widely used constructions in this context are the BGL parametrisation~\cite{Boyd:1997kz}, the BCL parametrisation~\cite{Bourrely:2008za}, the CLN parametrisation~\cite{Caprini:1997mu}, and the Dispersive Matrix (DM) method~\cite{Bourrely:1980gp,Lellouch:1995yv}.
Although each of these approaches has been widely used in the literature, a systematic comparison of their underlying assumptions, first-principles content, and implications for truncation uncertainties has so far been lacking.
A central contribution of this paper is to fill this gap by providing, to the best of my knowledge, the first unified and critical review of the BGL, BCL, CLN, and DM parametrisations for $B$-meson FFs.
This comparison also clarifies the relation between BGL and DM, showing that they are equivalent implementations of the same unitarity information.

A second central theme of this paper is an assumption that is usually left implicit in standard applications of unitarity bounds, namely that subthreshold branch cuts are neglected.
For local $B$-meson FFs, this assumption is strictly justified only for $B\to \pi$, while in generic channels rescattering effects can move the first branch point below the threshold used in the standard conformal map.
Once this happens, the usual BGL and DM bounds are no longer strictly rigorous.
This issue is not merely formal: depending on the channel, it can noticeably affect truncation-error estimates.
To address this issue, I further extend the framework developed together with A.~Gopal in Ref.~\cite{Gopal:2024mgb}, which I refer to as GG.
Ref.~\cite{Gopal:2024mgb} was restricted to $B\to K$ FFs; in the present work, I generalise the formalism to other $B$-meson decays and, in addition, formulate it within a DM-like approach.
The GG framework retains the practical advantages of unitarity-bounded parametrisations while incorporating the correct analytic structure.

A third contribution is crucial for phenomenological applications.
Using the GG framework, I provide precise FF predictions for the main $b\to u$, $b\to s$, and $b\to c$ channels, with machine-readable results intended to serve as practical inputs for phenomenological studies by the community.
\\

The rest of this paper is organised as follows.
In \refsec{analytic}, I define the local FFs, discuss their analytic structure, and derive the dispersion relations and unitarity bounds that underlie the rest of the analysis.
\refsec{review} reviews the main established parametrisations, with particular emphasis on simplified power-series parametrisations, BGL, DM, BCL, and CLN.
In \refsec{G}, I develop the GG construction for the case with subthreshold cuts and study its main theoretical ingredients.
In \refsec{fits}, I use the GG parametrization in combined analyses of $b\to u$, $b\to s$, and $b\to c$ FFs.
\refsec{concl} contains the summary and conclusions.
The appendices collect technical material on helicity amplitudes, weight functions, and outer functions in \refapp{outer}, a note on FF predictions in \refapp{FFcal}, the meson masses and thresholds used in this work in \refapp{meson-masses}, and the endpoint relations for GG coefficients in \refapp{endpointrel}.

\section{Analytic structure and unitarity bounds of form factors}
\label{sec:analytic}

In this section, I first define the local $B$-meson FFs and discuss their analytic properties.
I then show how these properties lead to specific dispersion relations, which provide one of the most powerful tools for extracting model-independent information on the FFs.
Finally, I use these dispersion relations to derive unitarity bounds, which constitute one of the main focuses of this article.

\subsection{Form-factor definitions and analytic structure}
\label{sec:FFdefs}

The prediction of all observables in (rare) semileptonic $B$-meson decays --- including branching ratios and angular observables --- relies on precise knowledge of the relevant $B$-to-meson matrix elements.
Depending on the number of space-time coordinates on which they depend, these matrix elements can be either local, e.g. 
\begin{align}
    \langle \pi(k) | \bar{u}\gamma^\mu b (0) | \bar{B}(p) \rangle
    \equiv
    \langle \pi(k) | \bar{u}\gamma^\mu b | \bar{B}(p) \rangle
    \,,
\end{align}
or non-local, e.g.
\begin{align}
    \label{eq:nonlocBK}
    \int d^4x\, e^{iq\cdot x}
    \bra{\bar{K}(k)} T\big
    \{ 
        \bar{c}\gamma^\mu c(x)
        (\bar c \gamma_\nu P_L c)(\bar s \gamma^\nu P_L b)(0) \big
    \} \ket{\bar{B}(p)} \, .
\end{align}
While charged-current semileptonic $B$ decays only depend on local matrix elements (at leading order in the electroweak interactions), rare $B$ decays also depend on non-local matrix elements.
In this paper, I restrict the discussion to local matrix elements. Treatments of the non-local contributions can be found in Refs.~\cite{Khodjamirian:2010vf,Khodjamirian:2012rm,Gubernari:2020eft,Feldmann:2023plv,Hurth:2025neo}.
\\

In general, hadronic matrix elements are functions of the momenta and polarisation vectors of the mesons involved in the considered process.
It is convenient to factor out the dependence on these four-vectors and work in terms of scalar functions of the squared momentum transfer between the mesons, usually denoted by $q^2$.
Such scalar functions are commonly called hadronic \emph{form factors} (FFs).
I adopt the same definitions as in Ref.~\cite{Ball:2004rg}. 
I define the $\bar{B} \to P$seudoscalar FFs as
\begin{align}
    \langle P(k) |\, J_V^\mu \,| \bar{B}(p) \rangle 
    &=
        \left[ (p + k)^\mu - \frac{m_B^2 - m_P^2}{q^2} q^\mu \right] f_+^{BP}
        + \frac{m_B^2 - m_P^2}{q^2} q^\mu f_0^{BP}, 
    \label{eq:def-fpf0}\\*
    \langle P(k) |\, J_T^\mu \,| \bar{B}(p) \rangle &=
        \frac{i f_T^{BP}}{m_B + m_P} \left[ q^2 (p + k)^\mu - (m_B^2 - m_P^2) q^\mu \right],
    \label{eq:def-fT}
\end{align}
and the $\bar{B} \to V$ector FFs as
\begin{align}
    \langle V(k, \eta) |\, J_V^\mu \,|\bar{B}(p) \rangle &=
        \epsilon^{\mu\nu\rho\sigma} \eta_\nu^* p_\rho k_\sigma \frac{2 \,V^{BV}}{m_B + m_V}, 
    \label{eq:def-V}\\
    \langle V(k, \eta) |\, J_A^\mu \,|\bar{B}(p) \rangle &=
        i \eta_\nu^* \bigg[ g^{\mu\nu} (m_B + m_V) A_1^{BV}
        - (p + k)^\mu q^\nu  \frac{A_2^{BV}}{m_B + m_V} \nonumber\\*
        & \hspace{0cm} - 2 m_V \frac{q^\mu q^\nu}{q^2} \left(\frac{m_B + m_V}{2 \, m_V} \, A_1^{BV} - \frac{m_B - m_V}{2 \, m_V} \, A_2^{BV} - A_0^{BV}\right) \bigg],
    \label{eq:def-A0A1A2}\\
    \langle V(k, \eta) |\, J_T^\mu \,| \bar{B}(p) \rangle &=
        2i \,\epsilon^{\mu\nu\rho\sigma} \eta_\nu^* p_\rho k_\sigma \,  T_1^{BV},
    \label{eq:def-T1} \\
    \langle V(k, \eta) |\, J_{AT}^\mu \,|\bar{B}(p) \rangle & =
         \eta_\nu^* \bigg[ \Big( g^{\mu\nu} (m_B^2 - m_V^2) - (p + k)^\mu q^\nu \Big) T_2^{BV} 
    \nonumber\\*
    & + q^\nu \left( q^\mu - \frac{q^2}{m_B^2 - m_V^2} (p + k)^\mu \right) T_3^{BV} \bigg].
    \label{eq:def-T2T3}
\end{align}
Here, $q_\mu = p_\mu - k_\mu$ and $\eta$ is the polarisation four-vector of the vector meson.
Throughout, I use the following conventions:
\begin{align}
    g_{\mu\nu}
    = {\rm diag} (+,-,-,-)
    \,,\quad
    \sigma_{\mu\nu}
    =
    \frac{i}{2}
    [\gamma_\mu,\gamma_\nu]
    \,,\qquad
    \gamma_5 = i\,\gamma^0 \gamma^1 \gamma^2 \gamma^3 \, ,
    \qquad
    \epsilon^{0123}=-1
    \,.
\end{align}
The quark currents $J_\Gamma^\mu$ --- with $\Gamma = V$ (vector), $A$ (axial vector), $T$ (tensor), and $AT$ (axial tensor) --- used in the FF definitions are
\begin{equation}
\begin{aligned}
    \label{eq:JGamma}
    J_V^\mu & = \bar{\psi}_f \, \gamma^\mu \, \psi_b    \,,&\qquad
    J_A^\mu & = \bar{\psi}_f \, \gamma^\mu \, \gamma_5 \psi_b      \,,\\
    J_T^\mu & = \bar{\psi}_f \, \sigma^{\mu \alpha}q_\alpha  \, \psi_b   \,,&
    J_{AT}^\mu & = \bar{\psi}_f \, \sigma^{\mu \alpha}q_\alpha \gamma_5  \, \psi_b
    \,.
\end{aligned}
\end{equation}
Here $\psi_b$ and $\psi_f$ denote the bottom and final-state quark fields.
Most of the results derived in this work are independent of the specific quark flavours entering the currents $J_{\Gamma}^\mu$, and in particular of the flavour $f$ of the quark field~$\psi_f$.
The notation
\begin{align}
    J_{\Gamma}^\mu (x;f) \equiv J_\Gamma^\mu(x) \,,
    \qquad
    \text{with } f = u,d,c,s ,
\end{align}
is used when the flavour of $\psi_f$ is relevant.
In the formulae above and in the rest of this article, the $q^2$ dependence of the FFs is not shown explicitly unless necessary, e.g. $f_+^{BP}(q^2)\equiv f_+^{BP}$.
In addition, the tensor FFs depend on the renormalisation scale because the tensor and axial-tensor currents have a non-vanishing anomalous dimension.
This scale argument is likewise suppressed in the notation.
Although the definitions in \refeqs{def-fpf0}{def-T2T3} are partly conventional, the number of independent Lorentz structures and FFs is fixed.
Furthermore, it is convenient to work with dimensionless FFs, even though some of the FFs in the commonly used basis of Ref.~\cite{Boyd:1997kz} carry dimensions.

For later use, I also introduce the helicity FFs $A_{12}^{BV}$ and $T_{23}^{BV}$:
\begin{align}
    \label{eq:def-A12}
    A_{12}^{BV} &= \frac{(m_B + m_V)^2 (m_B^2 - m_V^2 - q^2) \, A_1^{BV} - \lamkin(q^2) \, A_2^{BV}}{16 \, m_B m_V^2 (m_B + m_V)}
    \,, \\
    \label{eq:def-T23}
    T_{23}^{BV} &= \frac{(m_B^2 - m_V^2) (m_B^2 + 3 m_V^2 - q^2) \, T_2^{BV} -  \lamkin(q^2) \, T_3^{BV}}{8 \, m_B m_V^2 (m_B - m_V)}\,,
\end{align}
where
\begin{align}
    \lamkin(q^2) 
    := 
    (m_B^2 - m_M^2 - q^2)^2 - 4 \,q^2 m_M^2
    \equiv
    (s_+ - q^2) (s_- - q^2)
\end{align}
is the Källén function, while
\begin{align}
    \label{eq:spm}
    s_\pm \equiv s_\pm^{BM}
    :=
    (m_B \pm m_M)^2 \, ,
\end{align}
with $m_M$ denoting the mass of the final-state meson.
The relations~\eqref{eq:def-A12}-\eqref{eq:def-T23} may be inverted to express the FFs $A_2^{BV}$ and $T_3^{BV}$ in terms of $A_{12}^{BV}$ and $T_{23}^{BV}$.
Accordingly, throughout this work I choose
\begin{align}
    \label{eq:BVFFbasis}
    \left\{
        V^{BV}, A_0^{BV}, A_1^{BV}, A_{12}^{BV}, T_1^{BV}, T_2^{BV}, T_{23}^{BV}
    \right\}
\end{align}
as the basis of independent $B\to V$ FFs.
This choice is motivated in \refapp{outer}.

The absence of kinematic singularities at $q^2=0$ implies three endpoint relations among the FFs: 
\begin{equation}
    \label{eq:endp0}
    f_+^{BP}(0) = f_0^{BP}(0), \quad
    A_{12}^{BV}(0) = \frac{m_B^2 - m_V^2}{8 \, m_B m_V} \, A_0^{BV}(0) , \quad
    T_1^{BV}(0) = T_2^{BV}(0)\,.
\end{equation}
The symmetries of the helicity amplitudes at the kinematic endpoint $q^2 = s_-$ imply two additional endpoint relations~\cite{Hiller:2013cza}:
\begin{align}
    \label{eq:endpsm}
    A_{12}^{BV}(s_-)
    &=
    \frac{m_B^2 - m_V^2}{8 \, m_B m_V} \, A_1^{BV}(s_-)
    \,, \\
    T_{23}^{BV}(s_-)
    &=
    \frac{(m_B + m_V)^2}{4 \, m_B m_V} \, T_2^{BV}(s_-)\,.
\end{align}

The physical region for semileptonic decays, and hence for the $B\to M$ FFs, is
\begin{align}
    \label{eq:smreg}
    0 \le q^2 \le s_- \,,
\end{align}
where lepton masses have been neglected.
The FFs may, however, be analytically continued to the complex $q^2$ plane.
That is, they are well defined for arbitrary values of $q^2$, including complex ones, except at the singularities implied by unitarity and the QCD spectrum.
These singularities take the form of isolated simple poles and branch cuts.\footnote{ 
    In simple terms, a branch cut is a line or curve in the complex plane that ``cuts'' the function's domain in such a way that it becomes single-valued.
    See Ref.~\cite{Ahlfors1966} for a comprehensive introduction to the topic.
}

\begin{figure}[t!]
    \centering
    \includegraphics[scale=0.7]{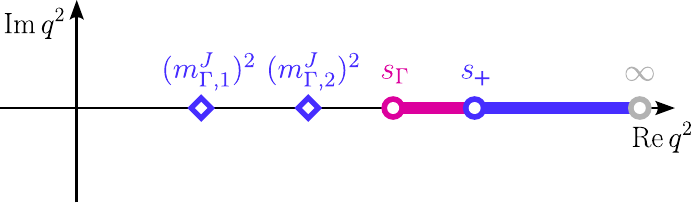}
    \caption{
        Analytic structure of a generic $B\to M$ FF in the complex $q^2$ plane. 
        The FF is analytic everywhere except for isolated poles (two are shown in this example), located at $q^2=(m_{\Gamma,i}^J)^2$, and for branch cuts along the positive real axis. 
        The first branch point occurs at the lowest multi-particle threshold $s_\Gamma$, which can lie below the threshold $s_+$ because of rescattering effects.
        The branch cut then extends to $+\infty$.
        }
    \label{fig:FFstruct}
\end{figure}

The isolated poles are associated with single-particle bound states carrying the same flavour quantum numbers and the same spin-parity quantum numbers as the current entering the FF under consideration.
These masses are denoted by $m_{\Gamma,i}^J$, where $\Gamma$ refers to one of the currents defined in \refeq{JGamma}, $J$ denotes the spin of the state, $i$ is an enumerative index, and the quark transition is understood implicitly.
The values of $m_{\Gamma,i}^J$ used throughout this work are listed in \reftab{states}.
For instance, the FF $V^{BD^*}$ has a pole at $q^2 = (m_{V,1}^1)^2 \equiv m_{B_c^*}^2$.

The branch cuts of the FFs originate from multi-particle intermediate states and correspond to a continuous spectrum starting at the relevant production threshold.
These cuts start at the lowest multi-particle threshold and extend to $+\infty$ along the positive real $q^2$ axis.
One could naively expect that the lowest threshold for a $B\to M$ FF is at $s_+$.
This expectation follows from crossing symmetry:
\begin{align}
    \label{eq:crossym}
    \langle M(k)\,|\,J_\Gamma^\mu\,|\,\bar{B}(p)\rangle
    \;\longleftrightarrow\;
    \langle 0\,|\,J_\Gamma^\mu\,|\,\bar{B}(p)\,\bar{M}(-k)\rangle \, .
\end{align}
In the crossed channel, $q^2$ corresponds to the invariant mass squared of the $\bar{B}\bar{M}$ state produced by the current $J_\Gamma^\mu$.
Consequently, this state can go on shell only for $q^2 \ge s_+$ and generate a branch cut.
However, the lowest threshold corresponds to $s_+$ only for $B\to \pi$ FFs.
In fact, FFs can develop \emph{subthreshold cuts} due to rescattering effects.
For instance, the FF $V^{BD^*}$ exhibits a subthreshold branch cut starting at $(m_{B_c} + m_{\pi})^2$, arising from the rescattering process $BD^* \to B_c \pi$.
This particular cut disappears in the exact isospin limit, because the $J_V^\mu(x;c)$ current is an isosinglet and therefore selects the $I=0$ component of the $BD^*$ channel, whereas the $B_c\pi$ state has 
isospin $I=1$.
Nevertheless, subthreshold cuts do not disappear altogether in the exact isospin limit, since isospin-conserving channels such as $BD$ still generate branch points below $s_+^{BD^*}$, namely at $s_+^{BD}$.

In general, FFs associated with a current $J_\Gamma^\mu(x;f)$ have their lowest multi-particle threshold, denoted by $s_\Gamma$, at
\begin{align}
    \label{eq:sGamma}
    s_V \equiv s_T 
    := 
    (m_{A,1}^0 + m_{\pi})^2
    \,,
    &&
    s_A \equiv s_{AT} := 
    \min\left(
        (m_{A,1}^0 + 2 m_{\pi})^2, (m_{V,1}^1 +  m_{\pi})^2
    \right)
    \,.
\end{align}
Note that $s_V \equiv s_T$ and $s_A \equiv s_{AT}$, since the two currents carry the same quantum numbers.
The analytic structure of a $B\to M$ FF is illustrated in \reffig{FFstruct}.

\subsection{Dispersion relations}
\label{sec:disprel}

\emph{Dispersion relations} have been used since the 1950s to study scattering amplitudes and correlators  in a broad range of contexts; see Refs.~\cite{Caprini:2019osi,Kubis:2025zji} for reviews.
Today, they remain a key tool in particle physics, particularly for the study of hadronic matrix elements.\footnote{
    The term ``dispersion relation'' may sound deceptive and uninformative in the context of quantum field theory. 
    This stems from its origin in an analogy with the Kramers--Kronig relations, which link dispersion and absorption through the real and imaginary parts of generalised indices of refraction~\cite{deL.Kronig:26,kramers1927diffusion,Kubis:2025zji}.
}
In simple terms, a dispersion relation is an identity that connects the real and imaginary parts of a correlator.
It is noteworthy that only a few fundamental principles --- namely causality, analyticity, unitarity and Lorentz invariance --- are used to derive them.

In what follows, I derive a dispersion relation using a simple two-point correlator, i.e., a correlator that depends on two spacetime coordinates.
As an explicit example, I consider the current $J_V^\mu(x;u)$ and define
\begin{align}
    \label{eq:corrDR}
    \Pi_V^{\mu\nu}(q;u) := i \int d^4 x \, e^{iq\cdot x} \bra{0} T\left\{
    J_V^\mu(x;u)\, J_V^{\nu,\dag}(0;u) 
    \right\} \ket{0} 
    \,.
\end{align}
This correlator can be decomposed into two scalar functions, as only two independent Lorentz structures with two indices can be constructed:
\begin{align}
    \label{eq:loreDR}
    \Pi_V^{\mu\nu}(q;u) 
    =
    \left(\frac{q^\mu q^\nu}{q^2} - g^{\mu\nu}\right)\,\Pi_V^1(q^2;u)
    +
    \frac{q^\mu q^\nu}{q^2} \,\Pi_V^0(q^2;u)
    \,.
\end{align}
Here the superscript $J=0,1$ denotes the total angular momentum of the intermediate states contributing to the corresponding correlator.
In other words, $\Pi_V^1$ receives only contributions from states with $J^P=1^-$ and $\Pi_V^0$ receives only contributions from states with $J^P=0^+$, where $P$ is the parity.
Replacing $J_V$ with $J_A$ generates contributions with opposite parity.

\begin{figure}[t!]
    \centering
    \includegraphics[scale=0.7]{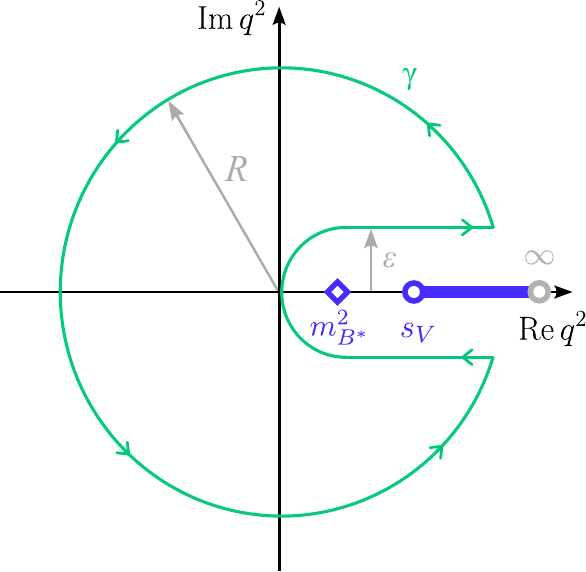}
    \caption{
        Representation of the domain of the correlator $\Pi_V^1(q^2;u)$.
        In blue the singularities of this correlator (a pole and a branch cut).
        In green the contour discussed in the text.
    }
    \label{fig:disp}
\end{figure}

I focus on the correlator $\Pi_V^1$ in this subsection; however, analogous considerations apply to $\Pi_V^0$.
From a mathematical point of view, $\Pi_V^1$ is a holomorphic (or analytic) function of $q^2$.
These correlators are mathematically well-defined at complex values of $q^2$, even though such values have no physical interpretation.
The domain of $\Pi_V^1(q^2;u)$ is the entire complex plane $\mathbb{C}$, excluding a simple pole at $q^2 =  m_{B^*}^2 \equiv (m_{V,1}^1)^2$ --- since the $B^*$ meson has quantum numbers $J^P = 1^-$ --- and a branch cut on the real axis, starting at $q^2= s_V := (m_B + m_{\pi})^2$ and extending to infinity, as shown in \reffig{disp}.
The correlator $\Pi_V^1(q^2;u)$ has the same general analytic structure as, e.g., the FFs $f_+^{B\pi}$ and $V^{B\rho}$, because it is determined by the current $J_V^\mu(x;u)$ (cf. \refsubsec{FFdefs}).
The branch point $s_V$ is only the first in an infinite sequence of branch points, with subsequent
ones appearing, for example, at $q^2= (m_B + 2\,m_{\pi})^2$ and $q^2= (m_B + m_\eta)^2$.
Nevertheless, only the position of the first branch point matters for the derivation of the dispersion relation.

Given the analytic structure outlined above, $\Pi_V^1$ can be represented using Cauchy's integral formula
\begin{align}
    \label{eq:predisp}
    \Pi_V^1(q^2;u) = \frac{1}{2\pi i} \oint_{\gamma} d s \, \frac{\Pi_V^1(s;u)}{s - q^2} \,.
\end{align}
Here, $\gamma$ is a positively oriented simple closed contour such that $\Pi_V^1$ is holomorphic on and inside $\gamma$, and $q^2$ is any point in its interior.
For simplicity, it is often chosen to be $q^2 = 0$.
The contour $\gamma$ can be deformed as long as this deformation does not cross any singularity.
In particular, it can be deformed into the green contour shown in \reffig{disp}, which defines $R$ and $\epsilon$.
If the correlator considered vanishes in the limit $|s| \to \infty$, \refeq{predisp} can be rewritten as
\begin{align}
    \label{eq:prerefl}
    \Pi_V^1(q^2;u) 
    & =
    \frac{1}{2\pi i} \int\limits_{0}^\infty d s \, \frac{\Pi_V^1(s+i\eps;u) - \Pi_V^1(s-i\eps;u)}{s - q^2} 
    \,.
\end{align}
The Schwarz reflection principle can be used to simplify  \refeq{prerefl}, since $\Pi_V^1$ is real-valued on the real axis for $q^2<s_V \land q^2 \neq (m_{V,1}^1)^2$:
\begin{align}
    \label{eq:disprel}
    \Pi_V^1(q^2;u) 
    & =
    \frac{1}{\pi} \int\limits_{0}^\infty d s \, \frac{\Im\,\Pi_V^1(s;u)}{s - q^2 - i\eps} 
    \,.
\end{align}
This identity is commonly referred to as the \emph{dispersion relation}.
In what follows, $\eps$ is omitted unless its presence is explicitly required.

If the correlator does not vanish in the limit $|s| \to \infty$, one encounters divergences and must modify the dispersion relation accordingly.
The simplest and most common way to take care of these divergences is to take derivatives with respect to $q^2$ of \refeq{disprel}:
\begin{align}
    \label{eq:subdisprelspec}
    \left.
        \frac{1}{k!} \left[\frac{\partial}{\partial q^2}\right]^k
        \Pi_V^1(q^2;u) 
    \right|_{q^2=-Q^2}
    & =
    \frac{1}{\pi} \int\limits_{0}^\infty d s \, \frac{\Im\,\Pi_V^1(s;u)}{(s + Q^2)^{k+1}} 
    \,,
\end{align}
where $q^2 = -Q^2$ denotes the subtraction point, whose choice is discussed in \refsubsec{Gdep}.
This procedure is also known as taking \emph{subtractions}, as it is equivalent to subtracting the first terms of the Taylor expansion of $\Pi_V^1$ around $Q^2$~\cite{Colangelo:2000dp,Kubis:2025zji}.
The number of subtractions $k$ may be chosen freely, provided that the r.h.s.\ of \refeq{subdisprelspec} converges.
The minimal value of $k$ required for convergence is not known a priori and depends on the correlator under consideration.
It can, however, be determined from a perturbative evaluation of the correlator.
For example, in the case of $\Pi_V^1$, one finds that the r.h.s.\ of \refeq{subdisprelspec} already converges for $k=2$.
In principle, one may also perform more subtractions than the minimal number required.

The above procedure to derive the dispersion relation for $\Pi_V^1$ can be easily extended to other correlators.
In particular, it can be shown that the following identity holds for the two-point correlators of the currents defined in \refeq{JGamma}:
\begin{mybox}
\begin{center}
    \textbf{(Subtracted) dispersion relation}
\end{center}
\vspace*{-0.5cm}
\begin{align}
    \label{eq:subdisprel}
    \chi_\Gamma^J(Q^2;k) 
    :=
    \left.
        \frac{1}{k!} \left[\frac{\partial}{\partial q^2}\right]^k
        \Pi_\Gamma^J(q^2) 
    \right|_{q^2=-Q^2}
    & =
    \frac{1}{\pi} \int\limits_{0}^\infty d s \, \frac{\Im\,\Pi_\Gamma^J(s)}{(s + Q^2)^{k+1}} 
    \,.
\end{align}
\end{mybox}
\noindent
For later convenience, I introduce the function $\chi_\Gamma^J$, which is commonly referred to in the literature as the \emph{susceptibility}.

Although \refeq{subdisprel} always holds for the currents in \refeq{JGamma} and for correlators of the form \refeq{corrDR}, two principal complications may arise when considering more general correlators.
First, the correlator may be complex-valued along the real axis, even in the region where it is analytic.
In this scenario, the Schwarz reflection principle cannot be invoked, and an expression analogous to \refeq{prerefl} must be employed instead of \refeq{disprel}.
Second, the correlator may exhibit a more intricate analytic structure.
For example, when considering a correlator that depends on more than two spacetime points, anomalous branch cuts may arise~\cite{Lucha:2006vc,Colangelo:2015ama,Mutke:2024tww}.
Unlike unitarity cuts, anomalous branch cuts are not constrained to lie along the real axis. In such cases, the contour $\gamma$ must be carefully defined to circumvent these singularities.
Neither of these situations is discussed further in this paper.

\subsection{Unitarity bounds}
\label{sec:UB}

\hi{Unitarity bounds, also known as dispersive bounds, constrain hadronic matrix elements and, consequently, the corresponding FFs.}
The first pioneering articles on unitarity bounds date back to the sixties~\cite{Meiman:1963,Geshkenbein1963}.
In the following decade, these bounds were applied to the study of kaon decays~\cite{Okubo:1971jf,Okubo:1972ih,Bourrely:1980gp}.
It was not until the nineties that they were extended to the analysis of $B$-meson decays (see, e.g., Refs.~\cite{deRafael:1992tu,deRafael:1993ib,Boyd:1994tt,Boyd:1995cf}). 

The unitarity bounds are derived from the same first principles used to obtain the dispersion relations in the previous subsection, i.e. causality, analyticity, unitarity and Lorentz invariance.
This makes them particularly valuable, as few results in the context of QCD have been derived from first principles.
\\

In the remainder of this subsection, I review the derivation of the unitarity bound for the FF $f_+^{B\pi}$ defined in \refeq{def-fpf0}.  
This example should allow readers unfamiliar with unitarity bounds to understand the general methodology and reasoning underlying their construction.
The derivation of the unitarity bounds for the other FFs is entirely analogous, and some details are given in \refapp{outer}.

The unitarity bounds for $f_+^{BP}$ and $f_0^{BP}$ can be derived starting from the following correlator
\begin{equation}
\begin{aligned}
    \label{eq:corrUB}
    \Pi_V^{\mu\nu}(q) & = 
    i \int d^4 x \, e^{iq\cdot x} \bra{0} T\left\{
    J_V^\mu(x)\, J_V^{\nu,\dag}(0) 
    \right\} \ket{0} 
    \\
    & =
    \left(\frac{q^\mu q^\nu}{q^2} - g^{\mu\nu}\right)\,\Pi_V^1(q^2)
    +
    \frac{q^\mu q^\nu}{q^2} \,\Pi_V^0(q^2)
    \,.
\end{aligned}
\end{equation}
This correlator is the same as the one defined in \refeqs{corrDR}{loreDR}, but here I keep the flavour of the non-$b$ quark generic, rather than fixing $f = u$.
As shown in the previous subsection, the scalar function $\Pi_V^1$ satisfies a doubly subtracted dispersion relation.
Similarly, one can show that $\Pi_V^0$ satisfies a singly subtracted dispersion relation, as given by \refeq{subdisprel} with $k = 1$.
Note that the second line of \refeq{corrUB} can be easily inverted
\begin{align}
    \label{eq:Pilambda}
    \Pi_\Gamma^J(q^2) 
    =
    \P_{\mu\nu}^{J}(q) \,
    \Pi^{\mu\nu}_{\Gamma}(q)
    \,,
\end{align}
where
\begin{align}
    \label{eq:proj}
    \P_{\mu\nu}^{1}(q) & = \frac{1}{D-1} \left(\frac{q_\mu q_\nu}{q^2} - g_{\mu\nu}\right)\,,&&
    \P_{\mu\nu}^{0}(q)  = \frac{q_\mu q_\nu}{q^2}\,,
\end{align}
and $D$ denotes the number of spacetime dimensions.

\hi{The general idea underlying the unitarity bounds is to calculate $\chi_\Gamma^J$ and, independently, to express the same quantity, $\chi_\Gamma^J$, in terms of hadronic matrix elements using unitarity.
By equating these two representations, the hadronic matrix elements can be constrained by the calculated value of $\chi_\Gamma^J$.}

\enlargethispage{\baselineskip} 

\paragraph{Calculation of $\bs{\chi_\Gamma^J}$} 
The function $\chi_\Gamma^J(Q^2;k)$ can be calculated using an operator product expansion (OPE) for values of $Q^2$ far below the threshold for bound state formation --- specifically, when~\cite{Boyd:1997kz}
\begin{align}
    \label{eq:Q2OPE}
    (m_b + m_f)^2 + Q^2
    \gg
    (m_b + m_f)\Lambda_\QCD
    \,.
\end{align}
Here $\Lambda_\QCD\sim 300\,\MeV$ is the QCD scale parameter.
This condition is typically satisfied at $Q^2 = 0$, which is the most common choice found in the literature.
This is because OPE calculations are simpler at $Q^2 = 0$, as they involve one fewer scale.
Alternative choices of $Q^2$ are explored in \refsubsec{Gdep}.
The OPE for $\chi_\Gamma^J$  takes the form~\cite{Wilson:1969zs,Wilson:1970ag,Wilson:1973jj}
\begin{align}
    \label{eq:OPEchi}
    \chi_\Gamma^J(Q^2;k) \Big|_\OPE
    = 
    \sum_i C_{\Gamma,i}^J (Q^2;k,\mu) \bra{0} O_i(\mu) \ket{0} 
    \,,
\end{align}
where $\mu$ denotes the renormalisation scale, $C_{\Gamma,i}^J$ are the Wilson coefficients, and the index $i$ labels operators $O_i$ according to their mass dimension.
The OPE in \refeq{OPEchi} offers a systematic method to separate contributions from short- and long-distance scales, with $\mu$ serving as the scale that distinguishes these regions. 
The scale $\mu$ should be chosen conveniently, as physical observables are independent of it. 
In the calculation of $\chi_\Gamma^J$ in $b$ physics, the most common choice is to set $\mu = m_b$.

The matrix elements $\bra{0} O_i(\mu) \ket{0}$ are known as \emph{QCD vacuum condensates} and correspond to the vacuum expectation values of normal-ordered local products of QCD fields, namely quark and gluon fields.
The first term in the sum of \refeq{OPEchi} is trivial, as the only operator with mass dimension zero is the identity:
\begin{align}
    \label{eq:O0}
    \bra{0} O_0(\mu) \ket{0} 
    \equiv
    \bra{0} \mathbbm{1} \ket{0}
    =1
    \,.
\end{align}
The next non-vanishing contributions arise from the dimension-three and dimension-four operators:
\begin{align}
    O_3 & = \bar{\psi}\psi
    \,,&
    O_4 & = \frac{\alpha_s}{\pi} G_{\mu\nu}^a G_a^{\mu\nu}\,,
\end{align}
where $G_a^{\mu\nu}$ denotes the gluon field strength tensor.
The vacuum expectation values of $O_3$ and $O_4$ correspond to the quark and gluon condensates, respectively.
Condensates of higher dimensional operators are discussed, for example, in Refs.~\cite{Reinders:1984sr,Colangelo:2000dp,Khodjamirian:2020btr}.

While the Wilson coefficients can be calculated perturbatively, obtaining reliable numerical values for the condensates is considerably more challenging, except for the trivial case of \refeq{O0}.
For this reason, the leading-power term in \refeq{OPEchi} is commonly referred to as the \emph{perturbative contribution}, since it is entirely determined within perturbation theory.
QCD vacuum condensates have been determined using lattice QCD and QCD sum rules, but their determinations remain affected by relatively large uncertainties~\cite{FLAG:2021npn,Ioffe:2002ee}.
Nevertheless, this does not compromise the accuracy of the calculation of the $\chi_\Gamma^J$ functions.
In fact, the contribution of operators with dimension three or higher to $\chi_\Gamma^J$ is usually small.
For example, it is below $2\%$ for $b\to u,d,s$ currents~\cite{Bharucha:2010im} and below $0.5\%$ for $b\to c$ currents~\cite{Boyd:1997kz}.
Therefore, the dominant source of theoretical uncertainty originates from the missing higher-order $\alpha_s$ corrections in the calculation of the leading-power Wilson coefficient $C_{\Gamma,0}^J$.
This Wilson coefficient has been computed at next-to-leading order (NLO) in $\alpha_s$ for all currents in \refeq{JGamma}, retaining the full $Q^2$ dependence~\cite{Djouadi:1993ss,Generet:2025hsv}.
For the vector and axial-vector currents, $J_V^\mu$ and $J_A^\mu$, it has furthermore been calculated at next-to-next-to-leading order (NNLO) at $Q^2=0$~\cite{Grigo:2012ji}.

Recently, lattice QCD computations of $\chi_\Gamma^J$ have been performed~\cite{Martinelli:2021frl,Melis:2024wpb,Harrison:2024iad}.
The most recent of these calculations, namely that of Ref.~\cite{Harrison:2024iad}, agrees with the OPE results.
This is expected, as $\chi_\Gamma^J$ is dominated by the perturbative contribution.
Minor tensions with the OPE results are observed in Refs.~\cite{Martinelli:2021frl,Melis:2024wpb}.

\enlargethispage{\baselineskip} 

\paragraph{Representation of $\bs{\Im\, \Pi_\Gamma^J}$ via unitarity}
The imaginary part of $\Pi_\Gamma^J$ can be expressed in terms of hadronic matrix elements using unitarity.
This can be achieved by inserting a complete set of hadronic states $H$ (including both single- and multi-particle configurations) into the correlator:
\begin{align}
    \label{eq:ImPi}
    \Im\,\Pi_\Gamma^{J}(s + i \eps) 
        = \frac{1}{2} 
        \P_{\mu\nu}^{J}(q)
        \sum \hspace*{-0.52cm} \int\limits_H d\tau_H (2\pi)^4 \delta^{(4)}(p_H - q)
            \braket{0 | J_\Gamma^{\mu} | H}\braket{H | J_\Gamma^{\dagger,\nu} | 0}\Big|_{q^2=s} 
    \,.
\end{align}
Here, $\sum_H \hspace*{-0.67cm} \int \hspace*{0.3cm} d\tau_H$ denotes the sum over all possible hadronic states $H$, together with the integral over their corresponding phase space.
Clearly, only the states $H$ with the same quantum numbers as the current $J_\Gamma^{\mu}$ give a nonzero contribution.
Equation~\eqref{eq:ImPi} follows from the optical theorem, and hence from the unitarity of the $S$-matrix.
A detailed derivation of this equation can be found in, e.g., Appendix~B of Ref.~\cite{Khodjamirian:2020btr}.

To illustrate how $\Im\,\Pi_\Gamma^J$ can be expressed in terms of hadronic matrix elements, I again use the correlator $\Pi_V^1(q^2;u)$ as an example.
I begin by considering the case in which $H$ is a one-particle state.
Only QCD-stable one-particle states, i.e. states with squared masses below $s_\Gamma$, contribute to the one-particle part of~\refeq{ImPi}.
Indeed, unstable resonances contribute through the multi-particle continuum and should not be added as independent one-particle terms, in order to avoid double counting.
For instance, the axial-vector $B$ meson listed by the PDG as $B_1(5721)$, with mass $m_{B_1}\equiv m_{A,1}^1 =5.726\,\GeV$, can decay into a $B^*\pi$ final state~\cite{PDG:2024cfk}.
Its contribution to this channel is therefore already contained in the $B^*\pi$ spectral density, and adding a separate one-particle term for the same resonance would amount to double counting.
For the correlator $\Pi_V^1(q^2;u)$, the only QCD-stable one-particle contribution is the $B^*$ state, whose mass satisfies $m_{B^*} \equiv m_{V,1}^1 < m_B + m_{\pi}$.
Defining the $B^*$ decay constant as
\begin{align}
    \braket{0 | J_V^\mu | \bar{B}^*(q,\eta)}
    & = m_{B^*} f_{B^*} \eta^\mu
    \,,
\end{align}
one readily finds
\begin{align}
    \chi_V^1(Q^2;k,u) \Big|_{\rm 1pt}
    = 
    \frac{m_{B^*}^2 f_{B^*}^2}{\left(m_{B^*}^2+Q^2\right)^{k+1}}
    \,.
\end{align}
Recall that the $B$ state contributes only to $\Im\, \Pi_A^0(q^2;u)$, as it is a pseudoscalar meson.

To express the generic one-particle part of \refeq{ImPi} in a notation consistent with the masses $m_{\Gamma,i}^J$ introduced in \refsubsec{FFdefs}, I denote by $B_{\Gamma,1}^J$ the lightest one-particle state in the channel labelled by $\Gamma\in\{V,A\}$ and spin $J$.
The corresponding decay constants are defined by
\begin{equation}
\begin{aligned}
    \label{eq:def-dec-const}
    \bra{0} J_V^\mu \ket{\bar{B}_{V,1}^0(q)}
    &=
    i \, q^\mu f_{V,1}^0 \,,
    &
    \bra{0} J_A^\mu \ket{\bar{B}_{A,1}^0(q)}
    &=
    i \, q^\mu f_{A,1}^0 \,,
    \\
    \bra{0} J_V^\mu \ket{\bar{B}_{V,1}^1(q,\eta)}
    &=
    m_{V,1}^1 \eta^\mu f_{V,1}^1 \,,
    &
    \bra{0} J_A^\mu \ket{\bar{B}_{A,1}^1(q,\eta)}
    &=
    m_{A,1}^1 \eta^\mu f_{A,1}^1 \,,
    \\
    \bra{0} J_T^\mu \ket{\bar{B}_{V,1}^1(q,\eta)}
    &=
    i \, (m_{V,1}^1)^2 \eta^\mu f_{T,1}^1 \,,
    &\qquad
    \bra{0} J_{AT}^\mu \ket{\bar{B}_{A,1}^1(q,\eta)}
    &=
    i \, (m_{A,1}^1)^2 \eta^\mu f_{AT,1}^1 \,.
\end{aligned}
\end{equation}
Substituting these matrix elements into \refeq{ImPi} gives the lowest one-particle contributions in the various channels:
\begin{equation}
\begin{aligned}
    \label{eq:chi1pt}
    \chi_V^0(Q^2;k) \Big|_{\rm 1pt}
    &=
    \frac{(m_{V,1}^0)^2 (f_{V,1}^0)^2}{\left((m_{V,1}^0)^2+Q^2\right)^{k+1}}
    \,,
    &
    \chi_A^0(Q^2;k) \Big|_{\rm 1pt}
    &=
    \frac{(m_{A,1}^0)^2 (f_{A,1}^0)^2}{\left((m_{A,1}^0)^2+Q^2\right)^{k+1}}
    \,,
    \\
    \chi_V^1(Q^2;k) \Big|_{\rm 1pt}
    &=
    \frac{(m_{V,1}^1)^2 (f_{V,1}^1)^2}{\left((m_{V,1}^1)^2+Q^2\right)^{k+1}}
    \,,
    &
    \chi_A^1(Q^2;k) \Big|_{\rm 1pt}
    &=
    \frac{(m_{A,1}^1)^2 (f_{A,1}^1)^2}{\left((m_{A,1}^1)^2+Q^2\right)^{k+1}}
    \,,
    \\
    \chi_T^1(Q^2;k) \Big|_{\rm 1pt}
    &=
    \frac{(m_{V,1}^1)^4 (f_{T,1}^1)^2}{\left((m_{V,1}^1)^2+Q^2\right)^{k+1}}
    \,,
    &
    \chi_{AT}^1(Q^2;k) \Big|_{\rm 1pt}
    &=
    \frac{(m_{A,1}^1)^4 (f_{AT,1}^1)^2}{\left((m_{A,1}^1)^2+Q^2\right)^{k+1}}
    \,.
\end{aligned}
\end{equation}
As discussed above, only those contributions that do not lead to double counting should be included.

The lightest two-particle state contributing to~\refeq{ImPi} in the case of $\Pi_V^1(q^2;u)$ is $H=\bar{B}\bar{\pi}$.
Using crossing symmetry~\eqref{eq:crossym} together with the FF definitions in~\refeq{def-fpf0} and \refeq{ImPi}, one finds
\begin{align}
    \label{eq:chiBpi}
    \chi_V^1(Q^2;k,u) \Big|_{B\pi}
    =&\, \frac{\kappa_I}{48 \pi^2} \int\limits_{s_+^{B\pi}}^\infty ds\, \frac{ \lamkin^{3/2}(s)}{s^2 (s + Q^2)^{k+1}}
        \, \left|f_+^{B\pi}(s)\right|^2 
    \,,
\end{align}
where $\kappa_I$ is the isospin Clebsch--Gordan factor, whose value is given below.
Further details on the derivation of \refeq{chiBpi} are given in \refapp{outer}.
The FF $f_0^{B\pi}$ contributes only to $\Im\, \Pi_V^0(q^2;u)$.
There is an infinite tower of two-particle states that gives a non-zero contribution, such as $\bar{B}^*\bar{\pi}$, $\bar{B}\bar{\rho}$, and $\Lambda_b \bar{p}$.
Again, care must be taken to avoid double counting; for instance, if the state $\bar{B}\bar{\pi}\pi$ is included, the contribution from $\bar{B}\bar{\rho}$ should not be considered separately.
In practice, this issue does not arise, since three-particle matrix elements are typically poorly known and therefore cannot be reliably included in \refeq{ImPi}.

It can be shown that any two-particle contribution to $\chi_\Gamma^J$ may be expressed as
\begin{align}
    \label{eq:defW} 
    \chi_\Gamma^J(Q^2;k) \Big|_{BM}
    & =
    \frac{1}{\pi}
    \sum_\F
    \int\limits_{s_+}^\infty ds\,
    W_\F(s)\,
    \left|\F(s)\right|^2
    \,,
\end{align}
where the sum runs over the FFs contributing to the $BM$ channel for the specified current $\Gamma$ and spin $J$.
The weight function $W_\F$ can be defined as
\begin{align} 
    \label{eq:WBGL}
    W_\F(s) :=
    \frac{\kappa_I }{K\pi} 
    \left|(s-s_+)^{\frac{a}{2}}(s-s_-)^{\frac{b}{2}} \right|
    s^{-(c+3)}
    \left(\frac{s}{s+Q^2}\right)^d 
    \,,
\end{align}
and the isospin Clebsch--Gordan factor takes the values~\cite{Boyd:1997kz}
\begin{align}
    \label{eq:kappaI}
    \kappa_I =
    \begin{cases}
    \dfrac{3}{2}, & \text{for } B \to \pi,\rho,\\[4pt]
    2, & \text{for } B \to K^{(*)},D^{(*)},\\[4pt]
    1, & \text{for } B_s \to K^{(*)},\phi,D_s^{(*)}.
    \end{cases}
\end{align}
The values of the parameters $K$, $a$, $b$, $c$, and $d$ depend on the FF considered and can be found in \reftab{outer} in \refapp{outer}.
For instance, comparing \refeq{chiBpi} with \refeqs{defW}{WBGL} for $f_+^{BP}$ with the minimal number of subtractions required $k=2$, one readily obtains $K=48$, $a=3$, $b=3$, $c=2$, and $d=3$.

\paragraph{Unitarity bound}
The unitarity bound is derived by equating the OPE calculation of $\chi_\Gamma^J$ with its representation in terms of hadronic matrix elements.
For the correlator $\Pi_V^1(q^2;u)$, one has
\begin{align}
    \label{eq:pre-Bound}
    \chi_V^1(Q^2;k,u)\Big|_{\OPE}
    =
    \chi_V^1(Q^2;k,u)\Big|_{\rm 1pt}
    +
    \chi_V^1(Q^2;k,u)\Big|_{B\pi}
    +
    \dots
    \,,
\end{align}
where the ellipsis denotes contributions from additional hadronic states.
Since these additional contributions arise from squared moduli of matrix elements, they are manifestly positive. As a result, \refeq{pre-Bound} can be recast as the inequality
\begin{align}
    \label{eq:impl-BoundBpi}
    \chi_V^1(Q^2;k,u)\Big|_{\OPE}
    \geq
    \chi_V^1(Q^2;k,u)\Big|_{\rm 1pt}
    +
    \chi_V^1(Q^2;k,u)\Big|_{B\pi}
    \,.
\end{align}
Taking, for example, $f_{B^*}$ from Ref.~\cite{Lubicz:2017asp} and the OPE value of $\chi_V^1$ from Ref.~\cite{Grigo:2012ji}, one obtains a bound on the FF $f_+^{B\pi}$.
Equation~\eqref{eq:impl-BoundBpi} can be generalised to the two-point correlators of the currents defined in \refeq{JGamma}:
\begin{mybox}
\begin{center}
    \textbf{Unitarity bound}
\end{center}
\vspace*{-0.5cm}
\begin{align}
    \label{eq:impl-Bound}
    \chi_\Gamma^J(Q^2;k)\Big|_{\OPE}
    \geq
    \sum_{\rm 1pt}
    \chi_\Gamma^J(Q^2;k)\Big|_{\rm 1pt}
    +
    \sum_{BM}
    \chi_\Gamma^J(Q^2;k)\Big|_{BM}
    \,.
\end{align}
\end{mybox}
\noindent
This bound is known as the unitarity bound.
A few comments on this inequality are in order:
\begin{itemize}
    
    \item 
    The sums extend over all one-particle states that may be included, as well as over all $B$-meson FFs that yield a non-vanishing contribution.
    Although one may in principle consider a single process at a time, it is important to emphasise that the unitarity bounds become increasingly stringent when additional hadronic contributions are included simultaneously, since the inequality then moves closer to saturation.

    \item 
    To further increase the saturation of the bound, one may also include $\Lambda_b$ decays and states of higher multiplicity, i.e. three-particle, four-particle, and higher contributions.
    Such contributions are beyond the scope of the present work; for related discussions, see Refs.~\cite{Blake:2022vfl,Amhis:2022vcd,Herren:2025cwv}.

    \item 
    Since strictly positive contributions on the r.h.s. have been neglected, the symbol $\geq$ could safely be replaced by $>$.
    Yet this distinction makes no practical difference for FF analyses.

    \item 
    The constraining power of the unitarity bound depends on both $Q^2$ and $k$; this point is discussed in \refsubsec{Gdep}.
    
\end{itemize}
In summary, analyticity and unitarity turn current-current correlators into positive, model-in\-de\-pen\-dent constraints on the $B$-meson FFs, encoded in \refeq{impl-Bound}.
However, this bound cannot be used directly in phenomenological applications, since it only constrains the modulus squared of the FFs integrated from $s_+$ to infinity.
To make practical use of \refeq{impl-Bound}, one must introduce specific conformal mappings and FF parametrisations.
These are discussed in the next section.

\section{Review of established form-factor parametrisations}
\label{sec:review}

The two principal QCD-based methods for predicting FFs are lattice QCD and QCD light-cone sum rules (LCSRs).
Their main features, limitations, and complementarity are discussed in \refapp{FFcal}.
Regardless of the method employed, FFs are typically determined only at discrete values of $q^2$.
Therefore, a parametrisation of the FFs is required to interpolate and extrapolate them consistently over the entire semileptonic region~\eqref{eq:smreg}.

In \refsubsec{SPPpar}, I introduce the simplified power-series parametrisations (SPPs), which, as their name suggests, constitute the simplest class of FF parametrisations.
The problem is that, in practice, only a finite number of parameters can be determined. 
As a consequence, the series must be truncated, and the associated truncation uncertainty of the SPPs cannot be quantified in a model-independent way.
This issue is resolved by \emph{unitarity-bounded} parametrisations, which permit a rigorous estimate of the truncation error.
I discuss bounded parametrisations in \refsubsec{UBpar}.
Finally, there exists a class of parametrisations that makes use of unitarity bounds but does not allow for a rigorous estimate of the truncation error, thereby introducing a degree of model dependence.
I call these parametrisations \emph{quasi-bounded} and present them in \refsubsec{NUBpar}.

A potentially confusing point is that SPPs and quasi-bounded parametrisations can have very similar functional forms.
For example, the SPP proposed in Ref.~\cite{Bharucha:2015bzk} closely resembles the quasi-bounded parametrisation of Ref.~\cite{Bourrely:2008za}.
The distinction in the terminology adopted here is primarily operational rather than algebraic: SPPs are intended to be used without imposing unitarity constraints, whereas quasi-bounded parametrisations supplement a similar series expansion with such constraints.
Thus, according to this classification, any SPP can in principle be turned into a quasi-bounded parametrisation by deriving and imposing the corresponding unitarity constraint on its coefficients.

I do not consider parametrisations based on relatively strong model assumptions, such as the simple ans\"atze of Refs.~\cite{Becirevic:1999kt,Ball:2004ye,Ball:2006jz} or parametrisations built on specific implementations of Omn\`es representations~\cite{Flynn:2006vr,Albertus:2005ud}.
Such parametrisations were especially useful in the early stages of FF analyses, as they typically involve only a small number of parameters.
However, in view of the much larger body of lattice QCD and LCSR results now available, modern analyses favour more flexible and systematically improvable $z$-expansion parametrisations~\cite{Bharucha:2010im,Gambino:2019sif,Bernlochner:2019ldg,Gubernari:2023puw,Simula:2025fft}.

\subsection{Simplified power-series parametrisations}
\label{sec:SPPpar}

SPPs provide a convenient and flexible description of the $q^2$ dependence of the FFs, while involving only a limited number of parameters and a compact analytical form.
For this reason, they are widely used in phenomenological analyses and often serve as a useful benchmark against which more sophisticated parametrisations can be compared.
At the same time, their simplicity comes at the price of reduced theoretical control, most notably because \hi{the truncation error associated with the SPPs cannot be quantified in a model-independent way.}
A more detailed discussion of this point is given in \refsubsec{NUBpar}.

\subsubsection{Simplified $\bs{q^2}$ parametrisation}

Since FFs are functions of $q^2$ (cf.~\refsubsec{FFdefs}) and can exhibit singularities only for $q^2 > s_-$, they admit a Taylor expansion around $q^2 = 0$ that converges throughout the semileptonic region:
\begin{align}
    \label{eq:SPPq2}
    \F(q^2) = \sum_{n=0}^\infty \alpha_{\F,n}^{\text{SPP}} \left( \frac{q^2}{m_B^2} \right)^n .
\end{align}
Here, $\F$ denotes a FF in $B$ decays.
It is convenient to divide $q^2$ by $m_B^2$ to have a dimensionless expansion parameter and coefficients of~$\O(1)$.
Clearly, this choice is purely conventional.
Technically, one could instead normalise the expansion parameter by the squared mass of the Sun, with $M_{\odot} \approx 1.116 \times 10^{57}\,\mathrm{GeV}$, at the cost of obtaining extremely large higher-order coefficients.

The coefficients of any parametrisation are determined through a statistical analysis that uses the theoretical predictions for the FFs as constraints, and in some cases also experimental data.
Accordingly, series such as the one in \refeq{SPPq2} must be truncated at a finite order $N$, since only a finite number of parameters can be determined in this manner.
As mentioned already, since the full form of $\F$ is unknown, it is not possible to accurately estimate the truncation error of SPPs.

SPPs in terms of $q^2$ have been widely used in kaon physics (see, e.g., Refs.~\cite{Bourrely:2005hp,Caprini:2019osi} and the references therein) and for non-local FFs in $B\to K^{(*)}\ell^+\ell^-$ decays (see, e.g., Refs.~\cite{Jager:2012uw,Ciuchini:2015qxb}).

\subsubsection{Simplified $\bs{z}$ parametrisations}

A common strategy in mathematics and physics is to apply transformations that map a problem to a different domain, where it may admit a simpler formulation or solution.
In the case of hadronic FFs, it has proven convenient to introduce the following conformal mapping~\cite{Meiman:1963,Okubo:1971jf,Caprini:2019osi}:
\newcommand{\zmapeq}{%
    q^2 \mapsto z(q^2) 
    \equiv 
    z(q^2; s_0, s_\th) 
    := 
    \frac{
        \sqrt{s_\th-q^2}-\sqrt{s_\th-s_0^{\phantom{1}}}
    }{
        \sqrt{s_\th-q^2}+\sqrt{s_\th-s_0^{\phantom{1}}}
    }
}
\begin{align}
    \label{eq:zmap}
    \zmapeq
    \,.
\end{align}
This transformation conformally maps the complex domain $\mathbb{C}\setminus[s_\th, \infty)$ (i.e. the complex \(q^2\)-plane cut from the branch point \(s_\th\) to \(+\infty\)) onto the open unit disc $|z|<1$ in the complex $z$-plane.
The parameter $s_\th$ is usually taken as $s_\th = s_+$, where $s_+$ is defined in \refeq{spm}.
In \refsubsec{Gpar}, I explain why the choice $s_\th=s_\Gamma$ is required for a strictly rigorous treatment when subthreshold cuts are present.
The parameter $s_0$ can be chosen freely, provided that it satisfies $s_0<s_\th$.
The choice of $s_0$ determines the point on the $q^2$ plane that maps to the origin of the $z$ plane, i.e. $z(s_0) = 0$.
By imposing the condition $z(s_-) = -z(0)$, one chooses $s_0$ such that the semileptonic region is mapped symmetrically around the origin in the $z$ plane. 
This minimises the maximum value of $|z|$ over this region and therefore improves the convergence properties of the truncated $z$ expansion presented below. 
One obtains
\begin{align} 
    \label{eq:s0opt}
    \sopt := 
    s_\th \left(
        1-\sqrt{1-\frac{s_-}{s_\th}} 
    \right) \, .
\end{align}
For $s_\th = s_+$, this can be written as $\sopt = (m_B + m_M) \left(\sqrt{m_B} - \sqrt{m_M}\right)^2$.

\begin{figure}[t!]
    \centering
    \includegraphics[width=0.75\textwidth]{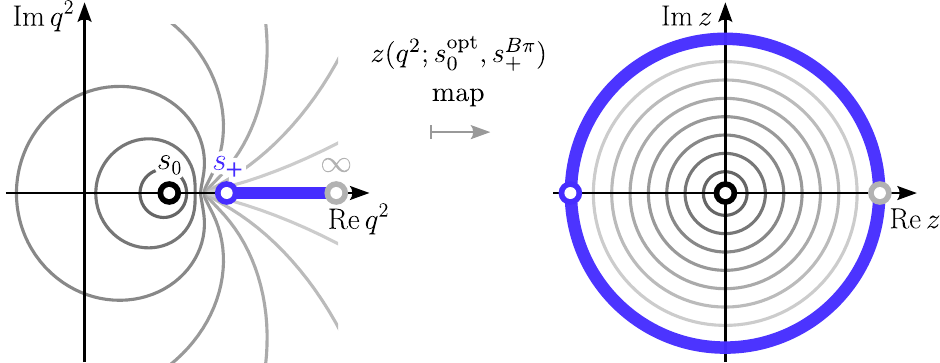}
    \caption{
        Illustration of the conformal transformation \(z(q^2;\sopt,s_+^{B\pi})\) defined in \refeq{zmap}. 
        The point \(q^2=\sopt\) is sent to the origin, \(z=0\), while the branch point \(q^2=s_+^{B\pi}\) maps to \(z=-1\) and the point at infinity maps to \(z=1\). 
        The grey curves represent families of contours in the \(q^2\)-plane and their images, which appear as circles in the \(z\)-plane.
        The branch cut is mapped to the boundary \(|z|=1\), highlighted in blue.
    }
    \label{fig:zmap1}
\end{figure}

An illustration of the mapping~\eqref{eq:zmap} for $B\to \pi$ FFs is shown in \reffig{zmap1}.
In this case, I choose $s_\th = s_+^{B\pi}$, which corresponds to the first branch point of these FFs (cf.~\refsubsec{FFdefs}), and~$s_0 = \sopt$.
\\

Based on these considerations, the FFs may be expressed as a Taylor series in $z$:
\begin{align}
    \label{eq:SPPz}
    \F(q^2) = \sum_{n=0}^\infty \beta_{\F,n}^{\text{SPP}} \, z(q^2; s_0, s_\Gamma)^n .
\end{align}
Neglecting simple poles for the moment --- which can be removed straightforwardly, see \refsubsec{UBpar} --- the series in the equation above converges for $|z|<1$.
At this point, the advantage of $z$ expansions over $q^2$ expansions should be evident.
The $z$ map~\eqref{eq:zmap} compresses the entire complex $q^2$ plane ---  excluding the branch cut --- onto the unit disc, thereby translating the analytic structure of the FFs into a particularly simple domain.
This implies that the FFs can be described by a single parametrisation over their entire domain, rather than only over a restricted part of it.
In addition, the semileptonic region is mapped onto a small interval in $z$, which improves the convergence of truncated series.

A widely used SPP is the one proposed by A. Bharucha, D. Straub, and R. Zwicky (BSZ) in Ref.~\cite{Bharucha:2015bzk}:
\begin{align}
    \label{eq:BSZ}
    \F(q^2) = \frac{1}{1 - \frac{q^2}{\left(m_{\Gamma,1}^J\right)^2}}\sum_{n=0}^\infty \alpha_{\F,n}^{\rm BSZ} \left( z(q^2; \sopt, s_+) -  z(0; \sopt, s_+) \right)^n .
\end{align}
For a given FF, the pole mass appearing in the prefactor is to be identified with the lowest-lying state carrying the corresponding quantum numbers $J$ and $\Gamma$ of that FF (cf. \reftab{outer}).
The masses $m_{\Gamma,1}^J$ are listed in \reftab{states}. 
The role of the prefactor is to factor out the contribution of the lowest-lying pole. This improves the analytic properties of the remaining function and leads to a faster convergence of the series.

It should be noted that the choice of expansion variable, $z(q^2; \sopt, s_+) - z(0; \sopt, s_+)$, is somewhat peculiar, as it does not amount to an expansion about the origin in the $z(q^2; \sopt, s_+)$ plane.
Consequently, the radius of convergence is in general smaller than $1$ (cf. \refsubsec{UBpar}).
While this does not invalidate the parametrisation, it partially defeats the purpose of introducing the conformal map in \refeq{zmap} and of adopting the choice $s_0=\sopt$, whose rationale is precisely to maximise the benefits of the mapping to the unit disc.
Put differently, the expansion variable $z(q^2; \sopt, s_+) - z(0; \sopt, s_+)$ is nearly equivalent to $z(q^2; 0, s_+)$ and is therefore a suboptimal choice. 

Motivated by the same underlying idea as Ref.~\cite{Bourrely:2008za} (see also \refsubsec{NUBpar}), I propose the following SPP:
\begin{mybox}
\begin{center}
    \textbf{Optimised SPP}
\end{center}
\vspace*{-0.5cm}
\begin{align}
    \label{eq:OP}
    \F(q^2) = \frac{1}{1 - \frac{q^2}{\left(m_{\Gamma,1}^J\right)^2}}\sum_{n=0}^\infty \alpha_{\F,n}^{\rm OP} 
    \, z(q^2; \sopt, s_\Gamma)^n .
\end{align}
\end{mybox}
\noindent
The threshold $s_\Gamma$ is defined in \refeq{sGamma}.
The expansion in \refeq{OP} is centred at $z=0$ and thus makes direct use of the conformal map~\eqref{eq:zmap} and the optimal choice $s_0=\sopt$.
The use of $s_\Gamma$ rather than $s_+$ ensures that the position of the relevant branch point is correctly encoded in the parametrisation.
Therefore, \refeq{OP} constitutes a particularly well-motivated choice of SPP.
\hi{I recommend its use in FF analyses for which a careful assessment of the truncation error is not required.
It is equally well suited for analyses in which unitarity bounds are not meant to be imposed, such as the presentation of lattice QCD results.}
In that context, it is generally preferable not to combine pure lattice-QCD results with unitarity bounds, because the latter introduce additional theoretical assumptions and correlations that may obscure the direct interpretation of the lattice calculation itself.
Furthermore, if such results are subsequently combined with other calculations in a FF analysis with unitarity bounds, double counting could arise.

\subsection{Unitarity-bounded parametrisations}
\label{sec:UBpar}

FF parametrisations depend on a finite number of parameters once the corresponding series expansions are truncated.
Unitarity bounds can provide a controlled way of estimating the effect of the neglected higher-order terms and may, in certain cases, further restrict the allowed parameter space.
In this sense, parametrisations based on unitarity bounds --- which I refer to as \emph{unitarity-bounded} --- are model-independent, since the truncation is a controlled approximation rather than a model assumption.
Neglecting, for the time being, the issue of subthreshold branch cuts (cf. \refsubsec{FFdefs}), which are addressed in \refsec{G}, there are only two unitarity-bounded parametrisations.
These are the Boyd--Grinstein--Lebed (BGL) parametrisation~\cite{Boyd:1997kz} and the Dispersive Matrix (DM) interpolation~\cite{Bourrely:1980gp,Lellouch:1995yv}.
The remainder of this subsection is devoted to a review of both approaches, followed by a comparison between them.

\subsubsection{BGL parametrisation}

\hi{The BGL parametrisation retains the main practical advantage of the SPPs, namely a compact expansion in a small variable, while adding a crucial ingredient: the coefficients are constrained by first principles.}
The BGL parametrisation reads
\begin{mybox}
\begin{center}
    \textbf{BGL parametrisation}
\end{center}
\vspace*{-0.5cm}
\begin{align}
    \label{eq:BGL}
    \F(q^2)
    =
    \frac{1}{\B_\Gamma^J(q^2;s_+)\,\phi_\F(q^2)}
    \sum_{n=0}^\infty \alpha_{\F,n}^{\rm BGL}\,
    z(q^2;s_0,s_+)^n .
\end{align}
\end{mybox}
\noindent
The $z$ map is defined in \refeq{zmap}, whereas the Blaschke product is defined as
\begin{align}
    \label{eq:Bprod}
    \B_\Gamma^J(q^2)
    \equiv
    \B_\Gamma^J(q^2;s_\th) 
    := 
    \prod_{i:\,(m_{\Gamma,i}^J)^2 < s_\th}
    z(q^2;(m_{\Gamma,i}^J)^2,s_\th)
    \,.
\end{align}
Each Blaschke factor $z(q^2;(m_{\Gamma,i}^J)^2,s_\th)$ is included to factor out a simple pole of $\F$ below $s_\th$, with $s_\th=s_+$ in the present case (cf.~\refsubsec{FFdefs}).
As a consequence, the product $\B_\Gamma^J(q^2;s_\th) \F(q^2)$ is free of simple poles for $|z|<1$.
The Blaschke product may equivalently be written as
\begin{align}
    \B_\Gamma^J(q^2;s_\th) 
    \equiv 
    \prod_{i:\,(m_{\Gamma,i}^J)^2 < s_\th}
    \frac{
        z(q^2;s_0,s_\th) 
        -
        z((m_{\Gamma,i}^J)^2;s_0,s_\th)
    }{
        1 - z(q^2;s_0,s_\th) z((m_{\Gamma,i}^J)^2;s_0,s_\th)
    }
    \,.
\end{align}
The masses $m_{\Gamma,i}^J$ are collected in \reftab{states}.

The \emph{outer function} $\phi_\F$ is a normalising function coming from the derivation of the unitarity bounds and it is defined as
\begin{equation}
\begin{aligned}
    \label{eq:outerBGL}
    \phi_\F (s)
    & := 
    \sqrt{\frac{\kappa_I^{\phantom{n}}}{K \pi \widetilde{\chi}_\Gamma^J(Q^2;d-1)}} \,
    \left( \frac{s_+ - s}{s_+ - s_0} \right)^{\frac 1 4}
    \left( \sqrt{s_+ - s} + \sqrt{s_+ - s_0} \right)
    \\* 
    & \times 
    \left( s_+ - s \right)^{\frac a 4} 
    \left( \sqrt{s_+ - s} + \sqrt{s_+ - s_-} \right)^{\frac b 2}
    \left( \sqrt{s_+ - s} + \sqrt{s_+} \right)^{-(c+3)}
    \\* 
    & \times
    \left(
    \frac{
        \sqrt{s_+ - s} + \sqrt{s_+}
    }{
         \sqrt{s_+ - s} + \sqrt{s_+ + Q^2}
    }
    \right)^d
    \,.
\end{aligned}
\end{equation}
The values of the isospin Clebsch--Gordan factor $\kappa_I$ can be found in~\refeq{kappaI}.
The values of the parameters $K$, $a$, $b$, $c$, and $d$ depend on the FF considered and can be found in \reftab{outer} in \refapp{outer}.
The values of $\chi_\Gamma^J$ at $Q^2=0$, evaluated with the minimal number of subtractions $d-1$, are given in \refsubsec{X}.
For generic $Q^2$ and number of subtractions, $\chi_\Gamma^J$ can be computed from the ancillary files attached to the arXiv version of Ref.~\cite{Generet:2025hsv}.

The BGL coefficients are subject to the unitarity bound
\begin{mybox}
\begin{center}
    \textbf{BGL unitarity bound}
\end{center}
\vspace*{-0.5cm}
\begin{align}
    \label{eq:UBBGL}
    \sum_{n=0}^\infty \left|\alpha_{\F,n}^{\rm BGL}\right|^2 \leq 1 \, .
\end{align}
\end{mybox}
\noindent
This conservative bound is obtained by retaining only the considered $BM$ contribution in \refeq{impl-Bound} and neglecting all other positive hadronic contributions.
Including several channels simultaneously would lead to a stronger correlated bound.
\hi{The BGL unitarity bound has an important consequence.
If the series in \mbox{\refeq{BGL}} is truncated at order $N$, the neglected remainder is bounded and can be estimated.}
Indeed, using the inequality~\mbox{\eqref{eq:UBBGL}} together with the Cauchy--Schwarz inequality, one obtains
\begin{align}
\begin{aligned}
    \label{eq:BGLtruncerr}
    \left|\sum_{n>N}^\infty\alpha_{\mathcal F,n}^{\rm BGL} z^n\right|
    & \le
    \left(\sum_{n>N}^\infty|\alpha_{\mathcal F,n}^{\rm BGL}|^2\right)^{1/2}
    \!\left(\sum_{n>N}^\infty|z|^{2n}\right)^{1/2}
    \\
    & \le
    \left(1-\sum_{n=0}^N|\alpha_{\mathcal F,n}^{\rm BGL}|^2\right)^{1/2}
    \!\left(\sum_{n>N}^\infty|z|^{2n}\right)^{1/2}
    \\
    & =
    \left(1-\sum_{n=0}^N|\alpha_{\mathcal F,n}^{\rm BGL}|^2\right)^{1/2}
    \frac{|z|^{N+1}}{\sqrt{1-|z|^2}}
    \le
    \frac{|z|^{N+1}}{\sqrt{1-|z|^2}}
    \,,
\end{aligned}
\end{align}
which tends to zero as $N\to\infty$ for any $|z|<1$.

The BGL parametrisation has been extensively used in the literature, both in phenomenological analyses, see e.g. Refs.~\cite{Grinstein:2017nlq,Bigi:2017njr,Bernlochner:2019ldg,Gambino:2019sif,Biswas:2022yvh}, and in experimental analyses, see e.g. Refs.~\cite{Belle:2018ezy,LHCb:2020cyw,Belle:2023xgj}.
This is due to the fact that this parametrisation is relatively simple to implement  while providing rigorous theoretical control within its domain of validity.
Moreover, it has served as the starting point for several other parametrisations, including the one introduced in \refsec{G} and some of those reviewed in \refsubsec{NUBpar}.
Nevertheless, the BGL parametrisation has an important limitation: it implicitly neglects subthreshold branch cuts.
In \refsec{G}, I show how this limitation can be overcome within a BGL-like framework.
\\

In what follows, I review the derivation of \refeq{BGL} and the unitarity bound~\eqref{eq:UBBGL}.
Starting from \refeq{impl-Bound}, one readily obtains
\begin{align}
    \label{eq:chit}
    \widetilde{\chi}_\Gamma^J(Q^2;d-1)
    :=
    \chi_\Gamma^J(Q^2;d-1)\Big|_{\OPE}
    -
    \sum_{\rm 1pt}
    \chi_\Gamma^J(Q^2;d-1)\Big|_{\rm 1pt}
    \geq
    \chi_\Gamma^J(Q^2;d-1)\Big|_{BM}
    \,.
\end{align}
Here, I have fixed the number of subtractions to the minimal one required, that is $k=d-1$, and, without loss of generality, considered only a single two-particle contribution.
Using \refeq{defW} and the $z$ mapping defined in \refeq{zmap}, one obtains
\begin{equation}
\begin{aligned}
    \label{eq:BGLder}
    1
    \geq
    \frac{1}{\pi}
    \int\limits_{s_+}^\infty ds\,
    \frac{W_\F(s)}{\widetilde{\chi}_\Gamma^J(Q^2;d-1)}\,
    \left|\F(s)\right|^2
    & =
    \frac{1}{\pi}
    \int\limits_{s_+}^\infty ds\,
    \left|\frac{dz(s;s_0,s_+)}{ds}\right|
    \left|\B_\Gamma^J(s) \phi_\F(s)\F(s)\right|^2
    \\
    & =
    \frac{1}{2\pi i}
    \oint\limits_{|z|=1}
    \frac{dz}{z}\,
    \left|
        \B_\Gamma^J(s)\,
        \phi_\F(s)\,
        \F(s)
    \right|_{s=s(z)}^2
    \\
    & =
    \frac{1}{2\pi }
    \int\limits_0^{2\pi}
    d\theta\,
    \left|
        \B_\Gamma^J(s)\,
        \phi_\F(s)\,
        \F(s)
    \right|_{s=s(z),\,z=e^{i\theta}}^2
    \,.
\end{aligned}
\end{equation}
The factor of $2$ in the second line arises because the physical cut $s \ge s_+$ is mapped onto only one semicircle of $|z|=1$, whereas the contour integral extends over the full unit circle, with the upper and lower semicircles giving equal contributions.
Note that the Blaschke product defined in \refeq{Bprod} satisfies $|\B_\Gamma^J(s)|=1$ on the cut.
The modulus squared of the outer functions has to satisfy
\begin{align}
    \label{eq:phiBGLconstr}
    \left|\phi_\F(s)\right|^2
    =
    \frac{W_\F(s)}{\widetilde{\chi}_\Gamma^J(Q^2;d-1)}
    \left|\frac{dz(s;s_0,s_+)}{ds}\right|^{-1}
    \qquad
    \text{for } s>s_+
    \,.
\end{align}
The outer functions must be constructed such that their absolute values satisfy this equation, while preserving the analytic properties of the corresponding FFs.
Details of the construction of the outer functions, i.e. of the derivation of \refeq{outerBGL}, are given in \refapp{outer}.

Once the outer functions have been determined, it is natural to introduce
\begin{align}
    \label{eq:gBGL}
    \G_\F(z)
    :=
    \left.
        \B_\Gamma^J(s)\,
        \phi_\F(s)\,
        \F(s)
    \right|_{s=s(z)}
    \,.
\end{align}
By construction, $\G_\F$ is analytic for $|z|<1$.
Indeed, the Blaschke product cancels the subthreshold poles of $\F$, and the outer function is analytic and non-vanishing in the open unit disc.
Therefore, $\G_\F$ admits a Taylor expansion around the origin,
\begin{align}
    \label{eq:gBGLexp}
    \G_\F(z)
    =
    \sum_{n=0}^\infty \alpha_{\F,n}^{\rm BGL}\, z^n
    \,,
\end{align}
which is simply \refeq{BGL} rewritten in terms of $z$.

To derive the bound on the coefficients, it is convenient to equip the unit disc with the inner product
\begin{align}
    \label{eq:BGLinner}
    \langle f,g\rangle
    :=
    \frac{1}{2\pi}
    \int\limits_0^{2\pi} d\theta\,
    f(e^{i\theta})\, g(e^{i\theta})^*
    \,.
\end{align}
With this choice, the monomials in $z$ are orthonormal:
\begin{align}
    \label{eq:BGLortho}
    \langle z^n, z^m \rangle
    =
    \frac{1}{2\pi}
    \int\limits_0^{2\pi} d\theta\,
    e^{i(n-m)\theta}
    =
    \delta_{nm}
    \,,
\end{align}
where $z=e^{i\theta}$.
Hence the functions $1,z,z^2,\ldots$ form a complete set of orthonormal polynomials on the unit circle.
This is the reason why the BGL parametrisation naturally takes a monomial form.
Substituting \refeqa{gBGL}{gBGLexp} into \refeq{BGLder} and using \refeq{BGLortho}, one obtains the bound~\eqref{eq:UBBGL}:
\begin{align}
\begin{aligned}
    \label{eq:UBBGLder}
    1\geq\frac{1}{2\pi i}
    \oint\limits_{|z|=1}
    \frac{dz}{z}\,
    \left|\G_\F(z)\right|^2
    &=
    \frac{1}{2\pi}
    \int\limits_0^{2\pi} d\theta\,
    \left|
        \sum_{n=0}^\infty
        \alpha_{\F,n}^{\rm BGL}
        e^{i n \theta}
    \right|^2
    \\
    &=
    \sum_{n,m=0}^\infty
    \alpha_{\F,n}^{\rm BGL}
    \left(\alpha_{\F,m}^{\rm BGL}\right)^*
    \langle z^n, z^m \rangle
    \\
    &=
    \sum_{n=0}^\infty
    \left|\alpha_{\F,n}^{\rm BGL}\right|^2
    \,.
\end{aligned}
\end{align}

\subsubsection{Dispersive Matrix method}

The DM method provides an alternative to the BGL parametrisation for exploiting the unitarity bound~\eqref{eq:impl-Bound} together with the $z$ mapping~\eqref{eq:zmap}.
Although this approach is often attributed mainly to L. Lellouch~\cite{Lellouch:1995yv}, most of its key elements were already developed by C. Bourrely, B. Machet, and E. de Rafael~\cite{Bourrely:1980gp}.
It was subsequently first applied to $B$ decays by G. Boyd, B. Grinstein, and R. Lebed in Ref.~\cite{Boyd:1994tt} --- a recurring trio in the context of unitarity bounds --- and was further developed in Ref.~\cite{Lellouch:1995yv}.
After receiving comparatively limited attention for about 26 years, this approach was revived in Ref.~\cite{DiCarlo:2021dzg}, where it was introduced under the name ``Dispersive Matrix'' method.
That work was followed by a series of further studies by the same authors and close collaborators, who substantially developed and refined the method; see, for instance, Refs.~\cite{Martinelli:2021myh,Simula:2023ujs,Fedele:2023ewe,Martinelli:2023fwm}.

The DM approach is not a parametrisation in the traditional sense, but an interpolation framework. 
As in Lagrange interpolation, the DM interpolation is constructed using the FF values at selected $q^2$ points, from which the allowed behaviour of the FF in the semileptonic region is inferred.
The key difference from Lagrange interpolation is that the DM method is not simply a polynomial interpolation.
Analyticity and the unitarity bounds restrict the allowed interpolating functions, so instead of yielding a unique polynomial, DM defines an allowed band that collapses to the exact input values at the interpolation points.
\hi{Therefore, unlike BGL, whose parameters are expansion coefficients, DM is determined by the values of the FF at selected \hbox{$q^2$} points and uses analyticity and unitarity to constrain the allowed band elsewhere.}
\\

The conceptual starting point of the DM method is exactly the same as the BGL parametrisation: the relevant object is the normalised analytic function $\G_\F$ introduced in \refeq{gBGL}, which satisfies the unitarity bound~\eqref{eq:impl-Bound} that can be written as (cf. \refeqa{BGLder}{BGLinner}) 
\begin{align}
    \label{eq:boundcalG}
    \langle \G_\F,\G_\F\rangle \leq 1
    \,.
\end{align}
The difference appears only in the way this bound is exploited.
The BGL parametrisation expands $\G_\F$ in the orthonormal basis $1,z,z^2,\dots$ and therefore translates unitarity into a bound on expansion coefficients.
The DM method instead starts from the information that is usually available in practice, i.e. the values of the FF at a discrete set of kinematic points, and asks what values are still allowed by unitarity at another point.
This is why the DM method is a pointwise interpolation-and-bounding method rather than an expansion in free coefficients.
Its logic is easiest to follow in three steps: first one introduces a function --- namely a reproducing kernel --- that evaluates $\G_\F$ at a point, then one builds a positive matrix using the known inputs, and finally one solves the resulting quadratic inequality.

Throughout the derivation of the DM method, the variable $s$ is used in place of $q^2$ to simplify the notation.
For any point $s<s_+$, let
\begin{align}
    z_s := z(s;s_0,s_+)
    \,.
\end{align}
Because $s<s_+$, the variable $z_s$ is real.
Suppose also that the values of the FF are known at $N$ points $s_1,\dots,s_N$ below threshold.
In applications, these are typically the $q^2$ values at which lattice QCD or LCSR calculations are available.
At those points one knows the input values
\begin{align}
    z_i := z(s_i;s_0,s_+)
    \,,\qquad
    \G_\F(z_i)
    =
    \B_\Gamma^J(s_i)\,\phi_\F(s_i)\,\F(s_i)
    \qquad (i=1,\dots,N)
    \,.
\end{align}
To constrain the FF at another point $s$, one wants to determine the allowed values of
\begin{align}
    \G_\F(z_s)
    =
    \B_\Gamma^J(s)\,\phi_\F(s)\,\F(s)
    \,.
\end{align}
Following Refs.~\cite{Bourrely:1980gp,Lellouch:1995yv,DiCarlo:2021dzg}, I introduce the functions\footnote{
    Technically, $g_s$ is the reproducing kernel of the Hardy space.
    The Hardy space on the unit disc can be defined, in simple terms, as the space of analytic functions on $|z|<1$ whose norm induced by the inner product~\eqref{eq:BGLinner} is finite.
    The function $\G_\F$ belongs to this space because it is analytic in the disc and satisfies \refeq{boundcalG}.
    The kernel is the special function that evaluates a Hardy-space function at a point via the inner product.
    In general one has $K(z,w)=1/(1-z\bar w)$; since $z_s$ is real for $s<s_+$, this reduces here to \refeq{DMkernel}.
}
\begin{align} 
    \label{eq:DMkernel}
    g_s(z)
    &:=
    \frac{1}{1-z_s z}
    \,,
    &
    g_i (z)
    &:= 
    \frac{1}{1-z_i z}    
    \,.
\end{align}
Using the inner product~\eqref{eq:BGLinner} and Cauchy's theorem, one finds
\begin{align}
    \label{eq:DMrepr}
    \langle \G_\F,g_s\rangle = \G_\F(z_s)
    \,,\qquad
    \langle g_i,g_s\rangle = \frac{1}{1-z_i z_s}
    \,.
\end{align}
Therefore, taking the inner product between $\G_\F$ and $g_s$ simply extracts the value of $\G_\F$ at the point $z_s$.
This is why the DM method is naturally formulated in terms of the kernels $g_i$ and the corresponding values $\G_\F(z_i)$.
They are directly tied to pointwise input, whereas BGL is naturally written in terms of the monomials $1,z,z^2,\ldots$ and their expansion coefficients.

To proceed, one considers the set of vectors
\begin{align}
    \G_\F,\ g_s,\ g_1,\dots,g_N
    \,.
\end{align}
Positivity of the inner product implies that the corresponding Gram matrix, that is the matrix of all pairwise inner products of these vectors, is positive semidefinite and therefore has non-negative determinant:
\begin{align}
    \label{eq:DMGram}
    \det
    \begin{pmatrix}
        \langle \G_\F,\G_\F\rangle & \langle \G_\F,g_s\rangle & \langle \G_\F,g_1\rangle & \cdots & \langle \G_\F,g_N\rangle \\
        \langle g_s,\G_\F\rangle & \langle g_s,g_s\rangle & \langle g_s,g_1\rangle & \cdots & \langle g_s,g_N\rangle \\
        \langle g_1,\G_\F\rangle & \langle g_1,g_s\rangle & \langle g_1,g_1\rangle & \cdots & \langle g_1,g_N\rangle \\
        \vdots & \vdots & \vdots & \ddots & \vdots \\
        \langle g_N,\G_\F\rangle & \langle g_N,g_s\rangle & \langle g_N,g_1\rangle & \cdots & \langle g_N,g_N\rangle
    \end{pmatrix}
    \ge 0
    \,.
\end{align}
Moreover, since $\langle \G_\F,\G_\F\rangle \leq 1$, the unknown entry $\langle \G_\F,\G_\F\rangle$ may be replaced by its upper bound $1$ while preserving a valid inequality.
The reason is that the determinant is linear in its first entry, with coefficient given by the lower-right principal minor, which is non-negative and strictly positive for $s\neq s_i$.
After making this replacement and evaluating the scalar products (cf. \refeq{DMrepr}), one obtains
\begin{align}
    \label{eq:DMmatrix}
    \det
    \begin{pmatrix}
        1 & \G_\F(z_s) & \G_\F(z_1) & \cdots & \G_\F(z_N) \\
        \G_\F(z_s) & \dfrac{1}{1-z_s^2} & \dfrac{1}{1-z_s z_1} & \cdots & \dfrac{1}{1-z_s z_N} \\
        \G_\F(z_1) & \dfrac{1}{1-z_1 z_s} & \dfrac{1}{1-z_1^2} & \cdots & \dfrac{1}{1-z_1 z_N} \\
        \vdots & \vdots & \vdots & \ddots & \vdots \\
        \G_\F(z_N) & \dfrac{1}{1-z_N z_s} & \dfrac{1}{1-z_N z_1} & \cdots & \dfrac{1}{1-z_N^2}
    \end{pmatrix}
    \ge 0
    \,.
\end{align}
This determinant condition is the core of the DM method.
The first row and column contain the unknown function $\G_\F(z_s)$, while the remaining entries are fixed by the chosen input data.
The inequality~\eqref{eq:DMmatrix} therefore implies that $\G_\F(z_s)$, and hence $\F(s)$, must be compatible both with unitarity and with the input values $\{\G_\F(z_i)\}$.
The same inequality can also be derived by reformulating the problem as an extremal interpolation problem and solving it by means of Lagrange multipliers~\cite{Abbas:2010jc}.

The next step is simple in principle.
In \refeq{DMmatrix}, all quantities are known except $\G_\F(z_s)$.
Therefore the determinant is a quadratic polynomial in $\G_\F(z_s)$.
If that quadratic has no real solution, the input data are incompatible with analyticity and unitarity.
If it does have real solutions, the allowed values of $\G_\F(z_s)$ form an interval, which immediately translates into a bound on $\F(s)$.
%
%
In the physically relevant case considered here, the points $s$ and $s_i$ lie on the real axis below threshold and the corresponding quantities $\G_\F(z_s)$ and $\G_\F(z_i)$ are real.
The Gram matrix in \refeq{DMmatrix} is then real symmetric, as are its principal submatrices.
The lower-right block of \refeq{DMmatrix} is a Cauchy matrix, whose determinant is
\begin{align}
    \det\left(\frac{1}{1-z_i z_j}\right)_{i,j=1,\dots,N}
    =
    \prod_{i=1}^N \frac{1}{1-z_i^2}
    \prod_{1\le i<j\le N}
    \left(
        \frac{z_i-z_j}{1-z_i z_j}
    \right)^2
    >
    0
    \,.
\end{align}
This positivity guarantees that the quadratic coefficient is strictly negative, so the allowed region is indeed a finite interval.
The explicit evaluation of the minors then leads to compact closed-form expressions for the DM bounds.

Introducing the weights
\begin{align}
    \label{eq:DMweights}
    \delta(s)
    &:=
    \prod_{m=1}^N
    \frac{1-z_s z_m}{z_s-z_m}
    \,,\\
    \delta_j
    &:=
    \prod_{\substack{m=1\\m\neq j}}^N
    \frac{1-z_j z_m}{z_j-z_m}
    \qquad (j=1,\dots,N)
    \,,
\end{align}
together with the $s$-independent combination
\begin{align}
    \label{eq:DMchi}
    \Xi_\F
    :=
    \sum_{i,j=1}^N
    \G_\F(z_i)\,\G_\F(z_j)\,\delta_i \delta_j\,
    \frac{(1-z_i^2)(1-z_j^2)}{1-z_i z_j}
    \,,
\end{align}
it is possible to solve the quadratic inequality for $\G_\F(z_s)$.
After dividing by $\B_\Gamma^J\,\phi_\F$, one obtains the following compact expression valid for $q^2<s_-$:
\begin{mybox}
\begin{center}
    \textbf{DM method}
\end{center}
\vspace*{-0.5cm}
\begin{align}
    \label{eq:DMpar}
    &
    \F_{\rm lo}^{\rm DM}(q^2) \leq  \F(q^2) \leq  \F_{\rm up}^{\rm DM}(q^2)
    \,,
    &&
    \text{where}
    &&
    \F_{\rm lo/up}^{\rm DM}(q^2)
    =
    \frac{
        \beta_\F(q^2)
        \mp
        \sqrt{\gamma_\F(q^2)}
    }{
        \B_\Gamma^J(q^2)\,\phi_\F(q^2)
    }
    \,.
    &
\end{align}
\end{mybox}
\noindent
For convenience, I have introduced the functions
\begin{align}
    \beta_\F(s)
    &:=
    \frac{1}{\delta(s)}
    \sum_{j=1}^N
    \G_\F(z_j)\,\delta_j\,
    \frac{1-z_j^2}{z_s-z_j}
    \,,
    \label{eq:DMbeta}
    \\
    \gamma_\F(s)
    &:=
    \frac{1-\Xi_\F}{(1-z_s^2)\,\delta(s)^2}
    \,.
    \label{eq:DMgamma}
\end{align}

These formulae make the structure of the method transparent.
The quantity $\beta_\F(s)$ is the part fixed directly by the input points, while the square-root term measures how much freedom remains once unitarity has been imposed.
The condition
\begin{align}
    \Xi_\F \le 1
    \label{eq:DMfilter}
\end{align}
is the DM \emph{unitarity filter}.
If it is violated, no analytic function satisfying the unitarity bound can pass through the chosen input values.
Crucially, this filter depends only on the input data and not on the point $s$ where the FF is being evaluated.
From \refeqs{DMpar}{DMgamma} one sees that the DM framework is determined by the input values \(\G_\F(z_i)\), or equivalently by the values of the FF \(\F(s_i)\) at a chosen set of kinematic points.

A practical property of the DM construction is that it interpolates the inputs exactly.
Indeed, when $s\to s_j$, one has $\delta(s)\sim \delta_j(1-z_j^2)/(z_s-z_j)$, so that $\gamma_\F(s)\to 0$ and both bounds collapse to
\begin{align}
    \F_{\rm lo}^{\rm DM}(s_j)
    =
    \F_{\rm up}^{\rm DM}(s_j)
    =
    \F(s_j)
    \,.
\end{align}
This exact interpolation is often emphasised because it is not automatically guaranteed by a truncated BGL fit when the number of retained coefficients is smaller than the number of input points~\cite{DiCarlo:2021dzg}.
Although one might think that this is a strength of the DM method, in practice it can be a limitation of the framework. 
Exact interpolation is not a unique conceptual advantage of DM, but simply a consequence of choosing the input values themselves as the basic variables of the construction.
With a sufficiently flexible ansatz, a BGL representation can likewise be made to reproduce the same input points exactly as shown below.
This sample-by-sample exact interpolation can make the analysis highly inefficient, because the unitarity filter~\eqref{eq:DMfilter} rejects a large fraction of the samples.
As a result, special procedures may become necessary~\cite{Martinelli:2023fwm}.
For this reason, exact interpolation should be regarded as a technical feature of the DM construction rather than as a decisive practical advantage.

At this stage, it should be apparent that the DM method should not be interpreted as a first-principles method for calculating FFs. 
Rather, like the BGL parametrisation, it is a model-independent framework that exploits analyticity, unitarity, and external input values of the FFs to constrain them over the full kinematical range.

\subsubsection{Comparison between the BGL parametrisation and the DM method}

The BGL parametrisation and the DM interpolation are based on exactly the same ingredients: the conformal $z$ mapping to the unit disc~\eqref{eq:zmap}, the removal of subthreshold poles through a Blaschke product~\eqref{eq:Bprod}, the outer function~\eqref{eq:outerBGL} fixed by the weight function~\eqref{eq:WBGL}, and the same unitarity inequality for the normalised analytic function \(\G_\F\)~\eqref{eq:boundcalG}.
The difference lies only in how this common unitarity information is implemented.
BGL expands \(\G_\F\) in the orthonormal basis \(1,z,z^2,\dots\) and turns unitarity into the bound~\eqref{eq:UBBGL} on the expansion coefficients, whereas DM evaluates \(\G_\F\) at selected $q^2$ points and turns the same bound into a positivity condition on a Gram matrix~\eqref{eq:DMmatrix}.
Thus, BGL is naturally formulated in coefficient space, while DM is naturally formulated in terms of FF values at given points.

This makes the relation between the two methods conceptually straightforward.
DM does not exploit additional physical input beyond that used in BGL, nor does it lead to a stronger first-principles constraint.
Rather, it solves a different mathematical problem within the same framework: given a set of input values \(\F(s_i)\), it determines the range of values still allowed at another point \(q^2\).
The DM band may be viewed as the envelope of all admissible BGL-like representations that satisfy the same unitarity bound and pass through the chosen input points.
The distinction between the two approaches is not conceptual, but lies solely in the variables employed and in the way the construction is implemented.

\hi{The considerations above imply that the BGL parametrisation and the DM interpolation must yield equivalent results in FF analyses, provided that both are implemented consistently and based on the same physical input.}
Any differences observed in practice should then be traced not to a different first-principles content, but to the specific implementation, such as the truncation adopted in BGL, the choice of interpolation points in DM 
or the statistical treatment of the input uncertainties and correlations.
They may also depend on the fitting framework adopted in the analysis, such as a frequentist or Bayesian approach, and on the implementation of the unitarity filter.
Nevertheless, these differences should not lead to significant changes in the final FF results.

A direct analytical comparison between the two methods is not possible.
This is because they are not formulated in terms of the same variables.
A truncated BGL analysis is characterised by a finite set of coefficients \(\alpha_{\F,n}^{\rm BGL}\), whereas the DM method determines, point by point, the extremal range allowed by the input values and the bound~\eqref{eq:boundcalG}.
For a given set of inputs, many different sets of BGL coefficients can reproduce the same interpolation points, while DM retains only the corresponding extremal band.
As a result, a direct analytic comparison would require solving an optimisation problem over infinitely many BGL coefficients. 
What can be compared instead are specific numerical implementations, in particular once the BGL truncation order and the FF input points are fixed.
In what follows, I present a numerical comparison of the two methods based on a few examples.
\\

For the comparison between the BGL and DM approaches, I do not carry out a complete FF fit for a given process (see, e.g., Refs.~\cite{Simula:2023ujs,Martinelli:2023fwm,Juettner:2025dif}).
Instead, I adopt a simpler strategy that is sufficient for the present purpose and makes the comparison more transparent.
The aim is not to perform a state-of-the-art determination of the $B\to K$ FFs, but to compare how the BGL and DM constructions propagate exactly the same FF pointwise input.
As an illustrative example, I take the central values of the $B\to K$ FFs from Ref.~\cite{Parrott:2022rgu} at three representative points in the high-$q^2$ region:
\begin{align}
    \label{eq:BKq2points}
    \{q_1^2,q_2^2,q_3^2\}
    :=
    \{15.26~\GeV^2, 19.08~\GeV^2, s_-=22.90~\GeV^2 \}
    \,.
\end{align}
These values can be generated straightforwardly with the ancillary \texttt{Python} code accompanying the arXiv version of that work.

\begin{figure}[t!]
    \centering
    \begin{subfigure}{0.48\textwidth}
        \centering
        \includegraphics[width=\textwidth]{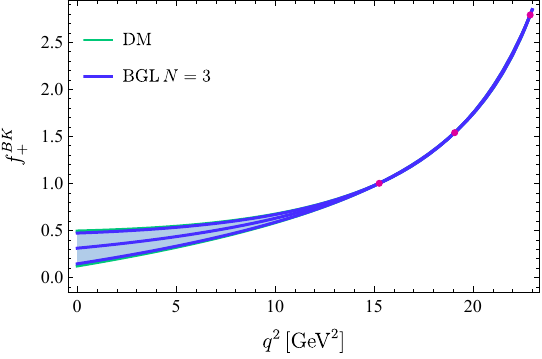}
    \end{subfigure}
    \hfill
    \begin{subfigure}{0.48\textwidth}
        \centering
        \includegraphics[width=\textwidth]{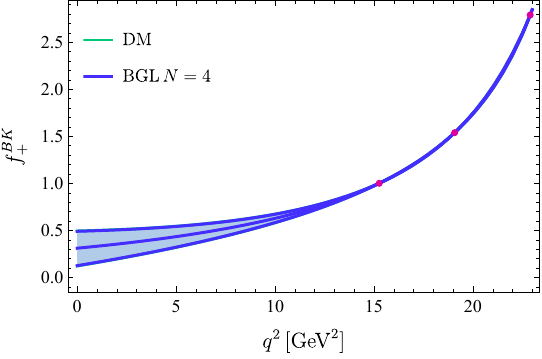}
    \end{subfigure}
    \caption{
        Comparison between the DM band and the optimised finite-order BGL band for $f_+^{BK}$.
        The green curves show the DM bounds obtained from the central values of $f_+^{BK}$ at three $q^2$ points~\eqref{eq:BKq2points}, while the blue curves show the extremal BGL intervals obtained by imposing exact interpolation at the same points together with the bound~\eqref{eq:UBBGL}.
        These central values, shown as magenta points, are taken from Ref.~\cite{Parrott:2022rgu}.
        The left and right panels correspond to truncation orders $N=3$ and $N=4$, respectively.
        For $N=4$, the two curves are indistinguishable.
    }
    \label{fig:fBK_DM_BGL}
\end{figure}

For the DM method, the problem is then fully specified by these inputs.
To construct the corresponding BGL comparison, I truncate the expansion~\eqref{eq:BGL} at order $N\geq 3$ and require the truncated series to interpolate the input values exactly.
For concreteness, consider $f_+^{BK}$. The matching conditions read
\begin{equation}
\label{eq:BGLmatchBK}
    f_+^{BK}(q_i^2)
    =
    \frac{1}{\B_\Gamma^J(q_i^2;s_+)\,\phi_{f_+}(q_i^2)}
    \sum_{n=0}^N \alpha_{f_+,n}^{\rm BGL}
    \left(z(q_i^2;s_0,s_+)\right)^n,
    \qquad i=1,2,3.
\end{equation}
Because the three points $q_i^2$ are distinct, this linear system determines three coefficients of the truncated BGL series.
For instance, one may solve for $\alpha_{f_+,0}^{\rm BGL}$, $\alpha_{f_+,1}^{\rm BGL}$, and $\alpha_{f_+,2}^{\rm BGL}$ in terms of the remaining coefficients $\alpha_{f_+,n}^{\rm BGL}$ with $3 \leq n \leq N$.
The residual freedom is then constrained by the unitarity bound~\eqref{eq:UBBGL}.
I therefore minimise and maximise $f_+^{BK}(0)$ over the remaining coefficients subject to this bound.
This yields, for each truncation order $N$, the extremal BGL interval compatible with the same three input points and the same unitarity constraint.
Comparing this interval with the DM result then provides a clean numerical comparison of the two methods.
Since the two constructions rely on the same pointwise information, any difference between the resulting intervals can only come from the finite BGL truncation.
The same construction can be applied without modification to the other FFs.
I repeat the exercise for $f_0^{BK}$ and $f_T^{BK}$, and for the seven $B\to D^*$ FFs in the basis~\eqref{eq:BVFFbasis}.
For the latter I use the nominal results of Ref.~\cite{Bordone:2025jur} at
\begin{align}
    \label{eq:BDstarq2points}
    \{q_1^2,q_2^2,q_3^2\}
    :=
    \{7~\GeV^2, 9~\GeV^2, s_-=10.71~\GeV^2 \}
    \,.
\end{align}
For each FF, I compare the width of the BGL interval at $q^2=0$ with the corresponding DM interval.
The smallest non-trivial truncation in the present setup is $N=3$.
At this order, the BGL interval for the $B\to K$ FFs is slightly narrower than the DM one, with a relative reduction of only a few per cent at $q^2=0$.
This is illustrated for $f_+^{BK}$ in \reffig{fBK_DM_BGL}.
The difference is small, especially given that the example is deliberately demanding: the three input points all lie in the high-$q^2$ region, while the comparison is made at the opposite endpoint of the semileptonic region.
For the $B\to D^*$ FFs the agreement is even better; the relative difference is already below the per-cent level for $N=3$.
This is expected, since the physical $q^2$ range is shorter than in $B\to K$ and the chosen input points are correspondingly closer to the point at which the interval is evaluated.
In the $B\to D^*$ case, the unitarity bound is also closer to saturation, leaving less residual norm for higher-order terms.
Increasing the truncation to $N=4$ makes the difference between the BGL and DM intervals numerically negligible in all the examples considered.
In the limit $N\to\infty$, the optimised BGL envelope reproduces the DM band, because the two constructions are different representations of the same unitarity-bounded Hardy-space problem.

FF calculations are usually provided as central values and a covariance matrix for the FFs evaluated at selected \(q^2\) points.
Although the illustrative example above uses only central values, the same procedure can be repeated for samples drawn from such a distribution.
For each sample, one repeats both the DM construction and the constrained BGL optimisation.
The sample is kept only if it is compatible with unitarity.
For DM this means that the sample must pass the filter~\eqref{eq:DMfilter}; for BGL it means that the interpolation constraints and the coefficient bound~\eqref{eq:UBBGL} must define a non-empty allowed region.
Samples that fail these tests are discarded, or equivalently the sampling can be restricted from the start to the region allowed by unitarity.

I therefore do not share the view put forward in Ref.~\cite{Simula:2025lpc} that the DM filter~\eqref{eq:DMfilter} should also be imposed on the BGL input data.
The reason is that \refeq{DMfilter} and the requirement that the BGL interpolation constraints admit a solution compatible with the bound~\eqref{eq:UBBGL} encode the same unitarity information, up to finite-truncation effects.
\\

\hi{This comparison shows that neither method has a theoretical advantage over the other.
Both are based on the same analytic structure and the same unitarity inequality, and therefore constrain the space of admissible extrapolated FF values in an equivalent way.}
The DM construction provides a particularly elegant implementation of this information, since it is formulated directly in terms of FF values at given points and yields the extremal range allowed at any other point.
By contrast, BGL realises the same constraint in coefficient space.

As the example illustrates, BGL can even be arranged to interpolate the input values exactly, provided that a sufficient number of coefficients is retained and the remaining freedom is optimised under the unitarity bound.
This exact-interpolation version is useful for comparison with DM, but it is not how BGL is usually employed in phenomenological analyses, although it could be.
In a standard BGL fit, the coefficients are inferred from a correlated likelihood built from lattice QCD, LCSR, and possibly experimental inputs.
Therefore, the theory points constrain the coefficients statistically, rather than being imposed as exact interpolation conditions on their central values.

The main difference between the two approaches is practical rather than conceptual.
\hi{DM is well suited to constructing unitarity filters and pointwise allowed bands, but the outcome of a DM analysis is not naturally encoded in a small set of parameters from which the FFs can be reconstructed over the full kinematical range and readily reused in other phenomenological analyses.
BGL, by contrast, provides precisely such a representation: central values and a covariance matrix for a finite set of coefficients are sufficient to evaluate the FFs, propagate uncertainties, and reuse the result in phenomenological applications.}

\subsection{Quasi-bounded parametrisations}
\label{sec:NUBpar}

In the previous subsection I reviewed the BGL parametrisation and the DM interpolation, under the temporary assumption that subthreshold branch cuts are neglected.
Both constructions start from the same object, namely the normalised analytic function $\G_\F$ defined in \refeq{gBGL}, and from the same norm bound~\eqref{eq:boundcalG}.
The introduction of the Blaschke product~\eqref{eq:Bprod} and of the outer function~\eqref{eq:outerBGL} is therefore not a matter of convention: it is what converts the unitarity inequality~\eqref{eq:impl-Bound} into a bound for $\G_\F$.
Once this has been done, the BGL basis $\{1,z,z^2,\ldots\}$ is special because it is orthonormal with respect to the inner product~\eqref{eq:BGLinner}.
This is why the BGL parametrisation has a \emph{diagonal bound}~\eqref{eq:UBBGL}, i.e. a sum of positive terms with no interference between different coefficients.

By contrast, most parametrisations expand the FF itself, or the FF after factoring out only a simple pole, rather than the normalised function $\G_\F$, as in the SPPs discussed in \refsubsec{SPPpar}.
Schematically, such parametrisations have the form
\begin{align}
    \label{eq:QBansatz}
    \F(q^2)
    =
    \frac{1}{R_\F(q^2)}
    \sum_{n=0}^\infty \beta_{\F,n}\,
    p_n\!\left(z(q^2)\right)
    \,,
\end{align}
where the $p_n$ are monomials or polynomials in a chosen $z$ variable, and $R_\F$ denotes the prefactor used in the parametrisation.
Multiplying \refeq{QBansatz} by the BGL factors gives
\begin{align}
    \label{eq:QBbasis}
    \G_\F(z)
    =
    \sum_{n=0}^\infty \beta_{\F,n}\,
    h_{\F,n}(z)
    \,,
    \qquad
    h_{\F,n}(z)
    :=
    \left.
    \frac{\B_\Gamma^J(s)\,\phi_\F(s)}{R_\F(s)}
    p_n\!\left(z(s)\right)
    \right|_{s=s(z)}
    \,.
\end{align}
Thus the coefficients $\beta_{\F,n}$ can always be related to the BGL coefficients, provided the full infinite expansion is kept.
Indeed, expanding each $h_{\F,n}$ in $z$ monomials,
\begin{align}
    h_{\F,n}(z)
    =
    \sum_{m=0}^\infty C_{mn}^{\F}\,z^m
    \,,
\end{align}
where $C_{mn}^{\F}$ are the Taylor-expansion coefficients of the functions $h_{\F,n}$.
Substituting this expansion into \refeq{QBbasis}, one obtains
\begin{align}
    \G_\F(z)
    =
    \sum_{n=0}^\infty
    \beta_{\F,n}
    \sum_{m=0}^\infty C_{mn}^{\F}\,z^m
    =
    \sum_{m=0}^\infty
    \left(
        \sum_{n=0}^\infty C_{mn}^{\F}\,\beta_{\F,n}
    \right)
    z^m
    \,.
\end{align}
Comparison with the BGL expansion~\eqref{eq:gBGLexp} then gives
\begin{align}
    \alpha_{\F,m}^{\rm BGL}
    =
    \sum_{n=0}^\infty C_{mn}^{\F}\,\beta_{\F,n}
    \,.
\end{align}
Using this relation in \refeq{UBBGL}, one obtains
\begin{align}
    \label{eq:QBmatrixbound}
    1
    \geq
    \sum_{m=0}^\infty
    \left|
        \sum_{n=0}^\infty C_{mn}^{\F}\,\beta_{\F,n}
    \right|^2
    =
    \sum_{n,r=0}^\infty
    \beta_{\F,n}\,
    \Omega_{nr}^{\F}\,
    \beta_{\F,r}^*
    \,,
    \qquad
    \Omega_{nr}^{\F}
    :=
    \sum_{m=0}^\infty
    C_{mn}^{\F}\,
    \left(C_{mr}^{\F}\right)^*
    \,.
\end{align}
This is a \emph{non-diagonal bound}.
Equivalently, $\Omega_{nr}^{\F}=\langle h_{\F,n},h_{\F,r}\rangle$, and the off-diagonal entries of $\Omega^{\F}$ are generally nonzero because the functions $h_{\F,n}$ are not, in general, orthogonal with respect to the inner product~\eqref{eq:BGLinner}.

Equation~\eqref{eq:QBmatrixbound} is fully equivalent to the BGL bound if no truncation is made.
However, any expansion must in practice be truncated at some finite order $N$, and this introduces a fundamental difference between the two bounds.
For BGL, the norm splits into two positive pieces,
\begin{align}
    \sum_{n=0}^\infty
    \left|\alpha_{\F,n}^{\rm BGL}\right|^2
    =
    \sum_{n=0}^N
    \left|\alpha_{\F,n}^{\rm BGL}\right|^2
    +
    \sum_{n=N+1}^\infty
    \left|\alpha_{\F,n}^{\rm BGL}\right|^2
    \,,
\end{align}
which is the reason why the remainder can be bounded as in \refeq{BGLtruncerr}.
For a generic basis, splitting the sums in \refeq{QBmatrixbound} gives instead
\begin{align}
    \label{eq:QBsplit}
    1
    \geq
    \sum_{n,r=0}^N
    \beta_{\F,n}\,
    \Omega_{nr}^{\F}\,
    \beta_{\F,r}^*
    +
    2\,\mathrm{Re}\!
    \left[
        \sum_{n=0}^N
        \sum_{r=N+1}^\infty
        \beta_{\F,n}\,
        \Omega_{nr}^{\F}\,
        \beta_{\F,r}^*
    \right]
    +
    \sum_{n,r=N+1}^\infty
    \beta_{\F,n}\,
    \Omega_{nr}^{\F}\,
    \beta_{\F,r}^*
    \,.
\end{align}
The cross term prevents one from interpreting
$1-\sum_{n,r=0}^N \beta_{\F,n}\,\Omega_{nr}^{\F}\,\beta_{\F,r}^*$ as a positive residual norm available for the omitted coefficients.
Consequently, the condition
\begin{align}
    \label{eq:NOTbound}
    \sum_{n,r=0}^N \beta_{\F,n}\,\Omega_{nr}^{\F}\,\beta_{\F,r}^*\leq 1
\end{align}
does not generally hold when some coefficients with $n>N$ are nonzero, and therefore it is not possible to obtain a rigorous estimate of the truncation error for the parametrisation~\eqref{eq:QBansatz}.
Moreover, unlike the BGL coefficients, the coefficients $\beta_{\F,n}$ are not individually constrained by a universal bound such as $|\alpha_{\F,n}^{\rm BGL}|\leq 1$.
A large coefficient in one non-orthogonal direction can be compensated by other coefficients through the off-diagonal terms in \refeq{QBmatrixbound}; therefore there is no basis-independent analogue of the simple BGL statement $|\alpha_{\F,n}^{\rm BGL}|\leq 1$ for each coefficient.

I therefore refer to parametrisations of the form \refeq{QBansatz}, supplemented by a matrix version of the unitarity bound, as \emph{quasi-bounded}.
They do use unitarity and can provide useful phenomenological constraints.
However, after truncation \hi{quasi-bounded parametrisations do not provide the model-independent control of the omitted terms that characterises genuinely unitarity-bounded parametrisations such as BGL or, equivalently, the DM construction.}

In the following, I review the BCL and CLN parametrisations and briefly discuss a few other representative quasi-bounded constructions.

\subsubsection{BCL parametrisation}

The Bourrely--Caprini--Lellouch (BCL) parametrisation~\cite{Bourrely:2008za} is a specific $z$ expansion for the $B\to\pi$ FF $f_+^{B\pi}$:
\begin{align}
    \label{eq:BCL}
    f_+^{B\pi}(q^2)
    =
    \frac{1}{1-q^2/(m_{V,1}^1)^2}
    \sum_{n=0}^{N-1}
    \beta_n^{\rm BCL}
    \left[
        z(q^2;\sopt,s_+)^n
        -
        (-1)^{n-N}\frac{n}{N}z(q^2;\sopt,s_+)^N
    \right]
    \,.
\end{align}
The BCL parametrisation aims to retain the simplicity of a polynomial expansion in~$z$, while avoiding two artefacts of a truncated BGL series: the incorrect large-$q^2$ scaling and the unphysical threshold singularity induced by the outer function in \refeq{outerBGL}.
More explicitly, Ref.~\cite{Bourrely:2008za} starts from a truncated expansion of the pole-factored FF,
\begin{align}
    \left(1-\frac{q^2}{(m_{V,1}^1)^2}\right)
    f_+^{B\pi}(q^2)
    =
    \sum_{n=0}^{N}
    \beta_n^{\rm BCL}\,
    z(q^2;\sopt,s_+)^n
    \,,
\end{align}
which already has the correct analytic structure in the complex $q^2$ plane and the expected asymptotic behaviour, namely $f_+^{B\pi}(q^2)\sim 1/q^2$ for $|q^2|\to \infty$, as expected from perturbative QCD~\cite{Lepage:1980fj,Akhoury:1994tnu}.
In addition, angular momentum conservation implies $\Im f_+^{B\pi}(q^2)\propto (q^2-s_+)^{3/2}$ near $q^2=s_+$.
Since $q^2=s_+$ corresponds to $z=-1$, this threshold behaviour translates into
\begin{align}
    \label{eq:BCLrel}
    \left.\frac{d f_+^{B\pi}}{dz}\right|_{z=-1}
    =
    0
    \qquad \Longleftrightarrow \qquad
    \sum_{n=1}^{N}
    (-1)^{n+1} n\,\beta_n^{\rm BCL}
    =
    0
    \,.
\end{align}
Eliminating the highest coefficient,
\begin{align}
    \beta_N^{\rm BCL}
    =
    -\frac{(-1)^N}{N}
    \sum_{n=0}^{N-1}
    (-1)^n n\,\beta_n^{\rm BCL}
    \,,
\end{align}
one obtains the finite-order parametrisation~\eqref{eq:BCL}, which depends on only $N$ free coefficients.
This relation implicitly assumes that all $\beta_n^{\rm BCL}$ coefficients with $n>N$ vanish.

At any fixed truncation order, BCL is a special case of the quasi-bounded ansatz~\eqref{eq:QBansatz}.
In the notation introduced in the beginning of this subsection, it is obtained by choosing
$R_{f_+}(q^2)=1-q^2/(m_{V,1}^1)^2$ and
$p_n(z)=z^n-(-1)^{n-N} n z^N/N$ for $n=0,\dots,N-1$.
After multiplication by the Blaschke product and the outer function, the corresponding functions $h_{f_+,n}$ are not orthogonal with respect to the inner product~\eqref{eq:BGLinner}.
Hence, the BCL coefficients satisfy a non-diagonal quadratic constraint of the form \refeq{QBmatrixbound}, rather than the diagonal BGL bound \refeq{UBBGL}.
Eq.~(22) of Ref.~\cite{Bourrely:2008za}, which is equivalent to \refeq{NOTbound}, provides a phenomenological constraint on the fitted coefficients, but it does not provide an estimate for the truncation error.
Appendix~A of Ref.~\cite{Bourrely:2008za} does derive a bound on the remainder, by relating the BCL coefficients to the BGL ones, combining that relation with \refeq{BGLtruncerr}, and applying additional estimates.
However, as the authors note in Sec.~VII, that bound is too conservative for practical use.
This is conceptually different from the prescription used in the main text of Ref.~\cite{Bourrely:2008za}, where the maximum value of the next BCL coefficient is used as the truncation-error estimate.
The latter should be viewed as a phenomenological ansatz for the size of the first omitted term, not as a rigorous bound on the full truncation error.

An important observation is that the branch point at $s_+$ lies above the $B^*$ pole at $q^2=(m_{V,1}^1)^2$.
Consequently, the region in which finite-order BGL artefacts would be expected to have the largest impact is already dominated by the nearby physical pole and is situated beyond the semileptonic endpoint $s_-$.
At the same time, the large-$q^2$ part of the physical region is precisely the region in which lattice-QCD calculations constrain $f_+^{B\pi}$ most strongly.
Another issue is that eliminating $\beta_N^{\rm BCL}$ through the threshold relation~\eqref{eq:BCLrel} amounts to assuming that the omitted tail does not contribute to that relation.
This assumption is not well motivated.
The condition is imposed at $z=-1$, i.e. on the boundary of the convergence disc, where the omitted terms are not suppressed by powers of $|z|<1$.
Moreover, the derivative weights the coefficients by an additional factor of $n$, which makes the threshold relation especially sensitive to higher-order terms.
For this reason, imposing the threshold condition at finite order is best regarded as a phenomenological ansatz rather than as a rigorous consequence of analyticity and unitarity, and it introduces an additional source of model dependence.
Nevertheless, the BCL parametrisation remains a well-motivated and numerically efficient choice for describing the FF $f_+^{BP}$ in phenomenological analyses of $B$-to-light-meson decays.
It is also worth recalling that the BCL parametrisation has served as the basis for several other parametrisations, such as the BSZ parametrisation discussed in \refsubsec{SPPpar} and the ``simplified series expansion'' (SSE) introduced in Ref.~\cite{Bharucha:2010im}.

\subsubsection{CLN parametrisation}

The Caprini--Lellouch--Neubert (CLN) parametrisation~\cite{Caprini:1997mu} is specifically designed for the $\bar B\to D^{(*)}$ FFs.
Contrary to BGL, the final CLN formulae are not derived from a generic expansion with free coefficients constrained only by unitarity.
Instead, Ref.~\cite{Caprini:1997mu} starts from the full set of $\bar B^{(*)}\to D^{(*)}$ FFs, uses heavy-quark symmetry together with the leading $\alpha_s$ and $1/m_Q$ corrections to relate them near zero recoil (i.e. at $q^2=s_-$), and then approximates the domain allowed by unitarity with one-parameter functions.

The natural kinematic variable in this construction is
\begin{align}
    w
    :=
    \frac{m_B^2+m_{D^{(*)}}^2-q^2}{2m_B m_{D^{(*)}}}
    \,,
\end{align}
for which $w=1$ corresponds to zero recoil.
It is convenient to denote the corresponding CLN conformal variable by
\begin{align}
    \label{eq:zCLN}
    z_w
    &:=
    \frac{\sqrt{w+1}-\sqrt{2}}{\sqrt{w+1}+\sqrt{2}}
    \equiv
    z(q^2;s_-,s_+)
    \,.
\end{align}
Unlike the choice $s_0=\sopt$, this maps the zero-recoil point to the origin of the $z$ plane rather than the centre of the semileptonic region.
This implies a slower convergence of the resulting series at $q^2=0$.

Ref.~\cite{Caprini:1997mu} uses the FF $V_1$ as a reference FF.
This FF is proportional to $f_+^{BD}$:
\begin{align}
    \label{eq:V1fplus}
    V_1(w)
    =
    \frac{2 \sqrt{m_B m_D}}{m_B+m_D}\,
    f_+^{BD}(q^2)
    \,.
\end{align}
Therefore the normalised CLN expression may be written directly in terms of the FF basis adopted in this work:
\begin{align}
    \frac{V_1(w)}{V_1(1)}
    =
    \frac{f_+^{BD}(q^2)}{f_+^{BD}(s_-)}
    \,.
\end{align}

The derivation of the CLN parametrisation then proceeds in three steps.
First, the reference FF is expanded around zero recoil:
\begin{align}
    \label{eq:CLN-wexp}
    \frac{V_1(w)}{V_1(1)}
    =
    1
    -
    \rho_1^2 (w-1)
    +
    c_1 (w-1)^2
    +
    d_1 (w-1)^3
    +
    \O\!\left((w-1)^4\right)
    \,,
\end{align}
where $\rho_1^2$, $c_1$, and $d_1$ are the slope, curvature, and third-order coefficient at zero recoil.
Second, using \refeq{zCLN}, one has
\begin{align}
    \label{eq:wofzCLN}
    w
    =
    2\left(\frac{1+z_w}{1-z_w}\right)^2 - 1
    =
    1 + 8 z_w + 16 z_w^2 + 24 z_w^3 + \O(z_w^4)
    \,.
\end{align}
The coefficients $8$, $16$, and $24$ are exact, since they follow solely from the algebraic inversion of \refeq{zCLN} and the Taylor expansion around $z_w=0$.
Substituting \refeq{wofzCLN} into \refeq{CLN-wexp} gives
\begin{align}
    \label{eq:CLN-intermediate}
    \frac{V_1(w)}{V_1(1)}
    =
    1
    - 8 \rho_1^2 z_w
    + (64 c_1 - 16 \rho_1^2) z_w^2
    + (512 d_1 + 256 c_1 - 24 \rho_1^2) z_w^3
    + \O(z_w^4)
    \,.
\end{align}
Third, each $\bar B^{(*)}\to D^{(*)}$ FF is expressed, using heavy-quark effective theory (HQET), in terms of the reference function $V_1(w)$ as
\begin{align}
\begin{aligned}
    \frac{\F(q^2)}{V_1(w)}
    & =
    A_\F
    \left[
        1 + B_\F (w-1) + C_\F (w-1)^2 + D_\F (w-1)^3 + \O\!\left((w-1)^4\right)
    \right]_{w=w(q^2)}
    \\
    & + \O\!\left(\alpha_s^2,\frac{\alpha_s}{m_Q},\frac{1}{m_Q^2}\right)
    \,,
\end{aligned}
\end{align}
where the coefficients $A_\F,\ldots,D_\F$ are estimated using HQET, including the leading $\alpha_s$ and $1/m_Q$ corrections, together with QCD sum-rule input for the subleading Isgur-Wise functions.
When inserted into the four unitarity inequalities for the four spin-parity channels, these HQET relations turn the unitarity bounds into correlated constraints on the reference-shape parameters $(\rho_1^2,c_1,d_1)$.
In particular, the unitarity constraints lead to a highly elongated ellipse in the $(\rho_1^2,c_1)$ plane and to a nearly linear dependence of $d_1$ on $\rho_1^2$ and $c_1$:
\begin{align}
    (\rho_1^2-\bar \rho_1^2)^2
    +
    S
    \left[
        c_1 - \bar c_1 - T (\rho_1^2-\bar \rho_1^2)
    \right]^2
    < K^2
    \,,
    \qquad
    d_1
    =
    \alpha \rho_1^2 + \beta c_1 + \gamma \pm \Delta
    \,.
\end{align}
Here $(\bar\rho_1^2,\bar c_1)$ is the ellipse centre, $S$, $T$, and $K$ determine its shape and size, $\alpha$, $\beta$, and $\gamma$ specify the central relation for $d_1$, and $\Delta$ is its allowed half-width.
These quantities follow from the unitarity constraints and HQET inputs; see Eqs.~(25)--(27) and Table~3 of Ref.~\cite{Caprini:1997mu}.
The variation of the form-factor shape over the allowed ranges of $c_1$ and $d_1$ is sufficiently small that CLN approximate these coefficients by their values along the major axis of the allowed domain:
\begin{align}
    c_1
    &\approx
    1.05\rho_1^2 - 0.15
    \,, \qquad
    d_1
    \approx
    0.45\rho_1^2 - 1.35c_1 + 0.03
    \approx
    -0.97\rho_1^2 + 0.23
    \,,
\end{align}
where the second relation is obtained from a third-order fit.
Substituting these correlations into \refeq{CLN-intermediate} gives
\begin{align}
    \label{eq:CLN-V1}
    \frac{V_1(w)}{V_1(1)}
    \approx
    1
    -
    8\rho_1^2 z_w
    +
    (51\rho_1^2 - 10) z_w^2
    -
    (252\rho_1^2 - 84) z_w^3
    \,,
\end{align}
with the unitarity bound giving the constraint $-0.17<\rho_1^2<1.51$.

Ref.~\cite{Caprini:1997mu} also derives analogous one-parameter formulae for $\bar B\to D^*\ell\bar\nu$ in a heavy-quark basis.
In particular, the relations for the $B\to D^*$ decay rate and for the corresponding ratios of FFs are given in Eqs.~(35)--(40) of Ref.~\cite{Caprini:1997mu}.
The procedure used to derive them is completely analogous to the one just described for $f_+^{BD}$:
one chooses a reference FF, expands it around zero recoil, rewrites the result as a series in $z_w$, and finally uses the HQET and unitarity bounds to reduce the allowed shapes to a one-parameter form.

The economy of CLN comes at the price of additional theoretical input and assumptions.
The final coefficients in \refeq{CLN-V1} are not fixed by analyticity and unitarity alone.
They emerge only after selecting a reference FF, relating the complete set of heavy-to-heavy FFs through HQET, truncating both the $(w-1)$ expansion and the HQET expansion of the FF ratios, and finally replacing $c_1$ and $d_1$ by the correlations inferred from the unitarity analysis.
Therefore, CLN should be regarded as a theory-driven constrained ansatz, not as a model-independent unitarity-bounded parametrisation in the sense of BGL or the DM method.
More specifically, contributions of order $\alpha_s^2$, $\alpha_s/m_Q$, and $1/m_Q^2$ are neglected, where the most significant ones are clearly the $1/m_c^2$ corrections~\cite{Bordone:2019vic,Bernlochner:2022ywh,Bordone:2025jur}.
Such an approximation was justified at the level of precision available in the late 1990s, but it is no longer acceptable given the present experimental and lattice-QCD accuracy.
\hi{For these reasons, the CLN parametrisation should not be used in present-day precision analyses.
Historically, \hbox{Ref.~\cite{Caprini:1997mu}} nevertheless represented a major breakthrough, as it demonstrated how unitarity bounds and HQET could be combined into compact phenomenological formulae.
Its impact on exclusive \hbox{$B\to D^{(*)}$} analyses was profound for many years, and it also inspired several subsequent FF \hbox{parametrisations~\cite{Bordone:2019vic,Bernlochner:2022ywh}}.}

\enlargethispage{\baselineskip} 

\subsubsection{Other quasi-bounded parametrisations}

There is a large number of FF parametrisations, and reviewing all of them is beyond the scope of this work.
In what follows I limit myself to mentioning a few representative examples and encourage the interested reader to consult the original references.
None of the following parametrisations has the unitarity bound in the diagonal BGL form, but they still provide very powerful FF ans\"atze by incorporating additional analytic or dynamical information.

Ref.~\cite{Balz:2025auk} introduces two novel FF parametrisations designed for situations in which the analytic structure is richer than in the standard single-channel problem.
In particular, they allow resonances and left-hand cuts on the second Riemann sheet to be studied, while keeping the connection to the underlying partial-wave amplitudes explicit.
These constructions are therefore especially useful when one wants to go beyond first-sheet descriptions and retain direct control over the singularities associated with hadronic rescattering.

Ref.~\cite{Herren:2025cwv} derives a parametrisation for the semileptonic decay $B\to\pi\pi\ell\bar\nu$.
The starting point is a partial-wave decomposition of the two-pion system, combined with series expansions in suitable kinematic variables.
The expansion coefficients are constrained by unitarity, left-hand-cut effects are incorporated with dispersive methods, and the $\pi\pi$ line shapes are described through Omn\`es functions.
This provides a framework for a genuinely multi-hadron final state.

Refs.~\cite{Bordone:2019vic,Bernlochner:2022ywh} provide two different HQET-based parametrisations for $\bar B\to D^{(*)}$ FFs that go beyond CLN by including terms of order $1/m_c^2$.
Ref.~\cite{Bordone:2019vic} studies the full set of ten FFs within the heavy-quark expansion to $\O(\alpha_s,1/m_b,1/m_c^2)$, combining heavy-quark symmetry with unitarity information.
Ref.~\cite{Bernlochner:2022ywh} instead introduces a supplemental power counting that leads to a smaller and more constrained set of second-order HQET corrections, yielding a particularly predictive ansatz.
Compared with CLN, these parametrisations are much better suited to present-day precision analyses.
This improved theoretical control, however, comes at the price of a substantially larger parameter space, and their predictive power still derives from the assumed HQET truncation.

Taken together, these examples show that quasi-bounded parametrisations cover a wide range of phenomenologically useful strategies.
They can encode analytic and dynamical information that would be cumbersome to incorporate in a plain BGL series, but they do so at the price of additional assumptions and of losing a rigorous truncation bound.
\\

In summary, the appropriate FF parametrisation depends on its intended use.
SPPs provide efficient descriptions but do not offer rigorous control of truncation errors.
BGL and DM implement the same analyticity and unitarity information rigorously, thereby providing model-independent constraints on the allowed FF shapes and extrapolations.
Quasi-bounded parametrisations gain predictive power by incorporating additional analytic or dynamical inputs, thereby introducing some degree of model dependence.
The common limitation of the standard BGL and DM constructions is their neglect of subthreshold cuts, which motivates the framework developed in the next section.

\enlargethispage{\baselineskip} 

\section{A parametrisation for form factors with subthreshold cuts}
\label{sec:G}

In the previous section I reviewed the main FF parametrisations and argued that there are only two genuinely model-independent frameworks that implement the unitarity bounds: the BGL parametrisation and the DM method.
These two approaches are not different in principle, but rather two equivalent ways of exploiting the same unitarity information.
\hi{However, both the BGL parametrisation and the DM method are formulated under the simplifying assumption that subthreshold branch cuts can be neglected.
This assumption is more restrictive than is often recognised.
As discussed in \hbox{\refsubsec{FFdefs}}, subthreshold cuts are absent only for the $B\to\pi$ FFs, whereas all other $B$-meson FFs (and also $\Lambda_b$-baryon FFs) can receive rescattering contributions below the pair-production threshold.
Therefore, strictly speaking, the plain BGL and DM constructions are rigorous only for $B\to\pi$ FFs.}

In this section, I further develop the parametrisation proposed with A.~Gopal in Ref.~\cite{Gopal:2024mgb}, which I refer to as the GG parametrisation. 
While Ref.~\cite{Gopal:2024mgb} focused on $B\to K$ FFs, here I generalise the formalism to other $B$-meson decays.
The GG parametrisation extends the BGL construction to the case in which subthreshold cuts are present and therefore supersedes the plain BGL framework.
Because the DM method is based on the same normalised analytic function, the same reasoning can be translated straightforwardly into a DM-like interpolation.
For definiteness, I formulate the construction in a BGL-like language.
In \refsubsec{Gpar} I derive the parametrisation, and in \refsubsec{X} I discuss the calculation of the quantities entering it.
In \refsubsec{Gdep}, I then study its dependence on various choices, such as the number of subtractions and the choice of $Q^2$, while in \refsubsec{GBGL} I compare it with the plain BGL parametrisation.
The main conceptual implications of these results are summarised in \refsubsec{remark}, together with a discussion of related approaches in the literature.

\subsection{GG parametrisation}
\label{sec:Gpar}

The concept of a subthreshold branch cut has been introduced in \refsubsec{FFdefs}.
Rescattering effects can lower the first multi-particle threshold from $s_+$ to $s_\Gamma$, where $s_+$ and $s_\Gamma$ are defined in \refeqa{spm}{sGamma}, respectively.
It follows that, apart from isolated poles, a generic $B\to M$ FF is analytic in $\mathbb{C}\setminus[s_\Gamma,\infty)$ rather than in $\mathbb{C}\setminus[s_+,\infty)$ (see \reffig{FFstruct}).

\begin{figure}[t!]
    \centering
    \includegraphics[width=0.75\textwidth]{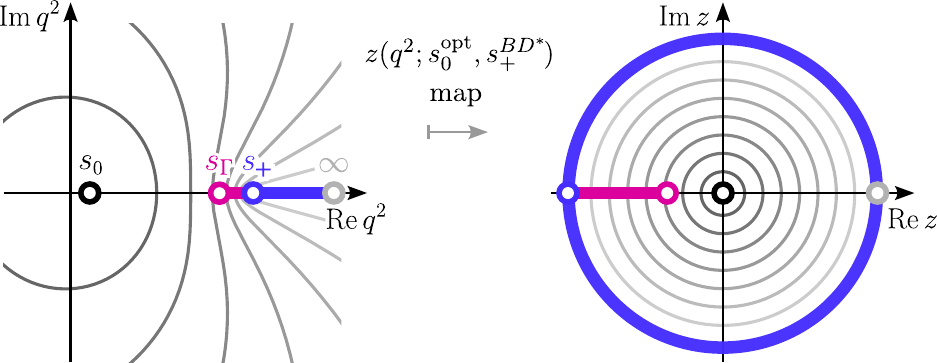}
    \caption{
        Illustration of the conformal transformation $z(q^2;\sopt,s_+^{BD^*})$ for the $B\to D^*$ FFs.
        The subthreshold branch point $q^2=s_V$ is mapped to the interior point $z\simeq -0.33$, whereas the pair-production threshold $q^2=s_+^{BD^*}$ is mapped to $z=-1$.
        Consequently, the subthreshold interval $[s_V,s_+^{BD^*})$ is sent to the segment inside the unit disc highlighted in magenta.
        See also the caption of \reffig{zmap1} for further details on the conformal map.
        \label{fig:zmapBDstarBGL}
    }
    \vspace*{0.3cm}
    \includegraphics[width=0.75\textwidth]{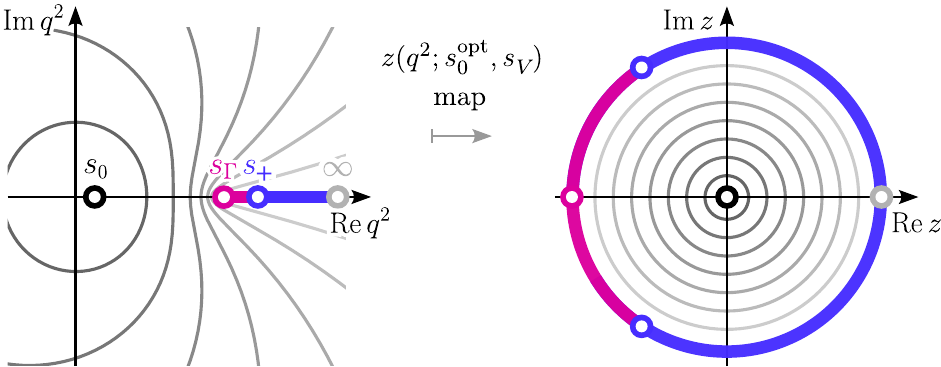}
    \caption{
        Illustration of the conformal transformation $z(q^2;\sopt,s_V)$ for the same $B\to D^*$ example.
        The subthreshold cut starting at $s_V$ is now mapped to the boundary of the unit disc rather than to its interior: the interval $[s_V,s_+^{BD^*})$ is represented by the magenta arc, while the continuation of the cut for $s\ge s_+^{BD^*}$ occupies the remaining blue part of $|z|=1$.
        %
        %
        See also the caption of \reffig{zmap1} for further details on the conformal map.
        \label{fig:zmapBDstarGG}
    }
    \includegraphics[width=0.77\textwidth]{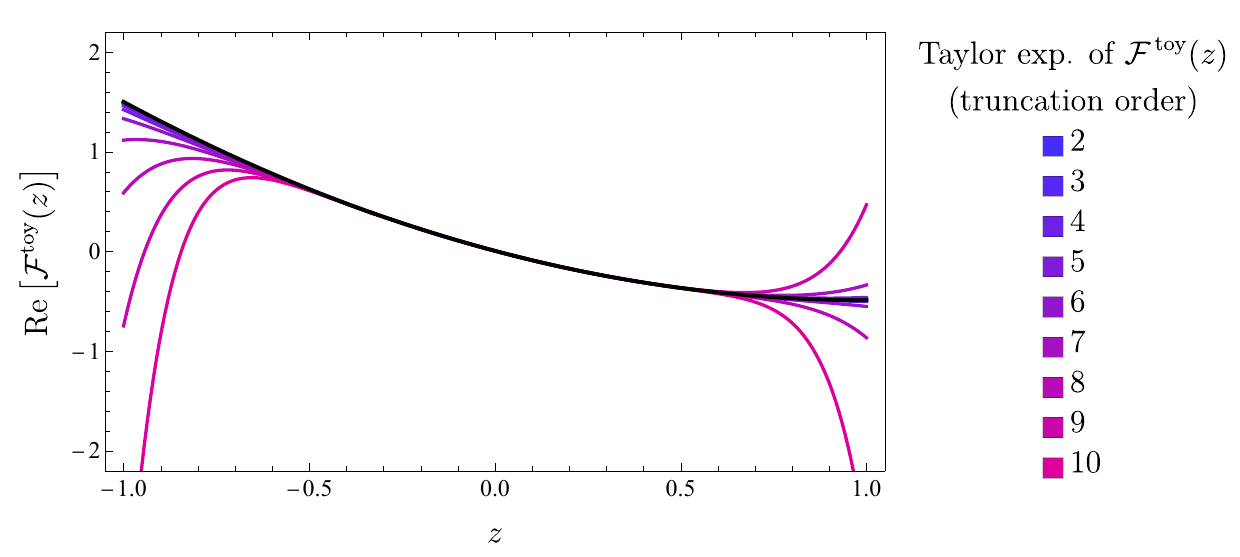}
    \captionsetup{format=plain} 
    \caption{
        Real part of the toy function $\F^{\rm toy}(z):=-z+\frac{1}{2}z^2+10^{-2}\sqrt{z + 0.33}$, together with its Taylor expansions around $z=0$ truncated at orders $N=2,\dots,10$ (coloured curves).
        Although the non-analytic contribution is small, the interior singularity controls the Taylor coefficients, so higher-order truncations worsen the approximation outside the convergence disc.
        This illustrates why a subthreshold cut cannot be neglected in a BGL-like expansion.
        \label{fig:Ftoy}
    }
    \vspace*{-0.3cm}
\end{figure}

To understand why the standard BGL and DM constructions fail to remain rigorous for all FFs except those of $B\to \pi$, it is useful to consider the usual BGL mapping $z(q^2;s_0,s_+^{BD^*})$ (cf. \refeq{zmap}) for the vector and tensor $B\to D^*$ FFs.
This mapping is displayed in \reffig{zmapBDstarBGL}, with the choice $s_0=\sopt$ defined in \refeq{s0opt}.
In this case the first branch point is at $s_V<s_+^{BD^*}$ and is mapped to
$
z(s_V;\sopt,s_+^{BD^*}) \simeq -0.33
$.
Therefore, the interval $[s_V,s_+^{BD^*})$ is mapped to a cut inside the unit disc.
Although this subthreshold interval is relatively small in the complex $q^2$ plane, its image in the $z$ plane is not far from the origin.
Hence the singularity that controls the convergence of the $z$ expansion lies much closer to the semileptonic region than one might naively expect.

This already shows that the basic assumption underlying the BGL parametrisation is violated.
The normalised function $\G_\F$ defined in \refeq{gBGL} must be analytic for $|z|<1$ to ensure the convergence of its Taylor expansion~\eqref{eq:gBGLexp} inside the unit disc.
If a branch cut lies inside the disc, however, the radius of convergence is reduced from $1$ to $|z(s_\Gamma;s_0,s_+)|<1$.
As a consequence, the large-order behaviour of the Taylor coefficients is controlled by this interior singularity, and the bound \refeq{UBBGL} no longer applies.
The same problem invalidates the DM method.
Its reproducing-kernel relations~\refeqa{DMkernel}{DMrepr} and the determinant inequality derived from them assume that $\G_\F$ is analytic on the whole unit disc.
Once a cut runs through the interior of the disc, the corresponding interpolation bounds cease to be rigorous.

One might be tempted to dismiss this issue when the discontinuity across the subthreshold cut is numerically small.
This is not sufficient.
The convergence of the Taylor coefficients is determined by the position of the nearest singularity, not by the size of its discontinuity.
Therefore, even a parametrically small cut contribution can induce exponentially growing large-order coefficients.
This is illustrated by the toy example shown in \reffig{Ftoy}.
\\

I therefore follow Ref.~\cite{Gopal:2024mgb} and adopt the conformal map
$
z(q^2;s_0,s_\Gamma)
$
as the starting point in order to remedy this issue.
For the $B\to D^*$ example, this mapping is shown in \reffig{zmapBDstarGG}.
With this choice, the first physical branch point is mapped to the boundary of the unit disc rather than to a point in its interior.

In the presence of subthreshold branch cuts, the BGL parametrisation is generalised as follows:
\begin{mybox}
\begin{center}
    \textbf{GG parametrisation and unitarity bound}
\end{center}
\vspace*{-0.5cm}
\begin{align}
    \label{eq:GG}
    &
    \F(q^2)
    =
    \frac{1}{\B_\Gamma^J(q^2;s_\Gamma)\,\widehat\phi_\F(q^2)}
    \sum_{n=0}^\infty \alpha_{\F,n}^{\rm GG}\,
    z(q^2;s_0,s_\Gamma)^n ,
    &&
    \text{with}
    &&
    \sum_{n=0}^\infty \left|\alpha_{\F,n}^{\rm GG}\right|^2 \leq 1
    \,.
    &
\end{align}
\end{mybox}
\noindent
The Blaschke product $\B_\Gamma^J(q^2;s_\Gamma)$ is defined in \refeq{Bprod}.
The outer function is defined as
\begin{equation}
\begin{aligned}
    \label{eq:outerGG}
    \widehat\phi_\F (s)
    & :=
    \sqrt{\frac{\kappa_I^{\phantom{n}}}{K \pi \Chi_\Gamma^J(Q^2;d-1)}} \,
    \left( \frac{s_\Gamma - s}{s_\Gamma - s_0} \right)^{\frac 1 4}
    \left( \sqrt{s_\Gamma - s} + \sqrt{s_\Gamma - s_0} \right)
    \\*
    & \times
    \left( s_+ - s \right)^{\frac a 4}
    \left( \sqrt{s_\Gamma - s} + \sqrt{s_\Gamma - s_-} \right)^{\frac b 2}
    \left( \sqrt{s_\Gamma - s} + \sqrt{s_\Gamma} \right)^{-(c+3)}
    \\*
    & \times
    \left(
    \frac{
        \sqrt{s_\Gamma - s} + \sqrt{s_\Gamma}
    }{
         \sqrt{s_\Gamma - s} + \sqrt{s_\Gamma + Q^2}
    }
    \right)^d
    \prod_{i:\,s_\Gamma < (m_{\Gamma,i}^J)^2 < s_+}
    \frac{
        (m_{\Gamma,i}^J)^2 - s
    }{
        \left( \sqrt{s_\Gamma - s} + \sqrt{s_\Gamma + Q^2} \right)^2
    }
    \,.
\end{aligned}
\end{equation}
The masses $m_{\Gamma,i}^J$ are collected in \reftab{states}, while the values of the isospin Clebsch--Gordan factor $\kappa_I$ can be found in~\refeq{kappaI}.
The values of the parameters $K$, $a$, $b$, $c$, and $d$ depend on the FF considered and can be found in \reftab{outer} in \refapp{outer}.
Compared with \refeq{outerBGL}, the changes are transparent: $s_+$ is replaced by $s_\Gamma$ in the conformal factors, $\widetilde{\chi}_\Gamma^J$ is replaced by $\Chi_\Gamma^J$, and any resonances in the interval $[s_\Gamma,s_+]$ are encoded by the last product.
If no such resonances are present, the product is absent and \refeq{outerGG} reduces to the generic-threshold outer function discussed in \refapp{outer}.
The function $\Chi_\Gamma^J$ is defined below, while the discussion of its numerical evaluation is deferred to the next subsection.

As is already clear from \refeq{impl-Bound} and the related discussion, the unitarity bound in \refeq{GG} becomes more effective when more FFs and channels are considered simultaneously, since their combined contributions drive the inequality closer to saturation.
An explicit example of this mechanism is given below in \refeq{satV1}.
\\

The remainder of this subsection is devoted to the derivation of the parametrisation~\eqref{eq:GG}.
Since the construction is BGL-like, the derivation follows the same overall logic, with two essential modifications: the conformal map is now $z(q^2;s_0,s_\Gamma)$, and one must explicitly account for the part of the cut between $s_\Gamma$ and $s_+$ in the calculation of $\Chi_\Gamma^J$.
I therefore focus only on the steps that differ from the standard BGL derivation discussed in \refsec{UBpar}.

The starting point is again the bound \refeq{chit}.
With the map $z(q^2;s_0,s_\Gamma)$, however, the last two lines of \refeq{BGLder} no longer involve an integral over the full unit circle, as is also clear from \reffig{zmapBDstarGG}.
Consequently, an expansion of $\G_\F$ in monomials of $z$ does not directly lead to a diagonal bound, because the monomials are orthonormal on the full circle, not on the arc corresponding to $s\ge s_+$.
Assuming for the moment that no poles are present in the interval \([s_\Gamma,s_+]\), I introduce the positive quantity~\cite{Gopal:2024mgb}
\begin{align}
    \label{eq:defDeltaChiGG}
    \Delta\chi_\Gamma^J(Q^2;d-1)
    &:=
    \frac{1}{\pi}
    \int\limits_{s_\Gamma}^{s_+} ds\,
    W_\F(s)\,
    \left|\F(s)\right|^2
    \,.
\end{align}
Here $W_\F$ is the weight function defined in \refeq{WBGL}.
For notational simplicity, I display the contribution of a single FF.
Nevertheless, the contributions of all relevant FFs included in the analysis must be added.
In \reffig{zmapBDstarGG}, $\Delta\chi_\Gamma^J$ is represented by the magenta arc.
Adding \refeq{defDeltaChiGG} to both sides of \refeq{chit} yields
\begin{align}
\begin{aligned}
    \Chi_\Gamma^J(Q^2;d-1)
    &:=
    \chi_\Gamma^J(Q^2;d-1)\Big|_{\OPE}
    -
    \sum_{\rm 1pt}
    \chi_\Gamma^J(Q^2;d-1)\Big|_{\rm 1pt}
    +
    \Delta\chi_\Gamma^J(Q^2;d-1)
    \\
    &\geq
    \frac{1}{\pi}
    \int\limits_{s_\Gamma}^{\infty} ds\,
    W_\F(s)\,
    \left|\F(s)\right|^2
    \,.
\end{aligned}
\end{align}
At this stage the integral extends over the full cut, and the derivation of \refeq{GG} proceeds exactly as in the BGL case.

So far I have assumed that no pole lies in the interval \([s_\Gamma,s_+]\).
This assumption fails, for instance, for $f_0^{BK}$ and all $B\to D^{(*)}$ FFs.
In the conformal variable, such poles lie on the unit circle and therefore cannot be removed by a Blaschke factor~\eqref{eq:Bprod}.
More precisely, they lie on the second Riemann sheet, i.e. just outside the unit disc.\footnote{
    Once $(m_{\Gamma,i}^J)^2 > s_\Gamma$, the state can decay into the continuum and is no longer stable.
    Hence the pole appears only after analytic continuation across the cut, i.e. on the second Riemann sheet.
}
Their proximity, however, can enhance $\Delta\chi_\Gamma^J$ and should therefore be treated explicitly.
Following Ref.~\cite{Gopal:2024mgb}, one suppresses these resonance enhancements directly at the level of the dispersion relation by taking suitable linear combinations of moments such that the resulting spectral weight remains positive.
This step would be unnecessary if $\Delta\chi_\Gamma^J$ were known precisely, but no such determination is currently available.
To display this construction explicitly, it is convenient to temporarily reinstate a generic subtraction index $k$, denoting the number of subtractions.
For a single pole at $q^2=(m_{\Gamma,1}^J)^2$, one then defines
\begin{align}
\begin{aligned}
    \label{eq:hDchi}
    \Delta\widehat\chi_\Gamma^J(Q^2;k)
    &:=
    \Delta\chi_\Gamma^J(Q^2;k)
    +
    \Big((m_{\Gamma,1}^J)^2+Q^2\Big)^2
    \Delta\chi_\Gamma^J(Q^2;k+2)
    \\&
    -
    2 \Big((m_{\Gamma,1}^J)^2+Q^2\Big)
    \Delta\chi_\Gamma^J(Q^2;k+1)
    \,.
\end{aligned}
\end{align}
In the present case, where $k=d-1$, this amounts to replacing $\Delta\chi_\Gamma^J$ by
\begin{align}
\begin{aligned}
    \Delta\widehat\chi_\Gamma^J(Q^2;d-1)
    &=
    \frac{1}{\pi}
    \int\limits_{s_\Gamma}^{s_+} ds\,
    W_\F(s)\,
    \left(
        \frac{(m_{\Gamma,1}^J)^2-s}{s+Q^2}
    \right)^2
    \left|\F(s)\right|^2
    :=
    \frac{1}{\pi}
    \int\limits_{s_\Gamma}^{s_+} ds\,
    \widehat{W}_\F(s)\,
    \left|\F(s)\right|^2
    \,,
\end{aligned}
\end{align}
which suppresses these resonance enhancements.
The same construction is applied to the OPE and one-particle contributions, yielding the corresponding hatted quantities.
If more than one pole is present, the generalisation is immediate: one multiplies the spectral weight by the product of the corresponding factors, as already encoded in \refeq{outerGG}.
Accordingly, when resonances with $s_\Gamma < (m_{\Gamma,i}^J)^2 < s_+$ are present, $\Chi_\Gamma^J$ is understood as
\begin{align}
    \label{eq:Chi-def}
    \Chi_\Gamma^J(Q^2;d-1)
    &:=
    \widehat\chi_\Gamma^J(Q^2;d-1)\Big|_{\OPE}
    -
    \sum_{\rm 1pt}
    \widehat\chi_\Gamma^J(Q^2;d-1)\Big|_{\rm 1pt}
    +
    \Delta\widehat\chi_\Gamma^J(Q^2;d-1)
    \,.
\end{align}
The analytic function in the unit disc is now
\begin{align}
    \label{eq:gGG}
    \widehat\G_\F(z)
    :=
    \left.
        \B_\Gamma^J(s)\,
        \widehat\phi_\F(s)\,
        \F(s)
    \right|_{s=s(z)}
    \,,
\end{align}
instead of $\G_\F$~\eqref{eq:gBGL}.
The outer function is then constrained by
\begin{align}
    \label{eq:phiGGconstr}
    \big|\widehat\phi_\F(s)\big|^2
    =
    \frac{\widehat{W}_\F(s)}{\Chi_\Gamma^J(Q^2;d-1)}
    \left|\frac{dz(s;s_0,s_\Gamma)}{ds}\right|^{-1}
    \qquad
    \text{for } s>s_\Gamma
    \,.
\end{align}
Once these replacements are made, the steps leading from \refeq{BGLder} to \refeq{UBBGLder} carry over verbatim and yield the GG expansion~\eqref{eq:GG} together with its diagonal unitarity bound.
\\

\hi{The discussion above shows that a truncated GG series satisfies the same type of rigorous unitarity bound, and therefore admits the same kind of remainder estimate, as the BGL expansion for $B\to\pi$ FFs.} 
The truncation error is controlled by Cauchy--Schwarz and by the fact that $|z(q^2;s_0,s_\Gamma)|<1$ throughout the semileptonic region (cf. \refeq{BGLtruncerr}).
To turn this construction into a numerical bound, one still needs an estimate of $\Delta\chi_\Gamma^J$, or of its hatted analogue when resonances with $s_\Gamma < (m_{\Gamma,i}^J)^2 < s_+$ are present.
It is important to stress that this estimate does not need to be precise: any upper bound is sufficient, and an overestimate merely weakens the resulting unitarity constraint without invalidating it.
In practice, $\Delta\chi_\Gamma^J$ is either expected to remain small or can be rendered small by an appropriate modification of the dispersion relation.
This point is discussed in the next subsection.

\hi{The same adaptation applies to the DM method (cf. \hbox{\refsubsec{UBpar}}).
One replaces the normalised function $\G_\F$ by $\widehat\G_\F$ and uses the conformal variable $z(s;s_0,s_\Gamma)$, such that the first branch cut is mapped to the boundary of the unit disc rather than to its interior.}
The reproducing kernel remains $g_i(z)=1/(1-z_i z)$, since it depends only on the Hardy-space structure of the unit disc, and the same Gram-matrix determinant construction then yields rigorous pointwise bounds.
When resonances with $s_\Gamma < (m_{\Gamma,i}^J)^2 < s_+$ are present, their effect is already encoded in $\widehat\G_\F$ through the modified outer function~\eqref{eq:outerGG}, so no further conceptual change is required.

Upon replacing $s_\Gamma$ by $s_+$, the GG parametrisation and the modified DM method reduce to the standard BGL parametrisation and the conventional DM method, respectively.
The reason is that the subthreshold interval $[s_\Gamma,s_+]$ then collapses to a point, implying that $\Delta\chi_\Gamma^J$ vanishes and $\Chi_\Gamma^J$ reduces to $\widetilde\chi_\Gamma^J$ defined in \refeq{chit}.
At the same time, $z(q^2;s_0,s_\Gamma)$ and $\B_\Gamma^J(q^2;s_\Gamma)$ reduce to their standard forms with threshold $s_+$, while the additional product in \refeq{outerGG} drops out.
As a result, $\widehat\phi_\F$ reduces to $\phi_\F$ and $\widehat\G_\F$ to $\G_\F$, so that \refeq{GG} becomes the usual BGL expansion and the associated DM construction reduces to the standard one.
\\

In this work, I do not consider non-local FFs, i.e. FFs that parametrise non-local hadronic matrix elements~\cite{Khodjamirian:2010vf,Khodjamirian:2012rm,Gubernari:2020eft,Feldmann:2023plv,Hurth:2025neo} (see, e.g., \refeq{nonlocBK}).
In addition to the subthreshold branch cuts discussed here, non-local FFs may exhibit \emph{anomalous branch cuts}~\cite{Mutke:2024tww,Gopal:2024mgb,Hoferichter:2026jlh}.
Unlike subthreshold cuts, anomalous thresholds are not associated with any physical particle-production channel.
Moreover, they are not constrained to lie on the real axis of the complex $q^2$ plane and may therefore occur at complex values of $q^2$.
Nevertheless, they can be treated in a way completely analogous to subthreshold cuts, as shown in Ref.~\cite{Gopal:2024mgb}.
If the anomalous thresholds are off the real axis, the conformal $z$ mapping~\eqref{eq:zmap} must be modified.
Its existence is, however, guaranteed by the Riemann Mapping Theorem, while the Schwarz--Christoffel formula provides a systematic method to construct it.
A potentially more challenging aspect is the estimate of $\Delta\chi_\Gamma^J$, which must be addressed on a case-by-case basis.

\subsection[Calculation of \texorpdfstring{$\Chi_\Gamma^J$}{X}]{Calculation of \texorpdfstring{$\bs{\Chi_\Gamma^J}$}{X} 
}
\label{sec:X}

A key ingredient of the parametrisation~\eqref{eq:GG} is $\Chi_\Gamma^J$, defined in \refeq{Chi-def}.
If no resonance with $s_\Gamma < (m_{\Gamma,i}^J)^2 < s_+$ is present, the hatted quantities reduce to the corresponding unhatted ones.
The numerical determination of $\Chi_\Gamma^J$ therefore requires the evaluation of its constituent terms, namely $\chi_\Gamma^J\big|_{\OPE}\,$, $\chi_\Gamma^J\big|_{\rm 1pt}\,$, and $\Delta\chi_\Gamma^J$.
This subsection is devoted to this task.

\subsubsection*{Calculation of $\bs{\chi_\Gamma^J\big|_{\OPE}}$}

The conceptual setup of the OPE calculation has been discussed in \refsubsec{UB}; here I only specify the numerical implementation.
I retain only the leading term in \refeq{OPEchi}, namely the Wilson coefficient of the identity operator.
The contributions of quark and gluon condensates are known to be very small~\cite{Boyd:1997kz,Bharucha:2010im}.
I therefore neglect them throughout.
With this choice, the dominant theoretical uncertainty of $\chi_\Gamma^J\big|_{\OPE}$ is perturbative and stems from uncalculated higher-order corrections in $\alpha_s$.

For the vector and axial-vector currents, I use the NNLO results of Ref.~\cite{Grigo:2012ji} whenever they are available, i.e. at $Q^2=0$ and for the moments relevant here.
For the tensor and axial-tensor currents, only NLO results are currently available.
I therefore use the recent calculation of Ref.~\cite{Generet:2025hsv}, carried out by T. Generet, E. Loisa, and myself.
A fully analytic evaluation of $\chi_\Gamma^J(Q^2;k)\big|_{\OPE}$ at arbitrary admissible $Q^2$ and number of subtractions $k$ is presently possible only at NLO, using Refs.~\cite{Djouadi:1993ss,Generet:2025hsv}.
Whenever the hatted quantities introduced in \refeq{Chi-def} are needed, I construct them from the same perturbative moments by applying the corresponding linear combinations (cf. \refeq{hDchi}).

Unless stated otherwise, I adopt the standard choice $Q^2=0$ together with the minimal number of subtractions, $k=d-1$.
I work in the $\overline{\rm MS}$ scheme and choose the renormalisation scale
\begin{align}
    \mu = \overline{m}_b(\overline{m}_b)\,.
\end{align}
For the final-state quark masses, I set $m_u=m_d=0$, while keeping the strange- and charm-quark masses finite.
Numerically, I use
\begin{align}
    \overline{m}_s(2\,\GeV)&=0.0935\,\GeV\,,
    &
    \overline{m}_c(\overline{m}_c)&=1.273\,\GeV\,,
    &
    \overline{m}_b(\overline{m}_b)&=4.183\,\GeV\,,
\end{align}
together with $\alpha_s(m_Z)=0.1180$~\cite{PDG:2024cfk}.
I then evolve the quark masses to the scale $\mu=4.183\,\GeV$ using the four-loop running implemented in \texttt{RunDec}~\cite{Chetyrkin:2000yt,Herren:2017osy}.
For $b\to u,d$ currents, the massless final-state limit implies
$
    \chi_V^0=\chi_A^0\,,
$
$
    \chi_V^1=\chi_A^1\,,
$
and
$
    \chi_T^1=\chi_{AT}^1\,,
$
which explains the repeated entries in the first block of \reftab{X}.

The resulting values of $\chi_\Gamma^J(0;d-1)\big|_{\OPE}$ and $\widehat\chi_\Gamma^J(0;d-1)\big|_{\OPE}$ are collected in \reftab{X}.
The entries with $J=0$ are dimensionless, whereas all entries with $J=1$ carry units of $[\GeV^{-2}]$.

\subsubsection*{Calculation of $\bs{\chi_\Gamma^J\big|_{\rm 1pt}}$}

The evaluation of $\widehat\chi_\Gamma^J\big|_{\rm 1pt}$ is straightforward using \refeqa{chi1pt}{hDchi}.
In channels in which no resonance is present with $s_\Gamma < (m_{\Gamma,i}^J)^2 < s_+$, the hatted and unhatted quantities are identical.
As discussed in \refsubsec{UB}, explicit one-particle states should be included only when their masses satisfy $(m_{\Gamma,i}^J)^2 < s_\Gamma$.
Otherwise, their effect overlaps with the contribution already encoded in the FFs.
Inspecting \reftab{states}, one sees that for all quark transitions considered in this work this criterion is met only by the pseudoscalar ground states in the $(A,0)$ channels (i.e. $B$, $B_s$, and $B_c$), and by the lowest-lying states in the $(V,1)$ channels (i.e. $B^*$, $B_s^*$, and $B_c^*$).
The same vector states also provide the lowest one-particle contribution in the tensor channel $(T,1)$, since $m_{T,1}^1 \equiv m_{V,1}^1$.
By contrast, the lightest states in the $(V,0)$, $(A,1)$, and $(AT,1)$ channels lie above the corresponding $s_\Gamma$ thresholds and therefore do not enter the explicit one-particle sum.
The corresponding decay constants are taken from Refs.~\cite{Colquhoun:2015oha,Lubicz:2017asp,Pullin:2021ebn,FLAG:2024oxs,Reina:2026pvo}.
Once these inputs are specified, the numerical evaluation is purely algebraic.
The resulting values of $\widehat\chi_\Gamma^J(0;d-1)\big|_{\rm 1pt}$ are collected in \reftab{X}.
As for the OPE contribution, the entries with $J=0$ are dimensionless, whereas those with $J=1$ carry units of~$[\GeV^{-2}]$.
Note that, when resonances with $s_\Gamma < (m_{\Gamma,i}^J)^2 < s_+$ are present in a given spin-parity channel, the corresponding one-particle contribution becomes suppressed.

\begin{table}[t!]
    \newcommand{\pp}{{\phantom{*}}}
    \centering
    \renewcommand{\arraystretch}{1.4}
    \begin{tabular}{
        c@{\hspace{0.5cm}}
        c@{\hspace{0.5cm}}
        c@{\hspace{0.5cm}}
        c@{\hspace{0.5cm}}
        c@{\hspace{0.5cm}}
        c@{\hspace{0.5cm}}
        c
    }
        \toprule
        Transition                                                  &
        $(\Gamma, J)$                                               &
        $\chi_\Gamma^J\big|_{\OPE}$                                 &
        $\widehat\chi_\Gamma^J\big|_{\OPE}$                         &
        $\widehat\chi_\Gamma^J\big|_{\rm 1pt}$                      &
        $\Delta\widehat\chi_\Gamma^J$                               &
        $\Chi_\Gamma^J$                                             \\
        \toprule
        \multirow{6}{*}{\makecell{
            $b\to u$  \\[0.1cm]
            $b\to d$
        }}                                                          &
        $(A,0)$                                                     &
        $1.507\cdot10^{-2}$                                         &
        $1.507\cdot10^{-2}$                                         &
        $1.295\cdot10^{-3}$                                         &
        $2.262\cdot10^{-3}$                                         &
        $1.603\cdot10^{-2}$                                         \\
                                                                    &
        $(V,0)$                                                     &
        $1.507\cdot10^{-2}$                                         &
        $5.506\cdot10^{-3}$                                         &
        $0$                                                         &
        $2.270\cdot10^{-5}$                                         &
        $5.529\cdot10^{-3}$                                         \\
                                                                    &
        $(V,1)$                                                     &
        $5.776\cdot10^{-4}$                                         &
        $5.776\cdot10^{-4}$                                         &
        $4.121\cdot10^{-5}$                                         &
        $4.622\cdot10^{-5}$                                         &
        $5.827\cdot10^{-4}$                                         \\
                                                                    &
        $(A,1)$                                                     &
        $5.776\cdot10^{-4}$                                         &
        $2.291\cdot10^{-4}$                                         &
        $0$                                                         &
        $5.324\cdot10^{-6}$                                         &
        $2.345\cdot10^{-4}$                                         \\
                                                                    &
        $(T,1)$                                                     &
        $4.058\cdot10^{-4}$                                         &
        $4.058\cdot10^{-4}$                                         &
        $4.975\cdot10^{-5}$                                         &
        $5.605\cdot10^{-5}$                                         &
        $4.121\cdot10^{-4}$                                         \\
                                                                    &
        $(AT,1)$                                                    &
        $4.058\cdot10^{-4}$                                         &
        $1.454\cdot10^{-4}$                                         &
        $0$                                                         &
        $2.453\cdot10^{-6}$                                         &
        $1.479\cdot10^{-4}$                                         \\
        \midrule
        \multirow{6}{*}{$b\to s$}                                   &
        $(A,0)$                                                     &
        $1.607\cdot10^{-2}$                                         &
        $1.607\cdot10^{-2}$                                         &
        $1.841\cdot10^{-3}$                                         &
        $5.733\cdot10^{-3}$                                         &
        $1.996\cdot10^{-2}$                                         \\
                                                                    &
        $(V,0)$                                                     &
        $1.406\cdot10^{-2}$                                         &
        $5.146\cdot10^{-3}$                                         &
        $0$                                                         &
        $2.074\cdot10^{-5}$                                         &
        $5.167\cdot10^{-3}$                                         \\
                                                                    &
        $(V,1)$                                                     &
        $5.969\cdot10^{-4}$                                         &
        $5.969\cdot10^{-4}$                                         &
        $5.852\cdot10^{-5}$                                         &
        $1.142\cdot10^{-4}$                                         &
        $6.526\cdot10^{-4}$                                         \\
                                                                    &
        $(A,1)$                                                     &
        $5.570\cdot10^{-4}$                                         &
        $2.206\cdot10^{-4}$                                         &
        $0$                                                         &
        $1.250\cdot10^{-5}$                                         &
        $2.331\cdot10^{-4}$                                         \\
                                                                    &
        $(T,1)$                                                     &
        $4.228\cdot10^{-4}$                                         &
        $4.228\cdot10^{-4}$                                         &
        $6.478\cdot10^{-5}$                                         &
        $1.379\cdot10^{-4}$                                         &
        $4.959\cdot10^{-4}$                                         \\
                                                                    &
        $(AT,1)$                                                    &
        $3.879\cdot10^{-4}$                                         &
        $1.390\cdot10^{-4}$                                         &
        $0$                                                         &
        $5.402\cdot10^{-6}$                                         &
        $1.444\cdot10^{-4}$                                         \\
        \midrule
        \multirow{6}{*}{$b\to c$}                                   &
        $(A,0)$                                                     &
        $2.365\cdot10^{-2}$                                         &
        $6.017\cdot10^{-3}$                                         &
        $1.902\cdot10^{-4}$                                         &
        $1.099\cdot10^{-4}$                                         &
        $5.936\cdot10^{-3}$                                         \\
                                                                    &
        $(V,0)$                                                     &
        $6.441\cdot10^{-3}$                                         &
        $2.292\cdot10^{-3}$                                         &
        $0$                                                         &
        $1.190\cdot10^{-4}$                                         &
        $2.411\cdot10^{-3}$                                         \\
                                                                    &
        $(V,1)$                                                     &
        $6.437\cdot10^{-4}$                                         &
        $1.804\cdot10^{-4}$                                         &
        $4.430\cdot10^{-6}$                                         &
        $4.020\cdot10^{-6}$                                         &
        $1.800\cdot10^{-4}$                                         \\
                                                                    &
        $(A,1)$                                                     &
        $3.912\cdot10^{-4}$                                         &
        $1.616\cdot10^{-4}$                                         &
        $0$                                                         &
        $2.113\cdot10^{-5}$                                         &
        $1.827\cdot10^{-4}$                                         \\
                                                                    &
        $(T,1)$                                                     &
        $4.864\cdot10^{-4}$                                         &
        $1.198\cdot10^{-4}$                                         &
        $5.951\cdot10^{-6}$                                         &
        $4.553\cdot10^{-6}$                                         &
        $1.184\cdot10^{-4}$                                         \\
                                                                    &
        $(AT,1)$                                                    &
        $2.462\cdot10^{-4}$                                         &
        $9.610\cdot10^{-5}$                                         &
        $0$                                                         &
        $4.658\cdot10^{-6}$                                         &
        $1.008\cdot10^{-4}$                                         \\
        \bottomrule
    \end{tabular}
    \caption{
        Numerical values of the quantities entering $\Chi_\Gamma^J$ defined in \refeq{Chi-def}, evaluated at $Q^2=0$ and with the minimal number of subtractions for each channel: $k=d-1$.
        In channels without resonances in the interval $[s_\Gamma, s_+]$, the hatted and unhatted quantities coincide.
        Entries with $J=0$ are dimensionless, whereas those with $J=1$ are given in units of $[\GeV^{-2}]$.
        \label{tab:X}
    }
\end{table}

\subsubsection*{Calculation of $\bs{\Delta\chi_\Gamma^J}$}

The quantity $\Delta\widehat\chi_\Gamma^J$ defined in \refeqa{defDeltaChiGG}{hDchi} depends on the FF along the subthreshold interval $s\in[s_\Gamma,s_+]$, where no direct first-principles determination is currently available.
For the numerical analysis, however, only an upper estimate is required.
Following Ref.~\cite{Gopal:2024mgb}, I approximate the FF modulus in this region by
\begin{align}
    \label{eq:DeltaChiAnsatz}
    |\F(s)|^2
    \simeq
    \mathcal{K}
    \left(\frac{s_\Gamma}{s}\right)^2
    \,.
\end{align}
This ansatz is inspired by the large-$q^2$ scaling behaviour obtained in perturbative QCD for $f_+^{BP}$ in heavy-to-light decays~\cite{Lepage:1980fj,Akhoury:1994tnu}.
Although this argument is not meant as a precise model for the FF in the interval $[s_\Gamma,s_+]$, it provides a convenient envelope for estimating the integral.
Moreover, $\mathcal{K}$ is expected to be of order unity~\cite{Boyd:1995sq,Caprini:1995wq,Beneke:2000ry,Becher:2005bg}.\footnote{
    This expectation is also qualitatively consistent with data on the electromagnetic pion and kaon FFs.
    Although these are different processes, the measured values above threshold and away from resonances are such that $|\F|^2$ is typically of order unity; see Ref.~\cite{PDG:2024cfk}.
}
To keep the estimate manifestly conservative, I nevertheless adopt $\mathcal{K}=100$ throughout.
With this choice, the subthreshold contribution is approximated as
\begin{align}
    \Delta\widehat\chi_\Gamma^J(Q^2;d-1)
    \simeq
    \frac{\mathcal{K}}{\pi}
    \int\limits_{s_\Gamma}^{s_+} ds\,
    \widehat W_\F(s)
    \left(\frac{s_\Gamma}{s}\right)^2
    \,.
\end{align}
This is expected to overestimate the true size of the integral.
This is mitigated by the fact that the weight function itself suppresses contributions from this region through the K\"all\'en function it contains.
In addition, in channels with subthreshold resonances, the hatted weight $\widehat W_\F$ contains explicit factors that effectively damp the contribution of $\Delta\widehat\chi_\Gamma^J$.
In general, one may introduce an auxiliary mass parameter $m_{\rm FP}$, which I refer to as a ``fictitious pole'', in \refeq{hDchi}, even if no physical resonance is present in the interval $[s_\Gamma,s_+]$.
This is legitimate because the corresponding spectral representation simply acquires the non-negative factor
$
\big((m_{\rm FP}^2-s)/(s+Q^2)\big)^2
$,
so the positivity argument used in the derivation of the bound remains unchanged.
The procedure is conservative as it reduces the impact of $\Delta\chi_\Gamma^J$.
An implication of this choice is that $\Chi_\Gamma^J$ receives contributions from higher moments of $\chi_\Gamma^J\big|_{\OPE}$, thereby yielding a less stringent unitarity bound (cf. \refsubsec{Gdep}).
From a technical point of view, one may introduce an arbitrary number of fictitious poles, although this progressively weakens the resulting unitarity bounds.

I use this option only for the channel $(\Gamma,J)=(V,0)$ in $b\to u,d$ transitions, where the estimate obtained from \refeq{DeltaChiAnsatz} would otherwise make $\Delta\chi_V^0$ comparable with $\chi_V^0\big|_{\OPE}$.
For definiteness, I choose to introduce
\begin{align}
    m_{\rm FP}
    \equiv
    m_{V,1}^0
    =
    \sqrt{\frac{s_\Gamma+s_+^{B_sK}}{2}}
    =
    5.642\,\GeV\,,
\end{align}
which lies near the midpoint of the interval relevant for $B_s\to K$.
Interestingly, quark-model expectations place the lowest scalar resonance in a similar mass region~\cite{Cheng:2017oqh}, although no such interpretation is required for the validity of the construction.

The resulting values are listed in \reftab{X}.
Even with the deliberately extreme choice $\mathcal{K}=100$, the contribution of $\Delta\widehat\chi_\Gamma^J$ never exceeds about one third of $\widehat\chi_\Gamma^J\big|_{\OPE}$ and is negligible in most channels.
Since $\Delta\widehat\chi_\Gamma^J$ enters $\Chi_\Gamma^J$ with a positive sign, overestimating it merely weakens the final unitarity bound; it does not invalidate the construction.
As discussed in \refsubsec{Gdep}, the numerical impact on the final FF constraints of such an increase of $\Chi_\Gamma^J$ is small compared to the typical FF uncertainties.
A more accurate calculation of $\Delta\chi_\Gamma^J$ would nevertheless be valuable, since it would strengthen the numerical constraining power of the unitarity bounds.

\subsection[Dependence on \texorpdfstring{$Q^2$}{Q2}, the number of subtractions, and \texorpdfstring{$\Chi_\Gamma^J$}{X}]
{Dependence on \texorpdfstring{$\bs{Q^2}$}{Q2}, the number of subtractions, and \texorpdfstring{$\bs{\Chi_\Gamma^J}$}{X}}
\label{sec:Gdep}

The GG construction derived above is rigorous for any admissible choice of the subtraction point $Q^2$ and of the number of subtractions $k$.
At the same time, its numerical constraining power depends on these choices, on the value of the positive quantity $\Chi_\Gamma^J$ entering the outer function~\eqref{eq:outerGG}, and eventually on the choice of the conformal parameter $s_0$.
This is already clear from \refeqa{subdisprel}{Chi-def}: varying $Q^2$ or $k$ changes the spectral weight in the dispersion relation, while increasing $\Chi_\Gamma^J$ weakens the normalisation of the bound.
Similarly, changing $s_0$ does not affect the validity of the construction, but it changes how fast the series in \refeq{GG} converges in the semileptonic region~\eqref{eq:smreg}.

To study these dependences in a transparent way, I follow the same strategy used in the comparison between BGL and DM in \refsubsec{UBpar}.
Rather than performing a complete FF fit for each choice, I consider fixed FF input at given points and compare the resulting allowed GG intervals.
For definiteness, I use the same representative high-$q^2$ points used in the $B\to K$ and $B\to D^*$ examples (see \refeqa{BKq2points}{BDstarq2points}), although the construction is completely general.
The choice of points in the high-$q^2$ region is deliberate, since it makes the numerical effect of the unitarity bound more visible. 
Indeed, if the input points were closer to the region where the FF is to be constrained, the interpolation problem would be easier and the resulting interval narrower.
For a given FF I fix a set of input values $\F(q_i^2)$ using the central values from Refs.~\cite{Parrott:2022rgu,Bordone:2025jur}.
The truncation order has to be large enough to interpolate them exactly.
The matching conditions then read
\begin{equation}
    \label{eq:GGmatch}
    \F(q_i^2)
    =
    \frac{1}{\B_\Gamma^J(q_i^2;s_\Gamma)\,\widehat\phi_\F(q_i^2)}
    \sum_{n=0}^N \alpha_{\F,n}^{\rm GG}
    \left(z(q_i^2;s_0,s_\Gamma)\right)^n,
    \qquad i=1,2,3.
\end{equation}
The extremal values allowed elsewhere are then determined from the truncated expansion~\eqref{eq:GG} together with the coefficient bound
$
\sum_{n=0}^N |\alpha_{\F,n}^{\rm GG}|^2 \leq 1
$.
The width of the resulting interval provides a direct measure of the constraining power of the unitarity bound for the values of $(Q^2,k,\Chi_\Gamma^J,s_0)$.

This fixed-input setup is adopted only for clarity; the same exercise can be carried out sample by sample when the input FF values are drawn from a multivariate distribution.
Each draw defines an interpolation problem of the same type and is retained only if it is compatible with the unitarity constraint.
For the present discussion, however, the fixed-input setup is sufficient, since it isolates the effect of varying one ingredient at a time without introducing additional statistical complications.

\subsubsection*{Dependence on $\bs{Q^2}$}

The dependence of the unitarity bound on $Q^2$ was already studied in Ref.~\cite{Caprini:1995wq} for $B\to D^{(*)}$ FFs and in Ref.~\cite{Simula:2023ujs} in the different context of the electromagnetic pion FF.
For decreasing $Q^2$, Ref.~\cite{Caprini:1995wq} concludes that ``The gain that can be achieved in this way is rather small [...]''.
I find the same qualitative behaviour in the present GG framework.
For the numerical illustrations presented in this subsection, I truncate the series at $N=6$.
With this choice, the effects of truncating the GG expansion are numerically negligible (cf. \refsubsec{UBpar}).

\begin{figure}[t!]
    \centering
    \begin{subfigure}{0.48\textwidth}
        \centering
        \includegraphics[width=\textwidth]{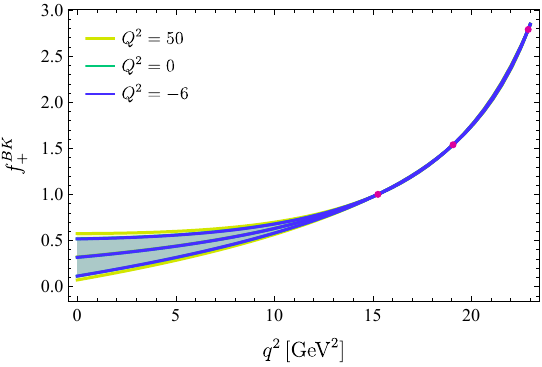}
    \end{subfigure}
    \hfill
    \begin{subfigure}{0.48\textwidth}
        \centering
        \includegraphics[width=\textwidth]{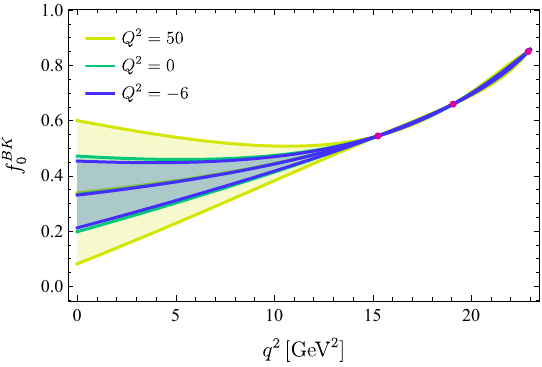}
    \end{subfigure}
    \caption{
        Dependence of the FF uncertainties on the subtraction point $Q^2$ for $f_+^{BK}$ (left) and $f_0^{BK}$ (right).
        The three magenta points indicate the fixed FF input values at the representative high-$q^2$ points given in \refeq{BKq2points}.
        The curves correspond to $Q^2=-6,\,0,$ and $50~\GeV^2$, while all other inputs are kept fixed.
        In both panels the series is truncated at $N=6$, for which the truncation error is numerically negligible on the scale shown.
        Lowering $Q^2$ strengthens the bound slightly.
        The effect is negligible for $f_+^{BK}$ and more visible for $f_0^{BK}$.
    }
    \label{fig:G_Q2}
\end{figure}

\reffig{G_Q2} illustrates this point for $f_+^{BK}$ and $f_0^{BK}$.
In both cases decreasing $Q^2$ leads to a slightly tighter interval, as expected from the modified spectral weight in the dispersion relation.
Recall that the minimal value of $Q^2$ is bounded by \refeq{Q2OPE}.
For $f_+^{BK}$, the effect is rather small: the curves for $Q^2=0$ and $Q^2=-6$ already lie very close to each other over most of the semileptonic region.
By contrast, the dependence is more visible for $f_0^{BK}$.
This is not surprising, since the scalar channel contains a resonance with $s_\Gamma < (m_{V,1}^0)^2 < s_+$.
As a consequence, the corresponding hatted construction introduces an additional $Q^2$ dependence both in the outer function and in $\Chi_V^0$.
Even in this more sensitive case, the overall effect remains moderate.
It is therefore well justified to adopt $Q^2=0$, as is done in most studies in the literature.

This qualitative behaviour can already be anticipated directly from \refeq{GG}, without the need for a full statistical analysis.
The $Q^2$ dependence enters only through the outer function $\widehat\phi_\F$, and hence through $\Chi_\Gamma^J$ and the explicit $Q^2$-dependent factors contained in \refeq{outerGG}.
Numerically, I find that $1/\widehat\phi_\F$ increases with $Q^2$ throughout the semileptonic region for the FFs and processes considered in this work. 
For fixed input values $\F(q_i^2)$, the matching conditions \refeq{GGmatch} therefore require smaller coefficients $\alpha_{\F,n}^{\rm GG}$ as $Q^2$ is increased.
The coefficient vector then lies further from saturating the bound
$
\sum_{n=0}^N |\alpha_{\F,n}^{\rm GG}|^2 \leq 1
$,
leaving more residual norm for deformations of the FF and hence leading to a weaker constraint, i.e. to a broader allowed interval.

\subsubsection*{Dependence on $\bs{k}$}

So far I have always taken the number of subtractions $k$ in the unitarity bound to be the minimal one, namely $k=d-1$ (see \refeqa{subdisprel}{impl-Bound} and \reftab{outer}).
Nevertheless, nothing prevents one from considering higher moments.
To parametrise this freedom, I write
\begin{align}
    k=d-1+e\,,
    \qquad e=0,1,2,\ldots
\end{align}
with $e$ denoting the number of additional subtractions.
To the best of my knowledge, this dependence has not been studied systematically in the literature.
In practice, this amounts to replacing $c\to c+e$ and $d\to d+e$ in \reftab{outer} and evaluating $\Chi_\Gamma^J(Q^2;d-1+e)$.
As in the discussion of the $Q^2$ dependence, the effect on the bound is then entirely encoded in the outer function $\widehat\phi_\F$.
A meaningful comparison of different values of $e$ requires some care, however.
If one increases $e$ at fixed small $Q^2$, $\Chi_\Gamma^J(Q^2;d-1+e)$ becomes progressively more sensitive to the near-threshold region and the corresponding OPE side is less cleanly dominated by short distances~\cite{Colangelo:2000dp}.
In other words, the QCD vacuum-condensate contributions can then become strongly enhanced.
It is therefore more natural to vary $Q^2$ simultaneously and keep
\begin{align}
    M^2:=\frac{Q^2}{e}
\end{align}
fixed for $e>0$, in analogy with the logic familiar from QCD sum rules~\cite{Colangelo:2000dp}; the case $e=0$ then corresponds to the standard choice $Q^2=0$.
For the numerical illustrations below, I choose $M^2=6\,\GeV^2$, which is a representative value for sum-rule applications involving a $b$ quark~\cite{Gelhausen:2014jea,Pullin:2021ebn}, and I truncate the GG series at $N=6$, so that truncation effects are again negligible on the scale of the plots.

\begin{figure}[t!]
    \centering
    \begin{subfigure}{0.48\textwidth}
        \centering
        \includegraphics[width=\textwidth]{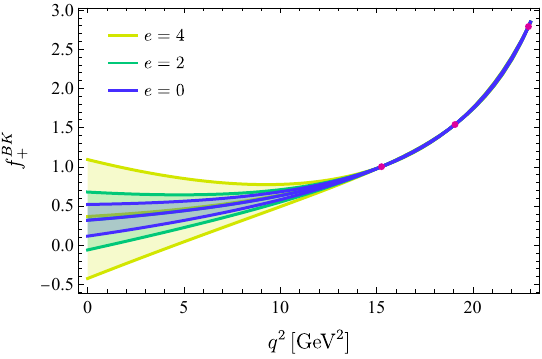}
    \end{subfigure}
    \hfill
    \begin{subfigure}{0.48\textwidth}
        \centering
        \includegraphics[width=\textwidth]{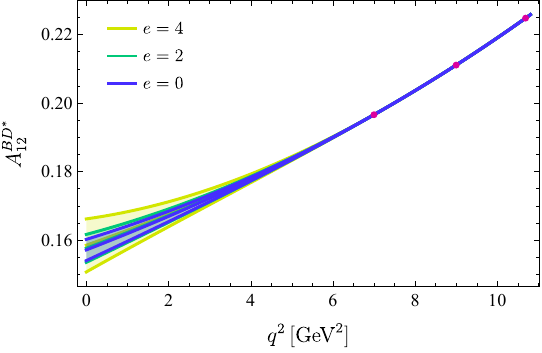}
    \end{subfigure}
    \caption{
        Dependence of the FF uncertainties on the number of additional subtractions $e$, defined through $k=d-1+e$, for $f_+^{BK}$ (left) and $A_{12}^{BD^*}$ (right).
        The three magenta points indicate the fixed FF input values at the representative high-$q^2$ points given in \refeqa{BKq2points}{BDstarq2points}.
        The comparison is performed by varying $e$ while keeping $M^2:=Q^2/e=6~\GeV^2$ fixed for $e>0$, with the standard choice $Q^2=0$ for $e=0$.
        All other inputs are kept fixed, and the series is truncated at $N=6$, for which truncation effects are numerically negligible on the scale shown.
        In these examples, increasing the number of subtractions weakens the bound.
    }
    \label{fig:G_e}
\end{figure}

Numerically, I find that $1/\widehat\phi_\F$ increases with $e$ throughout the semileptonic region for most FFs and processes considered in this work.
For fixed input values $\F(q_i^2)$, the matching conditions \refeq{GGmatch} then require smaller coefficients $\alpha_{\F,n}^{\rm GG}$ as $e$ is increased.
The coefficient vector therefore lies further from saturating the unitarity bound, leaving more residual norm for deformations of the FF and hence weakening the constraint.
This behaviour is illustrated in \reffig{G_e} for $f_+^{BK}$ and $A_{12}^{BD^*}$.
There are isolated exceptions, such as some $B\to\pi$ and $B_s\to K^{(*)}$ channels, in which taking $e>0$ yields a slight improvement.
Even in those cases, however, the numerical gain is marginal.

\subsubsection*{Dependence on $\bs{\Chi_\Gamma^J}$}

Although $\Chi_\Gamma^J$ is fixed once the OPE, one-particle, and subthreshold contributions have been specified, it is instructive to vary it artificially in order to understand its role in the bound.
From \refeq{outerGG}, a rescaling
$
\Chi_\Gamma^J \to \C\,\Chi_\Gamma^J
$
with $\C>0$ simply rescales the outer function as
$
\widehat\phi_\F \to \widehat\phi_\F/\sqrt{\C}
$.
Equivalently, one may keep the original outer function and rewrite the GG expansion in terms of
$
\beta_{\F,n}^{\rm GG}:=\sqrt{\C}\,\alpha_{\F,n}^{\rm GG}
$,
so that the coefficient bound becomes
\begin{align}
    \sum_n \left|\beta_{\F,n}^{\rm GG}\right|^2 \leq \C\,.
\end{align}
This makes explicit that increasing $\Chi_\Gamma^J$ can only weaken the bound.

\begin{figure}[t!]
    \centering
    \begin{subfigure}{0.48\textwidth}
        \centering
        \includegraphics[width=\textwidth]{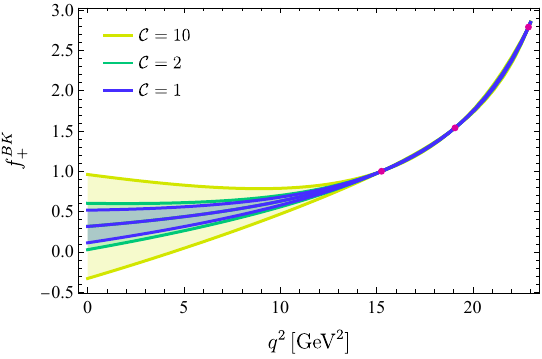}
    \end{subfigure}
    \hfill
    \begin{subfigure}{0.48\textwidth}
        \centering
        \includegraphics[width=\textwidth]{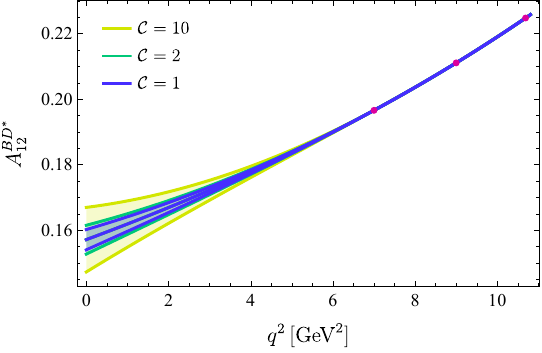}
    \end{subfigure}
    \caption{
        Dependence of the FF uncertainties on the normalisation factor $\Chi_\Gamma^J$ for $f_+^{BK}$ (left) and $A_{12}^{BD^*}$ (right).
        The three magenta points indicate the fixed FF input values at the representative high-$q^2$ points given in \refeqa{BKq2points}{BDstarq2points}.
        The curves correspond to representative rescalings $\Chi_\Gamma^J\to \C\,\Chi_\Gamma^J$, while all other inputs are kept fixed.
        Increasing $\Chi_\Gamma^J$ weakens the bound and therefore broadens the allowed interval.
    }
    \label{fig:G_Chi}
\end{figure}

For the numerical illustrations shown in \reffig{G_Chi}, I again truncate the GG series at $N=6$, so that truncation effects are negligible on the scale of the plots.
The behaviour is fully consistent with expectations: larger $\Chi_\Gamma^J$ values lead to wider allowed intervals for the extrapolated FFs.
The effect is nevertheless mild.
For instance, suppose that for a given set of FF inputs the unitarity bound is not close to saturation and implies an extrapolation uncertainty of $10\%$.
Taking $\C=1.2$, which corresponds to a $20\%$ increase in $\Chi_\Gamma^J$, would then enlarge this uncertainty only to about $11\%$, that is, by roughly one percentage point.

This exercise isolates only the effect of the normalisation entering the unitarity bound.
In particular, the FF input points are kept fixed throughout, so their uncertainties are not included in the present comparison.
In a full FF analysis, the uncertainty associated with the FF inputs must be combined with the unitarity-induced extrapolation uncertainty.
This provides another reason why varying $\Chi_\Gamma^J$ typically has only a mild effect in practice.
It also explains why conservative overestimates of $\Delta\chi_\Gamma^J$ reduce the final constraining power only moderately.
At the same time, this example highlights a conceptual distinction from SPPs, where $\C$ is, strictly speaking, infinite.
In a genuinely unitarity-bounded parametrisation, the extrapolation uncertainty is controlled by the unitarity bound.
By contrast, in an SPP the extrapolation error depends mostly on the chosen truncation order.

\subsubsection*{Dependence on $\bs{s_0}$}

Unlike $Q^2$, $k$, and $\Chi_\Gamma^J$, the choice of $s_0$ does not affect the unitarity bound~\eqref{eq:GG}.
It merely corresponds to a different conformal mapping~\eqref{eq:zmap} of the same analytic function.
Therefore, any visible $s_0$ dependence in a practical implementation can only originate from truncating the GG series at finite order.

This is consistent with the role of $s_0$ discussed in \refsubsec{SPPpar}.
The optimal choice $s_0=\sopt$, defined in \refeq{s0opt}, minimises the largest value of $|z(q^2;s_0,s_\Gamma)|$ over the semileptonic region and therefore maximises the convergence rate of the truncated $z$ expansion.
Truncated FF expansions constructed with different values of $s_0$ should therefore converge to the same FF values, but not necessarily at the same speed.

\begin{figure}[t!]
    \centering
    \begin{subfigure}{0.48\textwidth}
        \centering
        \includegraphics[width=\textwidth]{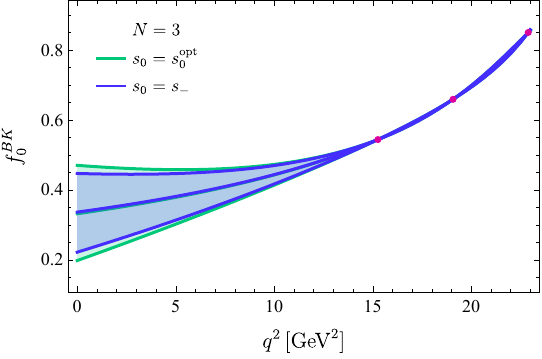}
    \end{subfigure}
    \hfill
    \begin{subfigure}{0.48\textwidth}
        \centering
        \includegraphics[width=\textwidth]{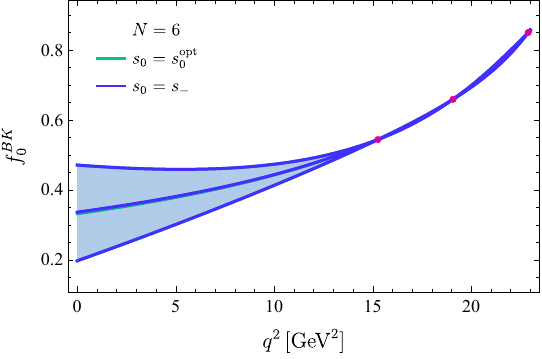}
    \end{subfigure}
    \caption{
        Dependence of the allowed GG interval for $f_0^{BK}$ on the conformal parameter $s_0$, for truncation orders $N=3$ (left) and $N=6$ (right).
        The three magenta points indicate the fixed FF input values at the representative high-$q^2$ points given in \refeq{BKq2points}.
        The curves correspond to representative choices of $s_0$, including the optimal value $s_0=\sopt$, while all other inputs are kept fixed.
        At low truncation order the optimal choice already reproduces the nearly converged interval, whereas $s_0=s_-$ gives a visibly narrower interval owing to larger truncation effects.
        For $N=6$ the curves become almost indistinguishable, showing that the exact bound is independent of $s_0$ and that the residual dependence at finite $N$ is purely a truncation effect.
    }
    \label{fig:G_s0}
\end{figure}

\reffig{G_s0} illustrates this point for $f_0^{BK}$.
For the low truncation order $N=3$, the resulting interval still exhibits a visible dependence on $s_0$.
The curve obtained with $s_0=\sopt$ is already close to the stable result, whereas more extreme choices of $s_0$, such as $s_0=0$ or $s_0=s_-$, lead to narrower intervals because the omitted higher-order terms are less suppressed.
Once the series is extended to $N=6$, however, the curves practically coincide.
This shows explicitly that the residual $s_0$ dependence is a truncation effect and that, for this FF, the choice $s_0=\sopt$ already yields a reliable uncertainty estimate at rather low order.

From a practical point of view, the conclusion is simple.
Choosing $s_0=\sopt$ does not modify the exact unitarity constraint, but it can substantially reduce the truncation order required for a stable implementation.
Since this choice is essentially cost-free, there is little reason to expand around a non-optimal point.
This provides yet another reason to prefer expansions centred at the origin of the optimal conformal map over BSZ-type variables, whose expansion parameter is effectively associated with a non-optimal choice.

\subsection{Comparison with BGL}
\label{sec:GBGL}

The plain BGL parametrisation is recovered from the GG construction by replacing $s_\Gamma$ by $s_+$, so that the interval $[s_\Gamma,s_+]$ collapses and the corresponding subthreshold contribution disappears.
At the same time, $\Chi_\Gamma^J$ reduces to $\widetilde\chi_\Gamma^J$, $\widehat\phi_\F$ reduces to the standard BGL outer function $\phi_\F$, and \refeq{GG} becomes the usual BGL expansion.
For FFs whose true analytic structure has $s_\Gamma<s_+$, this replacement amounts to neglecting the subthreshold cut.
The resulting parametrisation is then no longer rigorous, and the numerical size of the corresponding effect has to be assessed case by case.

To make this comparison transparent, I adopt exactly the same fixed-input setup as in the previous subsection.
In particular, I use the central values of Ref.~\cite{Parrott:2022rgu} at the three high-$q^2$ points given in \refeq{BKq2points}, choose $Q^2=0$ and $s_0=\sopt$, and truncate both expansions at the same order $N=6$.
For the GG parametrisation, this order is already high enough that truncation effects are numerically negligible on the scale of the plots.

\begin{figure}[t!]
    \centering
    \begin{subfigure}{0.48\textwidth}
        \centering
        \includegraphics[width=\textwidth]{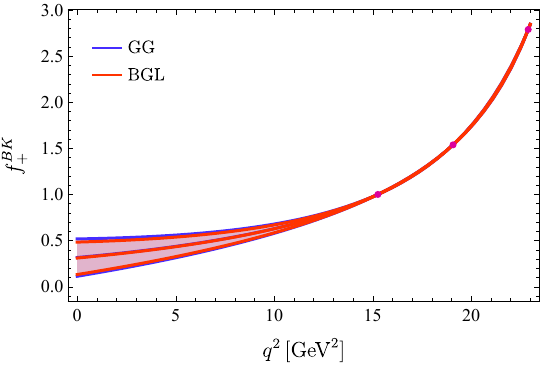}
    \end{subfigure}
    \hfill
    \begin{subfigure}{0.48\textwidth}
        \centering
        \includegraphics[width=\textwidth]{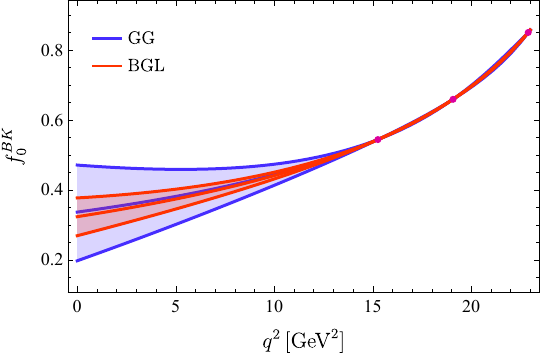}
    \end{subfigure}
    \caption{
        Comparison between the GG and plain BGL intervals for $f_+^{BK}$ (left) and $f_0^{BK}$ (right).
        The three magenta points indicate the fixed FF input values at the representative high-$q^2$ points given in \refeq{BKq2points}.
        Both expansions are truncated at the same order $N=6$, with $Q^2=0$ and $s_0=\sopt$.
        The difference between the BGL and GG parametrisations is moderate for $f_+^{BK}$ but substantially larger for $f_0^{BK}$, where the interval $[s_\Gamma,s_+]$ contains a resonance.
    }
    \label{fig:BK_BGL_GG}
\end{figure}

\reffig{BK_BGL_GG} shows the outcome for $f_+^{BK}$ and $f_0^{BK}$.
For $f_+^{BK}$ the difference between the two constructions is visible but moderate.
At $q^2=0$, the width of the plain-BGL interval is approximately $15\%$ smaller than that of the GG interval.
The plain BGL parametrisation therefore still gives a numerically similar result, even though it is not strictly rigorous.

For $f_0^{BK}$, by contrast, the discrepancy is substantially larger.
At $q^2=0$, the width of the plain-BGL interval is approximately $60\%$ smaller than that of the GG interval.
This enhanced effect originates from the presence, in the scalar channel, of a resonance with $s_\Gamma < (m_{V,1}^0)^2 < s_+$. Consequently, the modified construction must be expressed in terms of the hatted quantities introduced in \refsubsec{Gpar}.
As explained there, these quantities are built from linear combinations of higher moments and effectively encode the extra suppression needed to control the contribution from the interval $[s_\Gamma,s_+]$.
Neglecting this structure, as plain BGL does, leads to a noticeably different and typically more optimistic band.
The comparison therefore shows explicitly that the effect of the subthreshold cut need not be negligible.
This may occur even in situations where one might naively expect such contributions to be small, for instance when the corresponding channel is absent in the exact isospin limit.

\enlargethispage{\baselineskip} 

The numerical impact is generally largest in heavy-to-light channels such as $B\to K$, where the semileptonic region is comparatively wide and the unitarity bound is not close to saturation, so higher-order deformations have more room to contribute.
This point is illustrated again in \refsubsec{fitres} for a full $B\to K$ FF fit, where the same conclusion remains valid in a realistic phenomenological setting.
For $B\to D^*$ I find the corresponding differences to be smaller, since the physical $q^2$ range is shorter and the bound is already closer to saturation.

Furthermore, there is a conceptual issue with the BGL parametrisation that is absent in the GG parametrisation.
In fact, states with masses satisfying $s_\Gamma < (m_{\Gamma,i}^J)^2 < s_+$ are treated in the BGL parametrisation as poles on the first Riemann sheet, whereas they naturally belong to the second Riemann sheet.
Treating such states as ordinary subthreshold poles therefore imposes a pole-like structure on certain FFs which, especially for broad resonances, may bias their shape.
Such states are also often included in the BGL one-particle contribution to the bound, $\chi_\Gamma^J\big|_{\rm 1pt}$.
As discussed in \refsubsec{UB}, QCD-unstable states must not be included in $\chi\Gamma^J\big|_{\rm 1pt}$, in order to avoid double counting.
Their inclusion can lead to an artificially large one-particle contribution, in some cases exceeding $30\%$ of $\widehat\chi_\Gamma^J\big|_{\OPE}$, thereby reducing the remaining unitarity budget and driving the bound close to saturation.

The general lesson should be clear at this point.
Whenever $s_\Gamma<s_+$, the plain BGL parametrisation should be regarded at best as an approximation, whereas the GG construction preserves the rigour of the unitarity bound by incorporating the correct analytic structure.
The same conclusion applies to the modified DM method compared with the conventional DM construction discussed in \refsubsec{UBpar}.

\subsection{General remarks on parametrisations in the presence of subthreshold cuts}
\label{sec:remark}

\enlargethispage{\baselineskip} 

The GG parametrisation introduced in Ref.~\cite{Gopal:2024mgb} and developed here extends the rigorous content of the BGL and DM constructions to the generic situation in which the FF has a subthreshold cut.
From a practical point of view, the construction is nevertheless as simple to implement as plain BGL.
Once an estimate, or merely an upper bound, on $\Delta\chi_\Gamma^J$ (or on its hatted analogue) is available, the fit machinery, the diagonal coefficient bound, and the truncation-error estimate remain of the familiar BGL type.

It is useful to contrast this strategy with earlier treatments of subthreshold singularities.
Ref.~\cite{Caprini:1995wq} already shows that the additional cut can be incorporated exactly once an explicit ansatz for the discontinuity along the interval $[s_\Gamma,s_+]$ is chosen, and the numerical size of the effect is then estimated from a parametrisation of that cut whose normalisation is inferred from a resonance model.
Ref.~\cite{Boyd:1995sq} follows a closely related logic.
These analyses were important in showing that the problem is technically tractable.
However, in practice these approaches rely on a precise cancellation of the singular behaviour associated with the subthreshold cut.
Since the discontinuity along this cut is not known, such a cancellation cannot be implemented rigorously in phenomenological applications.
This is not analogous to removing an isolated pole with a Blaschke factor, because it depends on the full functional form of the cut discontinuity.
By contrast, the GG construction does not require one to model the detailed shape of the cut.
For the purpose of deriving a rigorous bound, any upper bound on $\Delta\chi_\Gamma^J$ is sufficient, and an overestimate just weakens the constraint without invalidating~it.

Together with D.~van Dyk and J.~Virto, I had already proposed a related idea in Ref.~\cite{Gubernari:2020eft}, i.e. prior to Ref.~\cite{Gopal:2024mgb}.
In that work, the conformal map is chosen as in the GG parametrisation, namely $z(q^2;s_0,s_\Gamma)$, and the expansion is written in polynomials orthonormal on the blue arc of \reffig{zmapBDstarGG}.
The problem is that, although the expansion coefficients satisfy a diagonal bound analogous to the usual BGL one, the corresponding polynomials are not uniformly bounded inside the unit disc.
For certain values of $z$ inside the unit disc, their absolute value grows exponentially with the polynomial degree.
Hence the omitted tail is not controlled by a suppressing factor such as $|z|^{N+1}$, and the construction does not yield a genuinely unitarity-bounded parametrisation in the BGL sense, nor a direct truncation-error estimate such as \refeq{BGLtruncerr}.

Ref.~\cite{Flynn:2023qmi} proposes a construction analogous to that of Ref.~\cite{Gubernari:2020eft}.
In that approach, one retains the standard BGL monomials in $z$ rather than introducing polynomials, and accounts for the subthreshold branch cut by restricting the unitarity integral to the relevant arc.
As a result, the diagonal BGL bound is replaced by a non-diagonal quadratic form in the monomial coefficients.
This avoids the exponentially growing arc-orthonormal polynomials of Ref.~\cite{Gubernari:2020eft}.
However, in the terminology of \refsubsec{NUBpar}, the resulting parametrisation is only quasi-bounded: after truncation, retained and omitted coefficients still mix, and therefore no direct rigorous truncation-error estimate can be derived.

Refs.~\cite{Simula:2025lpc,Simula:2025fft} adopt the same underlying construction as the GG parametrisation~\cite{Gopal:2024mgb}, but supplement it by splitting the bound~\eqref{eq:GG} into several separate bounds, in particular one for the magenta arc and one for the blue arc in \reffig{zmapBDstarGG}.
This is a conceptually interesting refinement, although these new bounds are non-diagonal and thus suffer from the limitations discussed in \refsubsec{NUBpar}.
Nonetheless, such constraints can improve the numerical precision in specific applications, especially if $\Delta\chi_\Gamma^J$ is known precisely.
A detailed analysis of this refinement lies beyond the scope of the present work.
\\

The GG construction also admits several natural extensions.
From a phenomenological perspective, the same reasoning should apply --- up to minor modifications --- to FFs in $\Lambda_b$ and lighter-hadron decays.
From a conceptual perspective, it would be worthwhile to combine the present framework with additional exact information on the FFs, such as their perturbative behaviour at large $|q^2|$~\cite{Lepage:1980fj,Akhoury:1994tnu}, their behaviour at threshold~\cite{Buck:1998kp,Bourrely:2008za}, and the indirect information encoded in resonances above threshold~\cite{Caprini:2017ins}.
The non-trivial aspect would be to incorporate these constraints without sacrificing the diagonal coefficient bound that makes the GG parametrisation practically convenient.
Preserving a diagonal coefficient bound would likely require  modifying the inner product defined in \refeq{BGLinner}.
In that case the natural basis would in general no longer be the monomials, but suitable orthogonal polynomials.
\\

In summary, the GG parametrisation incorporates the correct subthreshold cut without requiring a model for its detailed discontinuity, while retaining a diagonal coefficient bound and a rigorous truncation-error estimate.
It therefore provides a natural framework for the combined FF analyses presented in the next section.

\section{Form-factor analyses}
\label{sec:fits}

I carry out three combined analyses of $B$-meson FFs, grouped according to the underlying partonic transition:
\begin{mybox}
\begin{description}[
    labelindent=0pt,
    labelwidth=\widthof{$\bs{b\to u}$},
    leftmargin=!,
    labelsep=0.6em,
    font=\normalfont
    ]
    \item[$\bs{b\to u}$]
    The channels considered are $B\to\pi$, $B_s\to K$, and $B_s\to K^*$.
    In the isospin limit, this analysis is equivalent to that of $b\to d$ transitions.
    This is the first unitarity-based analysis combining these FFs.

    \item[$\bs{b\to s}$]
    The channels considered are $B\to K$, $B\to K^*$, and $B_s\to\phi$.
    This analysis supersedes that of Ref.~\cite{Gubernari:2023puw}.

    \item[$\bs{b\to c}$]
    The channels considered are $B\to D$, $B_s\to D_s$, $B\to D^*$, and $B_s\to D_s^*$.
    This is the first unitarity-based analysis combining these FFs without using HQET.
\end{description}
\end{mybox}
\noindent
For these analyses, I use the GG parametrisation introduced in the previous section.
All relevant inputs are provided there and in \refapp{outer} and \refapp{meson-masses}.
In addition, I adopt the optimal expansion point $\sopt$ defined in \refeq{s0opt}.
The implementation of the endpoint relations in \refeqa{endp0}{endpsm} within this parametrisation is discussed in \refapp{endpointrel}.

The rest of this section describes the analysis strategy (\refsubsec{fitstrat}), the datasets entering the likelihood (\refsubsec{FFinputs}), and the corresponding FF results (\refsubsec{fitres}).

\subsection{Analysis setup and strategy}
\label{sec:fitstrat}

I have implemented the GG parametrisation in \EOS~\cite{EOSAuthors:2021xpv}, starting from version 1.0.20~\cite{EOS:v1.0.20}, under the label \texttt{G2026}.
The code is publicly available on \href{https://github.com/eos/eos/}{\texttt{GitHub}}.
I have cross-checked the implementation against an independent \texttt{Mathematica} code.
The statistical analyses are performed in \EOS within a Bayesian framework.
Denoting by $\bs{\alpha}$ the set of fitted GG coefficients and by $\E$ the theoretical inputs, Bayes' theorem takes the form
\begin{align}
    \label{eq:bayes}
    p(\bs{\alpha}\,|\,\E)
    =
    \frac{p(\E\,|\,\bs{\alpha})\,p(\bs{\alpha})}{p(\E)}
    \, .
\end{align}
The likelihood $p(\E\,|\,\bs{\alpha})$ encodes the lattice-QCD and LCSR information, while the prior $p(\bs{\alpha})$ specifies the allowed ranges of the expansion coefficients and auxiliary parameters.
A practical advantage of bounded parametrisations is that the $z$-expansion coefficients are restricted to finite intervals.
More specifically, each coefficient must lie within the range $[-1,1]$, which removes the need to introduce arbitrarily large prior domains.
In practice, I use uniform priors and iteratively adjust their ranges so that they fully contain the posterior support, without leaving unnecessary extra room.

The fits are repeated for successive truncation orders, namely $N=2$, $N=3$ and $N=4$.
I regard a given truncation as sufficient only once the posterior FF bands are stable under the replacement $N\to N+1$.
Posterior samples are generated with dynamic nested sampling~\cite{Higson:2018}, using the \texttt{dynesty} implementation~\cite{Speagle:2020,dynesty:v3.0.0} interfaced to \EOS.
\\

Unitarity bounds are imposed in the combined analyses only after summing the contributions of all mesons carrying the same quantum numbers.
For example, in the $b\to c$ analysis, the saturation observable in the channel $(\Gamma,J)=(V,1)$ reads
\begin{align}
    \label{eq:satV1}
    \S_V^1
    =
    \sum_{n=0}^\infty \left|\alpha_{f_+^{BD},n}^{\rm GG}\right|^2
    +
    \sum_{n=0}^\infty \left|\alpha_{f_+^{B_sD_s},n}^{\rm GG}\right|^2
    +
    \sum_{n=0}^\infty \left|\alpha_{V^{BD^*},n}^{\rm GG}\right|^2
    +
    \sum_{n=0}^\infty \left|\alpha_{V^{B_sD_s^*},n}^{\rm GG}\right|^2
    \le 1 \,.
\end{align}
The expressions for the other saturation observables follow straightforwardly from \reftab{outer}.
The relevant saturation observables $\S_\Gamma^J$ are included in the analyses with upper bound $1$ and with a $5\%$ tolerance, in order to account for the residual uncertainty entering the $\Chi_\Gamma^J$ calculation.
The action of this constraint is simple:
\begin{align}
    \Delta \log p(\E\,|\,\bs{\alpha})\Big|_{\rm UB}
    =
    \begin{cases}
        0, &
        \S_\Gamma^J \le 1 \, ,\\[2mm]
        - \dfrac{1}{2}
        \left(
            \dfrac{\S_\Gamma^J - 1}{0.05}
        \right)^2, &
        \S_\Gamma^J > 1 \, ,
    \end{cases}
\end{align}
where $\Delta \log p(\E\,|\,\bs{\alpha})\big|_{\rm UB}$ denotes the additive unitarity-bound contribution to the log-likelihood.
Thus, in the actual analyses the bound is implemented as a soft one-sided likelihood penalty rather than as a strict cut.
In the formal limit of zero tolerance, the same constraint reduces to a hard exclusion of points with $\S_\Gamma^J > 1$.

I therefore impose the unitarity bounds directly in the statistical analysis, rather than applying the unitarity filter~\eqref{eq:DMfilter} to the lattice-QCD or LCSR input samples before the fit.
For noisy FF inputs, a sample drawn from the input distribution that violates the condition would indeed be non-unitary if interpreted as an exact FF.
Such a sample, however, is not the underlying physical FF, but just a statistical draw from an estimated distribution of FF values and correlations.
Rejecting such samples before the fit effectively means that the input data are no longer drawn from the original statistical distribution, but from a modified one in which all non-unitary fluctuations have been removed by hand.
The fit is therefore performed on an artificially constrained ensemble rather than on the distribution implied by the original lattice-QCD or LCSR input.
This can shift the central values, reduce the uncertainties, modify the correlations, and render the filtered ensemble non-Gaussian.
As a consequence, fitting only the surviving samples may produce biased FF results and may lead to underestimated uncertainties.\footnote{
    A simple analogy is provided by a quantity that is known on physical grounds to be positive, for instance the mass \(m>0\) of a given particle.
    Suppose that a noisy measurement yields \(m=0.10\pm0.15\).
    Samples drawn from a Gaussian distribution based on this measurement will then include some values with \(m<0\).
    If interpreted as exact masses, such samples are unphysical; at the same time, they are perfectly ordinary fluctuations of the measurement.
    Removing all samples with \(m<0\) before recomputing a new mean and uncertainty would shift the mean upward and artificially reduce the error.
    The same issue would also affect combinations with other measurements.
    For example, combining the filtered result with a more precise determination such as \(m=0.12\pm0.03\) would lead to a biased estimate, induced by the positivity cut.
    A statistically cleaner treatment is to impose \(m>0\) as a constraint on the true parameter directly in the likelihood fit or posterior, rather than filtering the inputs.
}
For this reason, unitarity should be imposed within the statistical analysis itself.
In this way, unitarity enters the fit as physical information, rather than being used as an external criterion to discard statistical fluctuations in the input data.

This statistical treatment is specific to coefficient-based parametrisations such as BGL or the GG form used here.
In a GG fit the primary variables are the expansion coefficients $\alpha_{\F,n}^{\rm GG}$, while the lattice-QCD and LCSR information enters through the likelihood.
By contrast, the DM construction requires a different treatment since the sampled quantities are the FF values at selected kinematic points.
Hence compatibility with analyticity and unitarity is naturally checked sample by sample through the filter~\eqref{eq:DMfilter}.
The physical content is the same in both cases; what changes is only the choice of variables and, accordingly, the practical realisation of the unitarity constraint.

\subsection{Form-factor datasets}
\label{sec:FFinputs}

A common way to use lattice QCD FF results is the following.
Lattice QCD collaborations typically publish the coefficients of a given parametrisation together with their covariance matrix.
From this information, one can generate FF data points at selected values of $q^2$.
The number of data points should not exceed the number of independent coefficients entering the published distribution; otherwise, the correlation matrix of the data points becomes singular.
Moreover, the chosen $q^2$ values should lie in the kinematic region directly constrained by the lattice simulation.
If this requirement is not met, the data points no longer reflect genuine lattice information alone, but also a parametrisation-dependent extrapolation into regions not determined by the simulation, thereby introducing an additional source of systematic uncertainty.

An analogous procedure can be applied to LCSR results when they are not provided directly at discrete $q^2$ points together with their correlations.
Throughout the discussion of the datasets below, all tensor FFs are taken at the common scale $\mu=4.183\,\GeV$, unless explicitly stated otherwise.

\subsubsection*{$\bs{b \to u}$ analysis}

The $b\to u$ analysis combines the channels $B\to\pi$, $B_s\to K$, and $B_s\to K^*$.
For each of them, I use the low-$q^2$ information from LCSRs and the high-$q^2$ information from lattice QCD, retaining the published correlations within each input set.

\begin{description}[
    labelindent=0pt,
    labelwidth=\widthof{$\bs{B_s\to K^*}$},
    leftmargin=!,
    labelsep=0.6em,
    font=\normalfont
    ]
\item[$\bs{B\to\pi}$]
    For the three FFs $f_+^{B\pi}$, $f_0^{B\pi}$, and $f_T^{B\pi}$, I include both the light-meson LCSR determination of Ref.~\cite{Leljak:2021vte} and the $B$-meson LCSR determination of Ref.~\cite{Gubernari:2018wyi}.
    As explained in \refapp{FFcal}, these are fully independent calculations and can therefore be included simultaneously.
    Ref.~\cite{Leljak:2021vte} provides correlated points at $q^2=\{-10,-5,0,5,10\}~\GeV^2$ for $f_+^{B\pi}$ and $f_T^{B\pi}$, and at $q^2=\{-10,-5,5,10\}~\GeV^2$ for $f_0^{B\pi}$.
    Ref.~\cite{Gubernari:2018wyi} provides the analogous low-$q^2$ information at $q^2=\{-15,-10,-5,0,5\}~\GeV^2$ for $f_+^{B\pi}$ and $f_T^{B\pi}$, and at $q^2=\{-15,-10,-5,5\}~\GeV^2$ for $f_0^{B\pi}$.
    At high $q^2$, I use the lattice-QCD average compiled by FLAG~2024~\cite{FLAG:2024oxs} for $f_+^{B\pi}$ and $f_0^{B\pi}$.
    From this result, I generate five data points, namely at $q^2=\{18,22,26\}~\GeV^2$ for $f_+^{B\pi}$ and at $q^2=\{18,26\}~\GeV^2$ for $f_0^{B\pi}$.
    The FLAG average is obtained from the computations of Refs.~\cite{FermilabLattice:2015mwy,Flynn:2015mha,Colquhoun:2022atw}.
    Because these determinations are not mutually compatible, the FLAG collaboration inflates the uncertainties of the averaged BCL coefficients by a factor of $\sqrt{\chi^2/\text{d.o.f.}}$.
    I omit the lattice-QCD determination of $f_T^{B\pi}$ from Ref.~\cite{FermilabLattice:2015cdh}, since its consistent inclusion together with the FLAG average would require propagating correlations with the Fermilab/MILC vector and scalar FF results of Ref.~\cite{FermilabLattice:2015mwy}.

\item[$\bs{B_s\to K}$]
    For $f_+^{B_sK}$, $f_0^{B_sK}$, and $f_T^{B_sK}$, I use both the light-meson LCSR determination of Ref.~\cite{Bolognani:2023mcf} and the $B$-meson LCSR determination of Ref.~\cite{Biswas:2025drz}.
    Ref.~\cite{Biswas:2025drz} provides correlated points at $q^2=\{-15,-10,-5,0,5\}~\GeV^2$ for $f_+^{B_sK}$ and $f_T^{B_sK}$, and at $q^2=\{-15,-10,-5,5\}\,\allowbreak\GeV^2$ for $f_0^{B_sK}$.
    Ref.~\cite{Bolognani:2023mcf} provides points at $q^2=\{-10,-5,0,5\}~\GeV^2$ for $f_+^{B_sK}$ and $f_T^{B_sK}$, and at $q^2=\{-10,-5,5\}~\GeV^2$ for $f_0^{B_sK}$.
    For the lattice QCD input, I use the FLAG~2024 average~\cite{FLAG:2024oxs} for $f_+^{B_sK}$ and $f_0^{B_sK}$, generating seven data points at $q^2=\{17.5,19.5,21.5,23.5\}~\GeV^2$ for $f_+^{B_sK}$ and at $q^2=\{17.5,19.5,23.5\}~\GeV^2$ for $f_0^{B_sK}$.
    This average is based on the computations of Refs.~\cite{Bouchard:2014ypa,FermilabLattice:2019ikx,Flynn:2023nhi}.
    Since these computations are not mutually consistent, the FLAG collaboration rescales the uncertainties of the averaged BCL coefficients by a factor of $\sqrt{\chi^2/\text{d.o.f.}}$.
    No final lattice-QCD determination of $f_T^{B_sK}$ with a complete uncertainty budget is currently available, although preliminary calculations have been reported~\cite{Jeong:2024glv,Roberts:2025bsp}.

\item[$\bs{B_s\to K^*}$]
    At low $q^2$, I include both the light-meson LCSR determination of Ref.~\cite{Bharucha:2015bzk} and the $B$-meson LCSR determination of Ref.~\cite{Biswas:2025drz}.
    Ref.~\cite{Biswas:2025drz} provides correlated points at $q^2=\{-15,-10,-5,0,5\}\, \allowbreak\GeV^2$ for $V^{B_sK^*}$, $A_0^{B_sK^*}$, $A_1^{B_sK^*}$, $T_2^{B_sK^*}$, and $T_{23}^{B_sK^*}$, and at $q^2=\{-15,-10,-5,5\}~\GeV^2$ for $A_{12}^{B_sK^*}$ and $T_1^{B_sK^*}$.
    The finite-width effects of the $K^*$ are taken into account according to Ref.~\cite{Descotes-Genon:2019bud}, i.e. by multiplying the central values of the $B_s\to K^*$ FFs by a factor of $1.10$.
    Regarding Ref.~\cite{Bharucha:2015bzk}, I adopt a more conservative treatment than in the original work.
    First, I use the calculation only up to $8\, \GeV^2$, rather than $14\, \GeV^2$, in order to avoid potentially large power corrections in the corresponding OPE.
    Second, I assign an additional $10\%$ uncorrelated systematic uncertainty to account for finite-width effects of the $K^*$, since no dedicated study of these effects has been carried out within light-meson LCSRs.
    Using the BSZ coefficients provided, I then generate data points at $q^2=\{0,4,8\}~\GeV^2$ for $V^{B_sK^*}$, $A_0^{B_sK^*}$, $A_1^{B_sK^*}$, $T_1^{B_sK^*}$, and $T_{23}^{B_sK^*}$, and at $q^2=\{4,8\}~\GeV^2$ for $A_{12}^{B_sK^*}$ and $T_2^{B_sK^*}$.
    At high $q^2$, the lattice-QCD computation of Ref.~\cite{Horgan:2015vla}, which updates Ref.~\cite{Horgan:2013hoa}, provides correlated results for the set $\{V^{B_sK^*},A_0^{B_sK^*},A_1^{B_sK^*},A_{12}^{B_sK^*}\}$ and, separately, for the set $\{T_1^{B_sK^*},T_2^{B_sK^*},T_{23}^{B_sK^*}\}$.
    Since the published information is split in this way, correlations between the \hbox{(axial-)}vector and tensor sectors are not included in the fit.
    Using the SPP coefficients given in Ref.~\cite{Horgan:2015vla}, I generate data points for each FF at $q^2=\{16,20\}~\GeV^2$.
    Taken together, all these inputs constrain the FFs at both low and high $q^2$.
\end{description}
The $B\to\rho$ FFs are particularly challenging because the $\rho$ is a broad resonance, with $\Gamma_\rho \simeq 150\,\mathrm{MeV}$, and therefore lies outside the domain of validity of the narrow-width approximation.
Consequently, existing lattice QCD determinations in Refs.~\cite{Bowler:2004zb,Flynn:2008zr}, which treat the $\rho$ as stable, are subject to uncontrolled systematics associated with finite-width effects.
Moreover, these results were obtained in the quenched approximation, i.e.\ neglecting sea-quark contributions, which introduces an additional bias that cannot be systematically improved.
For these reasons, I consider the quoted lattice QCD computations of the $B\to\rho$ FFs insufficiently reliable for precision phenomenology.
While the recent unquenched study of Ref.~\cite{Leskovec:2025gsw}, based on the $B\to\pi\pi$ amplitude, represents important progress, this result was obtained at unphysical pion masses and therefore cannot yet be used as a quantitative input for precision phenomenology.
LCSR calculations of the $B\to\rho$ FFs are also available, both in the narrow-width approximation~\cite{Bharucha:2015bzk,Gubernari:2018wyi} and with width effects taken into account~\cite{Kang:2013jaa,Cheng:2017smj,Cheng:2017sfk,Cheng:2025hxe}.
However, these calculations are available only in the low-$q^2$ region, and the errors are too large to obtain significant results in the whole semileptonic region.
For these reasons, I do not include $B\to\rho$ FFs in this analysis.

\subsubsection*{$\bs{b \to s}$ analysis}

The $b\to s$ analysis combines the channels $B\to K$, $B\to K^*$, and $B_s\to\phi$.
For the pseudoscalar channel I use lattice-QCD information only, while for the two vector channels I combine low-$q^2$ information from light-meson and $B$-meson LCSRs with the available high-$q^2$ lattice-QCD results.
Published correlations within each input set are retained throughout.

\begin{description}[
    labelindent=0pt,
    labelwidth=\widthof{$\bs{B\to K^*}$},
    leftmargin=!,
    labelsep=0.6em,
    font=\normalfont
    ]
\item[$\bs{B\to K}$]
    For the three FFs $f_+^{BK}$, $f_0^{BK}$, and $f_T^{BK}$, I do not include any LCSR input, since this channel is already very well constrained by lattice QCD.
    I use the lattice-QCD inputs of Refs.~\cite{Bouchard:2013eph,Bailey:2015dka,Parrott:2022rgu}, following Ref.~\cite{Gubernari:2023puw}.
    For Refs.~\cite{Bouchard:2013eph,Bailey:2015dka}, I take correlated points at $q^2=\{17,20,23\}~\GeV^2$ for $f_+^{BK}$, $f_0^{BK}$, and $f_T^{BK}$.
    For Ref.~\cite{Parrott:2022rgu}, I take correlated points at $q^2=\{0,12,22.9\}~\GeV^2$ for $f_0^{BK}$ and $f_T^{BK}$, and at $q^2=\{12,22.9\}~\GeV^2$ for $f_+^{BK}$.

\item[$\bs{B\to K^*}$]
    At low $q^2$, I include both the light-meson LCSR determination of Ref.~\cite{Bharucha:2015bzk} and the $B$-meson LCSR determination of Ref.~\cite{Gubernari:2018wyi}.
    Ref.~\cite{Gubernari:2018wyi} provides correlated points at $q^2=\{-15,-10,-5,0,5\}~\GeV^2$ for $A_1^{BK^*}$, $A_2^{BK^*}$, $V^{BK^*}$, $T_1^{BK^*}$, and $T_{23}^{BK^*}$, and at $q^2=\{-15,-10,-5,5\}~\GeV^2$ for $A_0^{BK^*}$ and $T_2^{BK^*}$.
    As for the $B_s\to K^*$ FFs, I account for the finite width of the $K^*$ by rescaling the central values of the $B\to K^*$ FFs by a factor of $1.10$.
    For Ref.~\cite{Bharucha:2015bzk}, I adopt the same conservative treatment described above.
    I therefore use the calculation only up to $8\,\GeV^2$ and assign an additional $10\%$ uncorrelated systematic uncertainty to account for finite-width effects of the $K^*$.
    Using the published BSZ coefficients, I then generate data points at $q^2=\{0,4,8\}~\GeV^2$ for $V^{BK^*}$, $A_0^{BK^*}$, $A_1^{BK^*}$, $T_1^{BK^*}$, and $T_{23}^{BK^*}$, and at $q^2=\{4,8\}~\GeV^2$ for $A_{12}^{BK^*}$ and $T_2^{BK^*}$.
    At high $q^2$, I use the 12 lattice-QCD data points in the range $11.9 \lesssim q^2 \lesssim 17.8~\GeV^2$ provided by the authors of Refs.~\cite{Horgan:2015vla,Horgan:2013hoa} for each of the seven FFs, as implemented in the \EOS software. 

\item[$\bs{B_s\to\phi}$]
    At low $q^2$, I include both the light-meson LCSR determination of Ref.~\cite{Bharucha:2015bzk} and the $B$-meson LCSR determination of Ref.~\cite{Gubernari:2020eft}.
    Ref.~\cite{Gubernari:2020eft} provides correlated points at $q^2=\{-15,-10,-5,0,5\}~\GeV^2$ for $A_1^{B_s\phi}$, $A_2^{B_s\phi}$, $V^{B_s\phi}$, $T_1^{B_s\phi}$, and $T_{23}^{B_s\phi}$, and at $q^2=\{-15,-10,\allowbreak-5,5\}~\GeV^2$ for $A_0^{B_s\phi}$ and $T_2^{B_s\phi}$.
    From Ref.~\cite{Bharucha:2015bzk} I generate data points at $q^2=\{0,4,8\}\,\allowbreak\GeV^2$ for $V^{B_s\phi}$, $A_0^{B_s\phi}$, $A_1^{B_s\phi}$, $T_1^{B_s\phi}$, and $T_{23}^{B_s\phi}$, and at $q^2=\{4,8\}~\GeV^2$ for $A_{12}^{B_s\phi}$ and $T_2^{B_s\phi}$.
    At high $q^2$, I use the lattice-QCD results of Ref.~\cite{Horgan:2015vla}.
    The situation is analogous to that of the $B_s\to K^*$ FFs discussed in the same reference.
    I therefore generate data points for each FF at $q^2=\{16.0,18.9\}~\GeV^2$ using the published SPP coefficients.
\end{description}

\subsubsection*{$\bs{b \to c}$ analysis}

The $b\to c$ analysis combines the channels $B\to D$, $B_s\to D_s$, $B\to D^*$, and $B_s\to D_s^*$.
In contrast to the $b\to u$ and $b\to s$ cases, I rely exclusively on lattice-QCD input, using essentially the same datasets as in Ref.~\cite{Bordone:2025jur}.
Whenever available, the published correlations within each input set are retained.

\begin{description}[
    labelindent=0pt,
    labelwidth=\widthof{$\bs{B_s\to D_s^*}$},
    leftmargin=!,
    labelsep=0.6em,
    font=\normalfont
    ]
\item[$\bs{B\to D}$]
    For $f_+^{BD}$ and $f_0^{BD}$, I include the results of Refs.~\cite{MILC:2015uhg,Na:2015kha}.
    From Ref.~\cite{MILC:2015uhg}, I generate correlated points at $q^2=\{0,8.49,10.07,11.64\}~\GeV^2$ for $f_+^{BD}$ and at $q^2=\{8.49,10.07,11.64\}~\GeV^2$ for $f_0^{BD}$.
    From Ref.~\cite{Na:2015kha}, I generate correlated points at $q^2=\{0,9.28,11.64\}~\GeV^2$ for $f_+^{BD}$ and at $q^2=\{9.28,11.64\}~\GeV^2$ for $f_0^{BD}$.
    In both cases, the point $f_0^{BD}(0)$ is not counted separately because of the kinematic relation $f_+^{BD}(0)=f_0^{BD}(0)$.
    For the tensor FF, I use the lattice determination of $f_T^{BD}(s_-)/f_+^{BD}(s_-)$ from Ref.~\cite{Atoui:2013zza}.

\item[$\bs{B_s\to D_s}$]
    For $f_+^{B_sD_s}$ and $f_0^{B_sD_s}$, I use the HPQCD result of Ref.~\cite{McLean:2019qcx}, from which I generate correlated points at $q^2=\{0,5.78,11.55\}~\GeV^2$ for $f_+^{B_sD_s}$ and at $q^2=\{5.78,11.55\}~\GeV^2$ for $f_0^{B_sD_s}$.
    The tensor FF is constrained through the lattice result for $f_T^{B_sD_s}(s_-)/f_+^{B_sD_s}(s_-)$ in Ref.~\cite{Atoui:2013zza}.
    In addition, the two pseudoscalar channels are linked by the lattice determination of the ratio $f_0^{B_sD_s}(m_\pi^2)/f_0^{BD}(m_\pi^2)$ from Ref.~\cite{Bailey:2012rr}.

\item[$\bs{B\to D^*}$]
    For the four FFs entering Standard-Model phenomenology, I include both the Fermilab/MILC and JLQCD lattice-QCD determinations of Refs.~\cite{FermilabLattice:2021cdg,Aoki:2023qpa}.
    Following the \EOS implementation, I convert the published results into the basis $\{V^{BD^*},A_0^{BD^*},A_1^{BD^*},\allowbreak A_{12}^{BD^*}\}$.
    This yields correlated points at $q^2=\{7.08,8.57,10.06\}~\GeV^2$ for each of these FFs from Ref.~\cite{FermilabLattice:2021cdg}, and at $q^2=\{8.562,9.411,10.154\}~\GeV^2$ from Ref.~\cite{Aoki:2023qpa}.
    In addition, I use the HPQCD result of Ref.~\cite{Harrison:2023dzh}, which provides the full FF set~\eqref{eq:BVFFbasis}.
    To avoid numerical instabilities in the covariance-matrix inversion, I discard the second-lowest published $q^2$ point and retain correlated points at $q^2=\{0,5.35,8.02,10.69\}~\GeV^2$ for $V^{BD^*}$, $A_1^{BD^*}$, and $T_1^{BD^*}$, at $q^2=\{5.35,8.02,10.69\}~\GeV^2$ for $A_0^{BD^*}$ and $T_2^{BD^*}$, and at $q^2=\{0,5.35,8.02\}~\GeV^2$ for $A_{12}^{BD^*}$ and $T_{23}^{BD^*}$.

\item[$\bs{B_s\to D_s^*}$]
    For this channel, I use the HPQCD determination of Ref.~\cite{Harrison:2023dzh}, which provides the full FF set~\eqref{eq:BVFFbasis}.
    As in the $B\to D^*$ case, I discard the second-lowest published $q^2$ point.
    I then retain correlated points at $q^2=\{0,5.30,7.95,10.59\}~\GeV^2$ for $V^{B_sD_s^*}$, $A_1^{B_sD_s^*}$, and $T_1^{B_sD_s^*}$, at $q^2=\{5.30,7.95,10.59\}~\GeV^2$ for $A_0^{B_sD_s^*}$ and $T_2^{B_sD_s^*}$, and at $q^2=\{0,5.30,7.95\}~\GeV^2$ for $A_{12}^{B_sD_s^*}$ and $T_{23}^{B_sD_s^*}$.
\end{description}

\subsection{Fit results and use of the posterior samples}
\label{sec:fitres}

Using the parametrisation~\eqref{eq:GG}, the analysis strategy discussed in \refsubsec{fitstrat}, and the FF calculations listed in \refsubsec{FFinputs}, I obtain posterior samples for the three combined FF analyses.
The \EOS fits are specified by YAML analysis files, whose usage is documented at \href{https://eoshep.org/doc/}{\texttt{eoshep.org/doc}}.
All numerical results, the \EOS analysis files, and the associated plots are collected in the supplementary material~\cite{suppl_unitb}.

The posterior samples are one of the main numerical results of this work.
For each fit, the supplementary material contains the nested-sampling output in machine-readable form through the files \texttt{nested/description.yaml} and \texttt{nested/dynesty\_results.npy}.
The former fixes the list and ordering of the varied parameters, while the latter stores the complete output of the nested-sampling run.
In addition, the directory \texttt{samples/} contains the weighted posterior samples of the fitted GG coefficients, whereas the \texttt{pred-*} directories contain the corresponding samples for the derived FF and saturation observables.
Taken together, these files are sufficient to reconstruct the posterior ranges and correlations, and to propagate them to any observable of interest.

In addition to the nested-sampling output, the supplementary material~\cite{suppl_unitb} also provides a lightweight \texttt{Mathematica} interface for practical use of the fits.
More specifically, the notebook \texttt{mathematica\_notebook/\allowbreak GG\_FFs.nb}, together with the files collected in \texttt{mathematica\_notebook/\allowbreak GG\_parameters/}, contains the posterior mean vectors and covariance matrices of the fitted GG coefficients in machine-readable form and can be used to generate correlated samples, extract central values and uncertainties, and reproduce the FF plots.
For the three combined analyses, the corresponding multivariate-Gaussian approximation reproduces the main features of the full posterior reasonably well and is therefore convenient for quick phenomenological applications.
At the same time, in order to keep the resampling procedure efficient, this interface retains only correlations within a given decay channel and not across different channels belonging to the same quark transition.
For precision applications, and in particular whenever the complete correlation structure is relevant, the full nested-sampling output described above should therefore be used.
\\

The global goodness-of-fit indicators are summarised in \reftab{fitres}.
The $b\to u$ and $b\to s$ analyses both yield global $p$-values above $99.99\%$, with lowest local $p$-values of about $74\%$ and $77\%$, respectively.
This indicates that none of the heavy-to-light datasets is in any visible tension with the combined fit.
At the same time, such very large global $p$-values should not be over-interpreted as evidence for an exceptionally constraining fit.
They also reflect the deliberately conservative setup of the analysis, which originates both from the large coefficient space and from the inflated uncertainties assigned to several inputs, including the FLAG averages and the additional $10\%$ systematic uncertainty applied to the $K^*$ LCSR constraints.
The $b\to c$ analysis also exhibits a good overall fit quality, with $\chi^2=27.51$ for 22 nominal degrees of freedom and a global $p$-value of $19.26\%$.
In that fit, the least compatible input is the ratio $f_0^{B_sD_s}(m_\pi^2)/f_0^{BD}(m_\pi^2)$, whose local $p$-value is about $54\%$, while the individual $B\to D^*$ lattice-QCD datasets have local $p$-values of approximately $70\%$ or higher.
I therefore find no evidence of significant tension among the input datasets within the adopted uncertainties and correlation assumptions.

\begin{table}[t!]
    \newcommand{\pp}{{\phantom{*}}}
    \centering
    \renewcommand{\arraystretch}{1.4}
    \begin{tabular}{
        c@{\hspace{0.4cm}}
        c@{\hspace{0.4cm}}
        c@{\hspace{0.4cm}}
        c@{\hspace{0.4cm}}
        c@{\hspace{0.4cm}}
        c
    }
        \toprule
        Analysis             &
        Truncation order $N$ &
        Free fit parameters  &
        $\chi^2$             &
        d.o.f.               &
        $p$-value [\%]       \\[0.1cm]
        \toprule
        $b\to u$             &
        4                    &
        59                   &
        29.87                &
        72                   &
        99.99                \\ 
        $b\to s$             &
        4                    &
        76                   &
        59.79                &
        152                  &
        99.99                \\ 
        $b\to c$             &
        3                    &
        70                   &
        27.50                &
        22                   &
        19.26                \\ 
        \bottomrule
    \end{tabular}
    \caption{
        Summary of the three combined GG fits discussed in this section.
        ``Free fit parameters'' counts the independent GG coefficients varied in the fit after imposing the endpoint relations of \refapp{endpointrel}.
        The quoted $\chi^2$ values refer to the best-fit point.
        The nominal number of degrees of freedom is $N_{\rm obs}-N_{\rm par}$; the six unitarity constraints included in each analysis are not counted as additional observations.
        The $p$-values use the corresponding $\chi^2$ reference distribution and should be regarded as indicative for these constrained fits.
        \label{tab:fitres}
    }
\end{table}

\begin{figure}[p]
    \centering
    \includegraphics[width=0.77\textwidth]{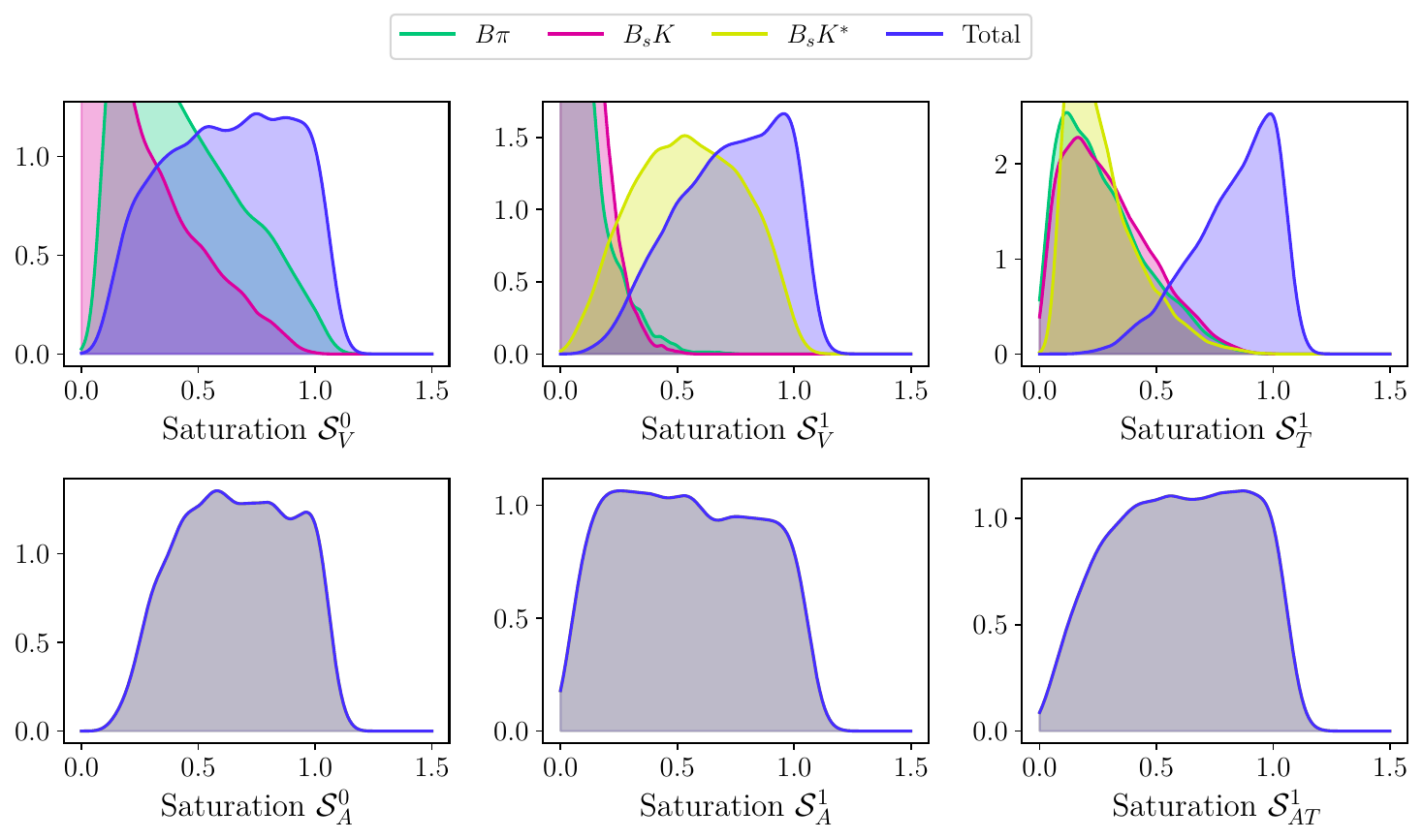}
    \\[0.3cm]
    \includegraphics[width=0.77\textwidth]{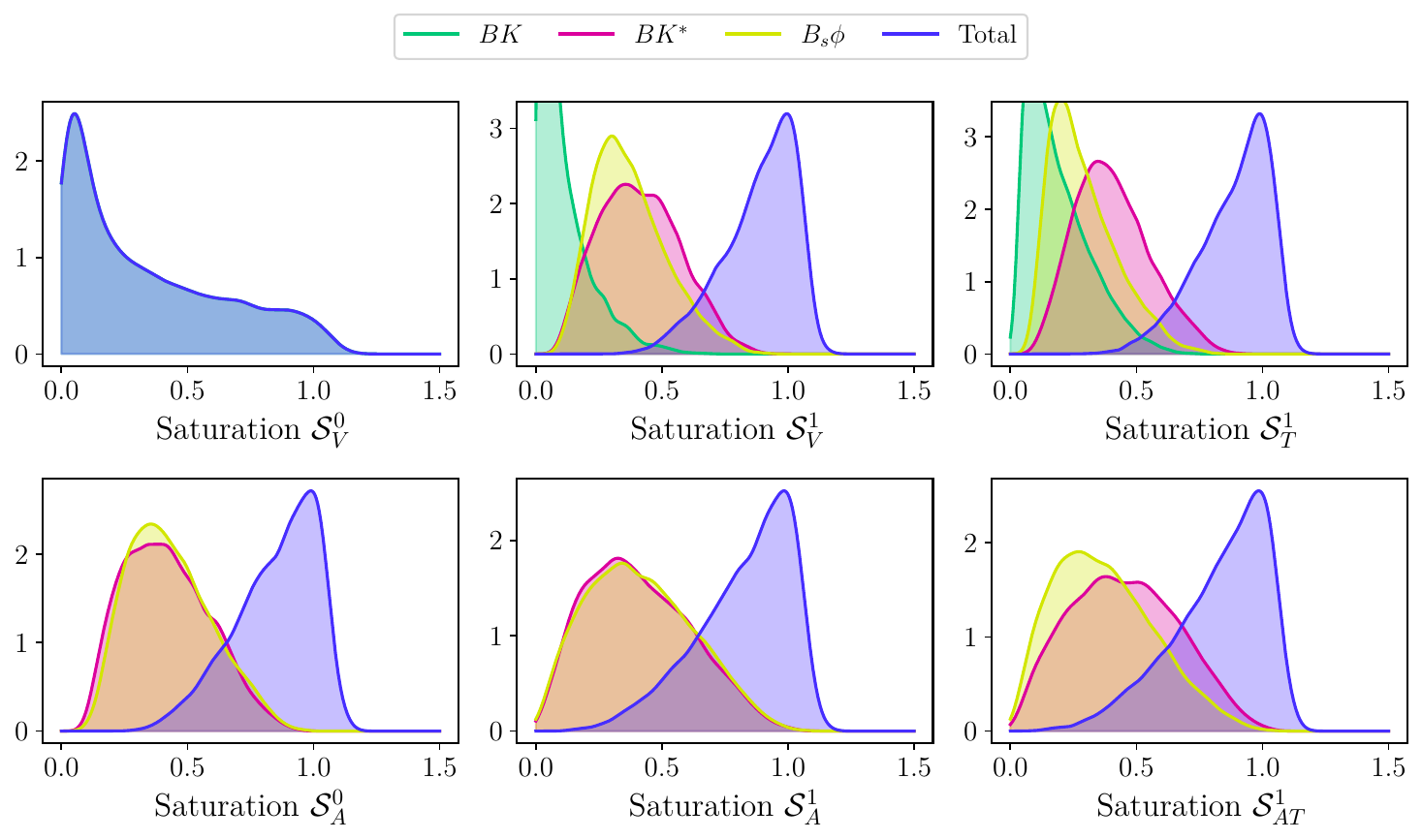}
    \\[0.3cm]
    \includegraphics[width=0.77\textwidth]{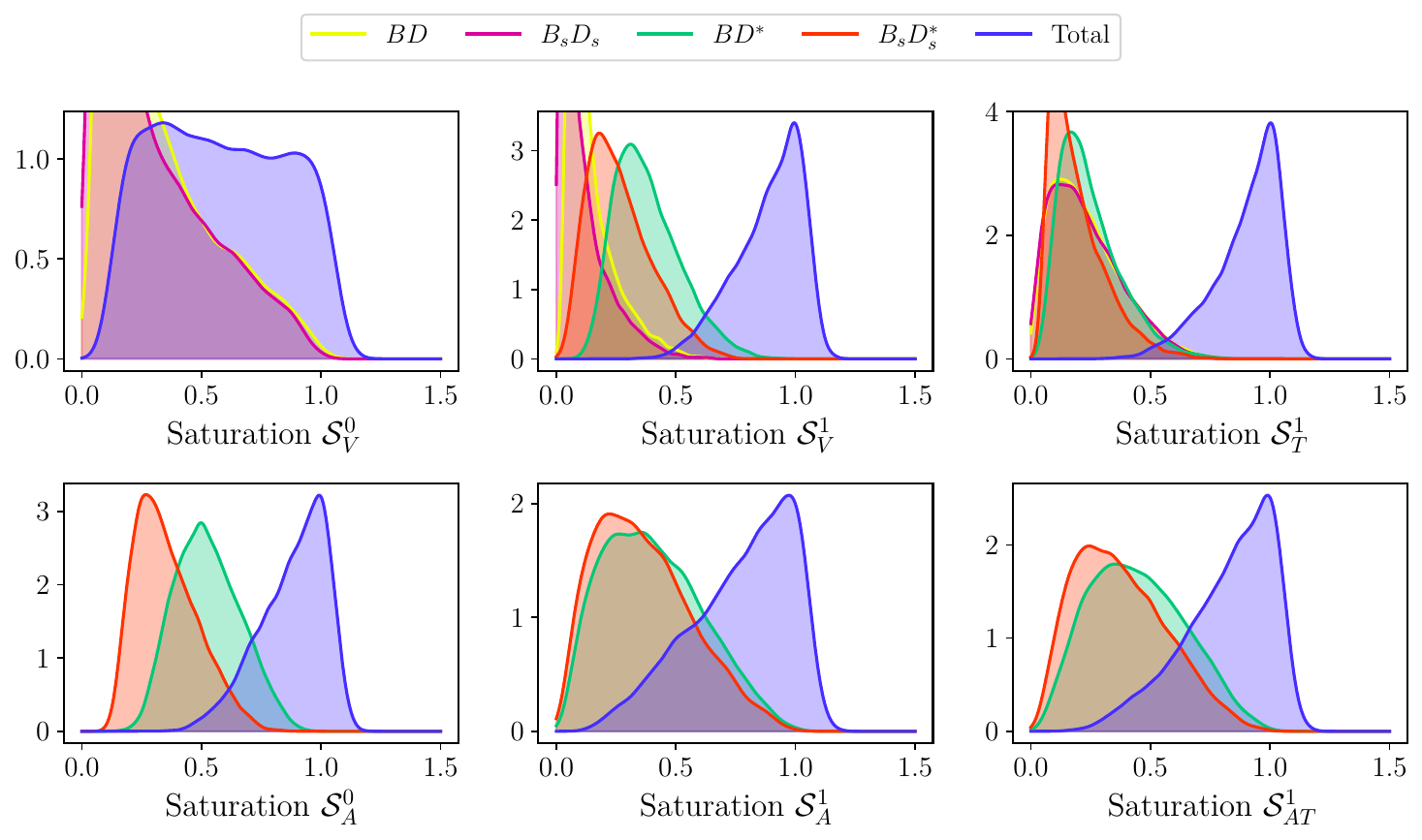}
    \captionsetup{format=plain} 
    \caption{
        Posterior probability densities of the saturation observables entering the three combined analyses (cf. \refeq{satV1}).
        \label{fig:sat}
    }
\end{figure}

For the $b\to u$ and $b\to s$ analyses, I retain the truncation at $N=4$.
Although the FF bands are already very stable at $N=3$, the total saturation observables shown in \reffig{sat} lie closer to the unitarity limit for $N=4$ and therefore provide a more reliable estimate of the residual truncation uncertainty.
By contrast, for the $b\to c$ channels the optimal GG mappings yield
\begin{align}
    0.037 \lesssim
    \max_{0\leq q^2\leq s_-}
    |z(q^2;\sopt,s_\Gamma)|
    \lesssim 0.042
    \,,
\end{align}
depending on the FF.
At truncation order $N=3$, the bound~\eqref{eq:BGLtruncerr} therefore limits the omitted tail of the normalised GG series to less than
\(
    |z|^4/\sqrt{1-|z|^2}<3.1\times10^{-6}
\)
throughout the semileptonic region.
Together with the lattice-QCD constraints, this strong suppression makes $N=3$ sufficient, and I therefore adopt this truncation order.

\reffig{sat} also displays the contributions of the individual decay channels to the saturation observables in the three FF analyses.
The decomposition is shown for all six $(\Gamma,J)$ combinations: $(V,0)$, $(V,1)$, $(T,1)$, $(A,0)$, $(A,1)$, and $(AT,1)$.
Since the bound is implemented with a $5\%$ tolerance, the total distributions may exhibit a small tail above one.
An important feature illustrated in \reffig{sat} is that several total saturation distributions have substantial support close to the unitarity limit, even though the individual decay channels generally give smaller contributions.
This demonstrates the advantage of imposing the bounds simultaneously across all channels with the same quantum numbers.
In particular, no single decay necessarily saturates the bound, whereas their combined contribution can make the unitarity constraint numerically relevant.
\\

\begin{figure}[p]
    \centering
    \begin{subfigure}{0.48\textwidth}
        \centering
        \includegraphics[width=\textwidth]{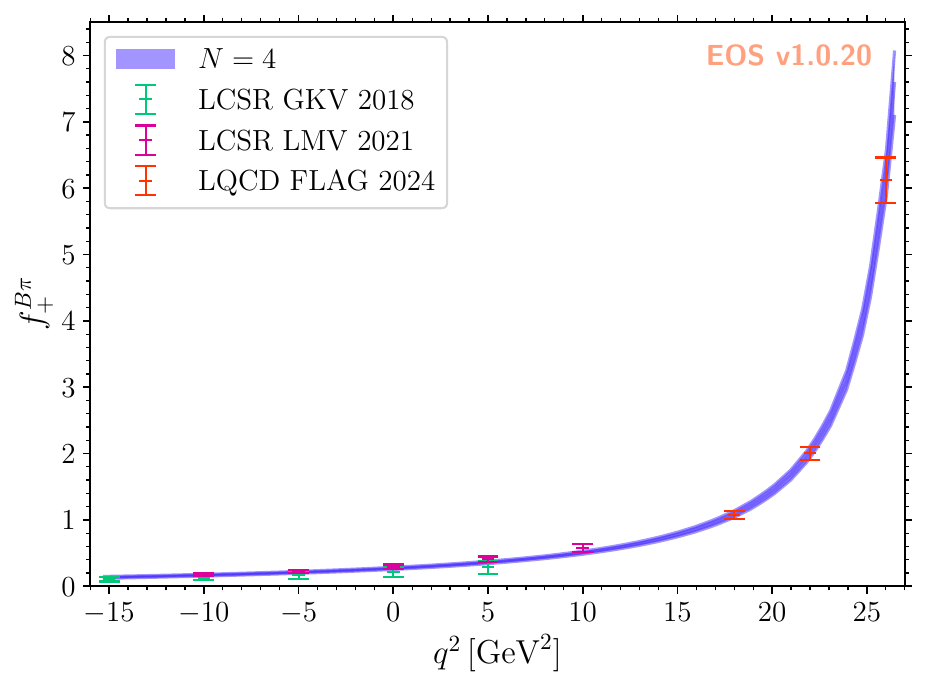}
    \end{subfigure}
    \hfill
    \begin{subfigure}{0.48\textwidth}
        \centering
        \includegraphics[width=\textwidth]{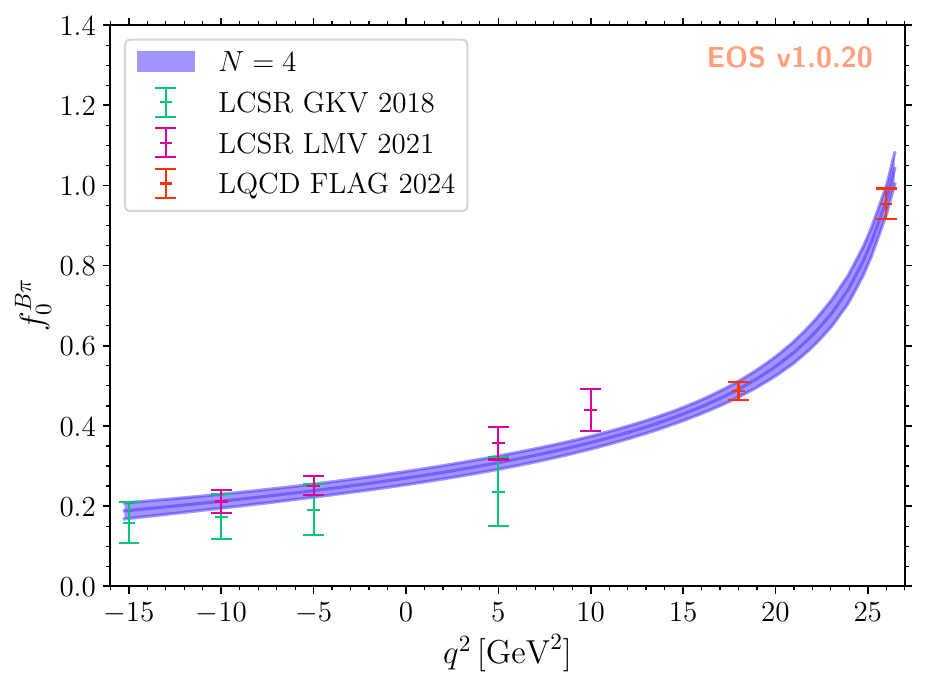}
    \end{subfigure}
    \\[0.7cm]
    \begin{subfigure}{0.48\textwidth}
        \centering
        \includegraphics[width=\textwidth]{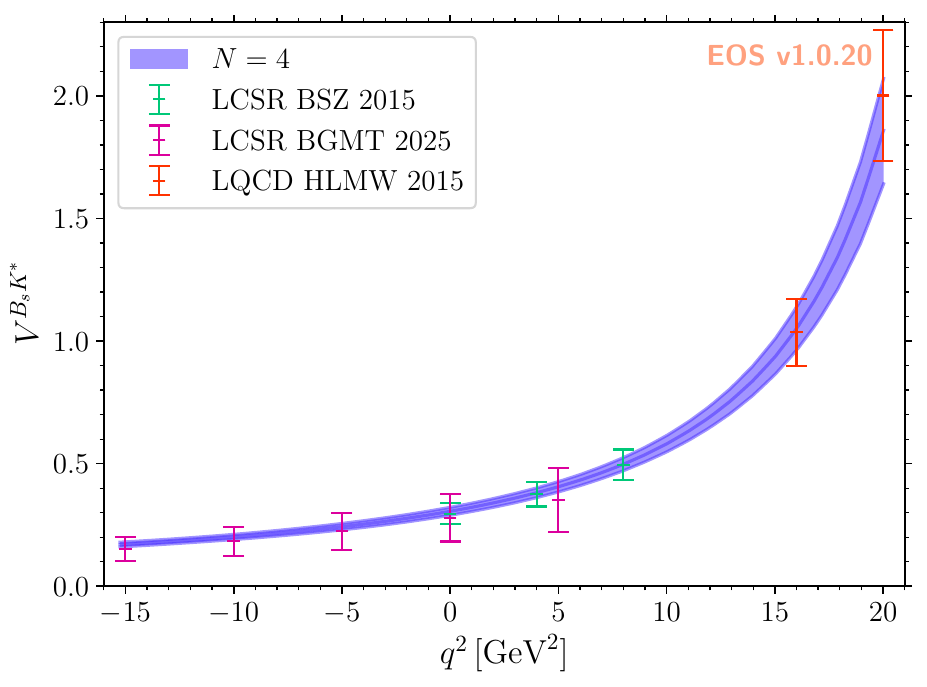}
    \end{subfigure}
    \hfill
    \begin{subfigure}{0.48\textwidth}
        \centering
        \includegraphics[width=\textwidth]{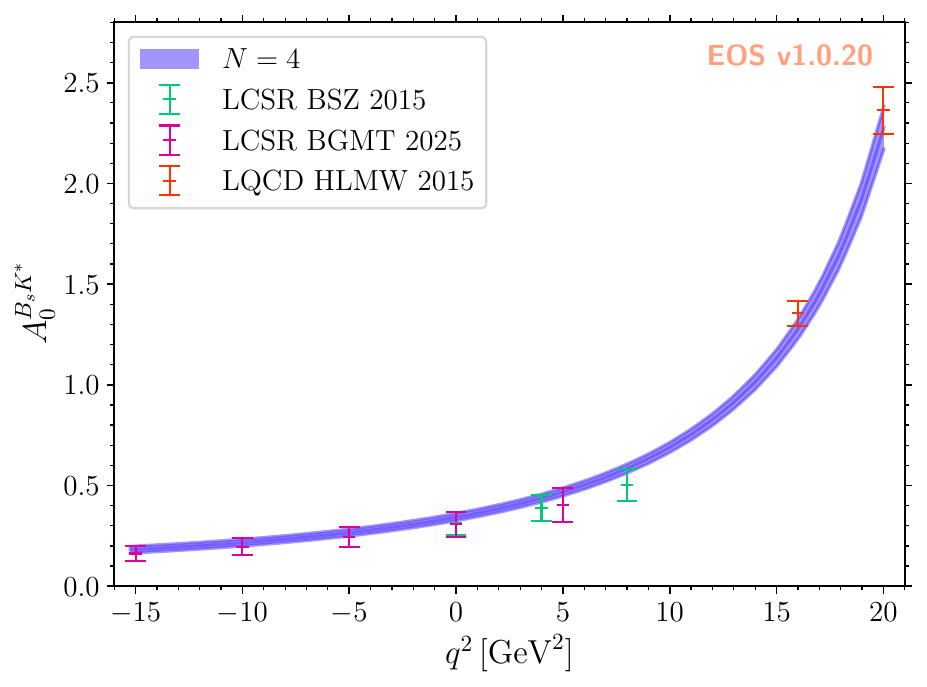}
    \end{subfigure}
    \\[0.7cm]
    \begin{subfigure}{0.48\textwidth}
        \centering
        \includegraphics[width=\textwidth]{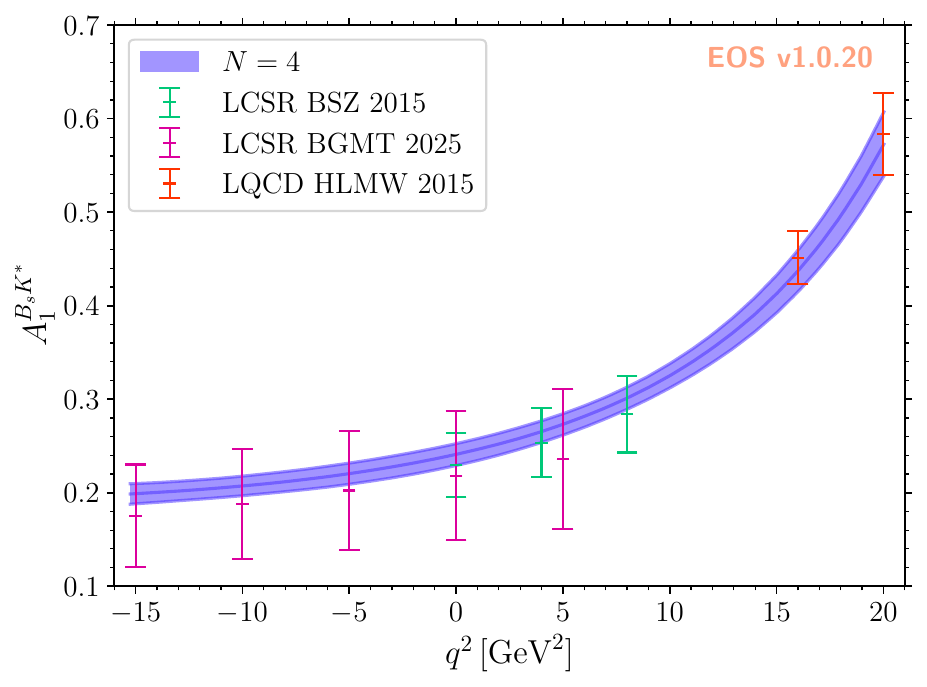}
    \end{subfigure}
    \hfill
    \begin{subfigure}{0.48\textwidth}
        \centering
        \includegraphics[width=\textwidth]{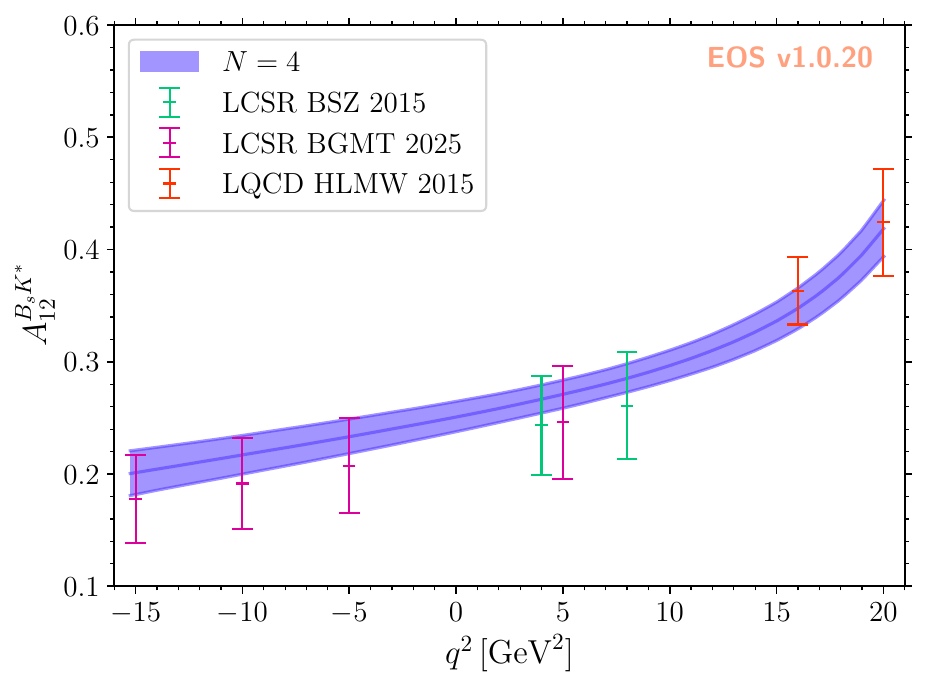}
    \end{subfigure}
    \captionsetup{format=plain} 
    \caption{
        Representative posterior FF results for the $b\to u$ analysis at truncation order $N=4$.
        The six panels show $f_+^{B\pi}$, $f_0^{B\pi}$, $V^{B_sK^*}$, $A_0^{B_sK^*}$, $A_1^{B_sK^*}$, and $A_{12}^{B_sK^*}$.
        In each panel, the blue band is the result of the combined GG fit, while the coloured points with error bars denote the LCSR and lattice-QCD inputs listed in the legend.
        The omitted FFs are provided in the supplementary material~\cite{suppl_unitb}.
        \label{fig:FF-bu}
    }
\end{figure}

\begin{figure}[p]
    \centering
    \begin{subfigure}{0.48\textwidth}
        \centering
        \includegraphics[width=\textwidth]{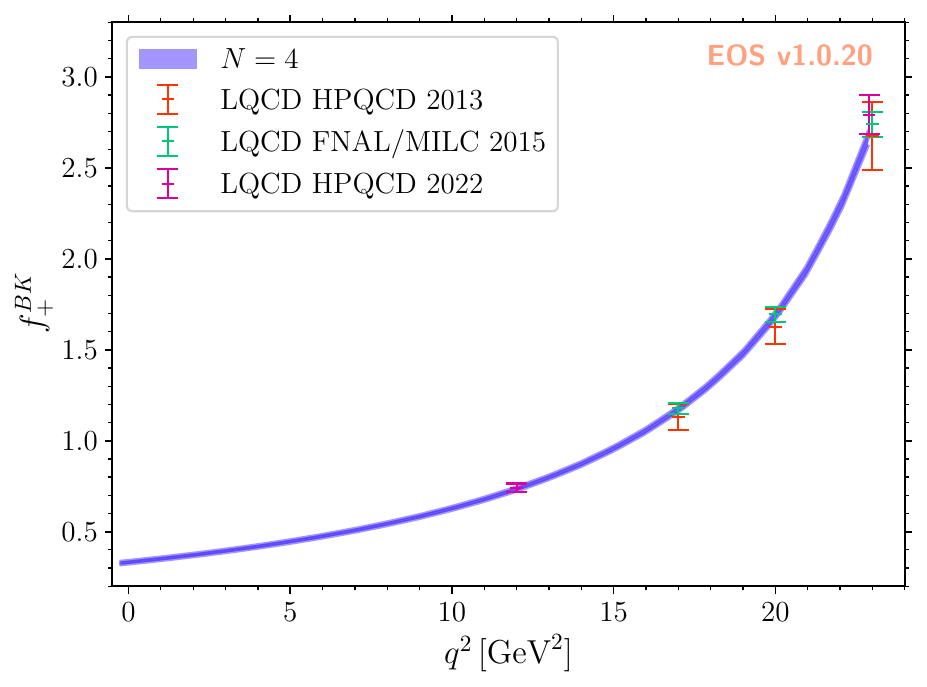}
    \end{subfigure}
    \hfill
    \begin{subfigure}{0.48\textwidth}
        \centering
        \includegraphics[width=\textwidth]{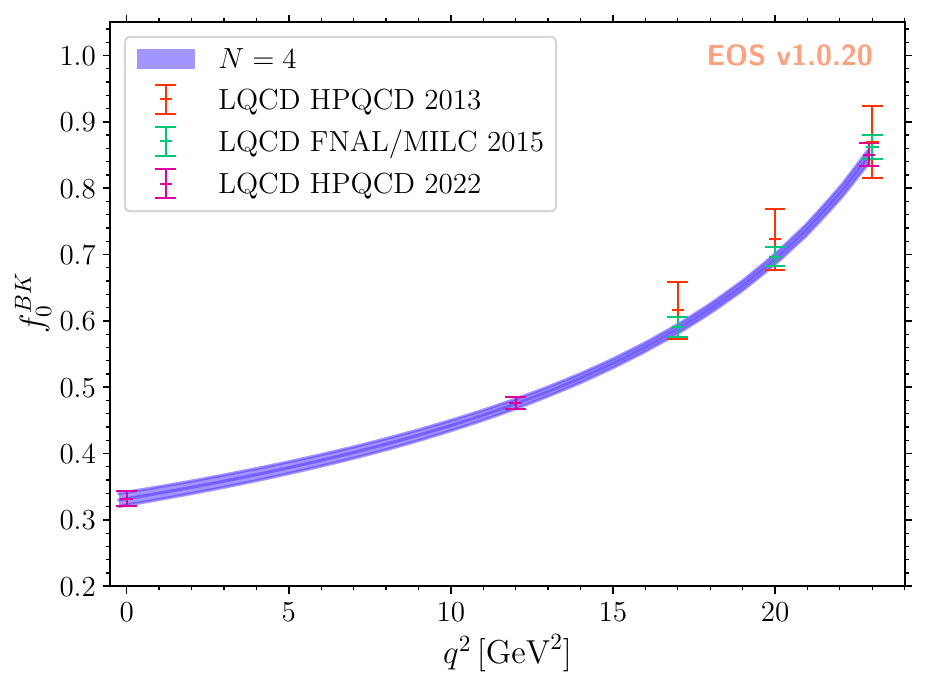}
    \end{subfigure}
    \\[0.7cm]
    \begin{subfigure}{0.48\textwidth}
        \centering
        \includegraphics[width=\textwidth]{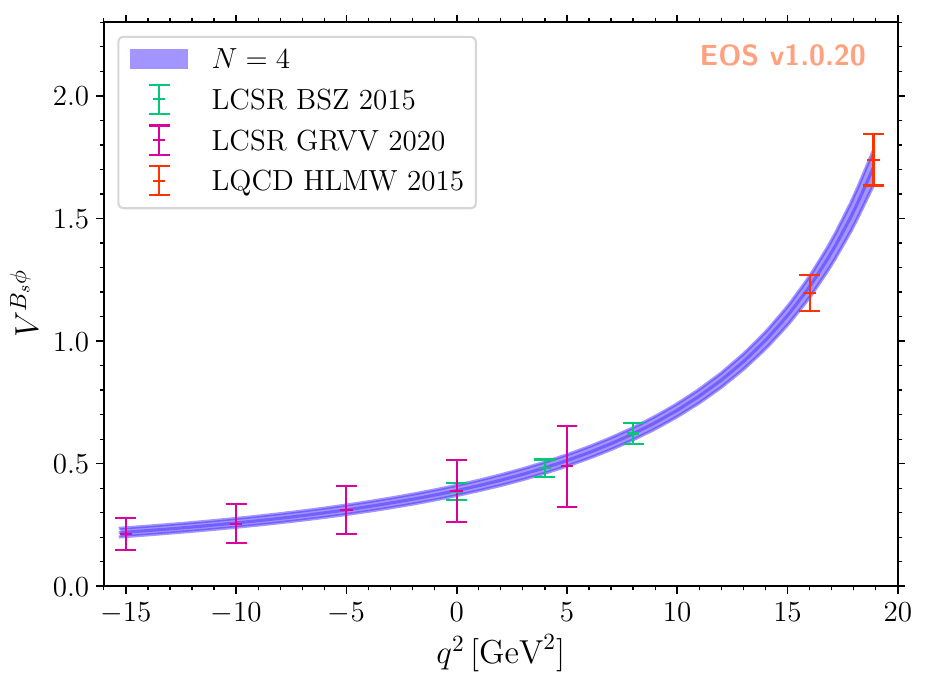}
    \end{subfigure}
    \hfill
    \begin{subfigure}{0.48\textwidth}
        \centering
        \includegraphics[width=\textwidth]{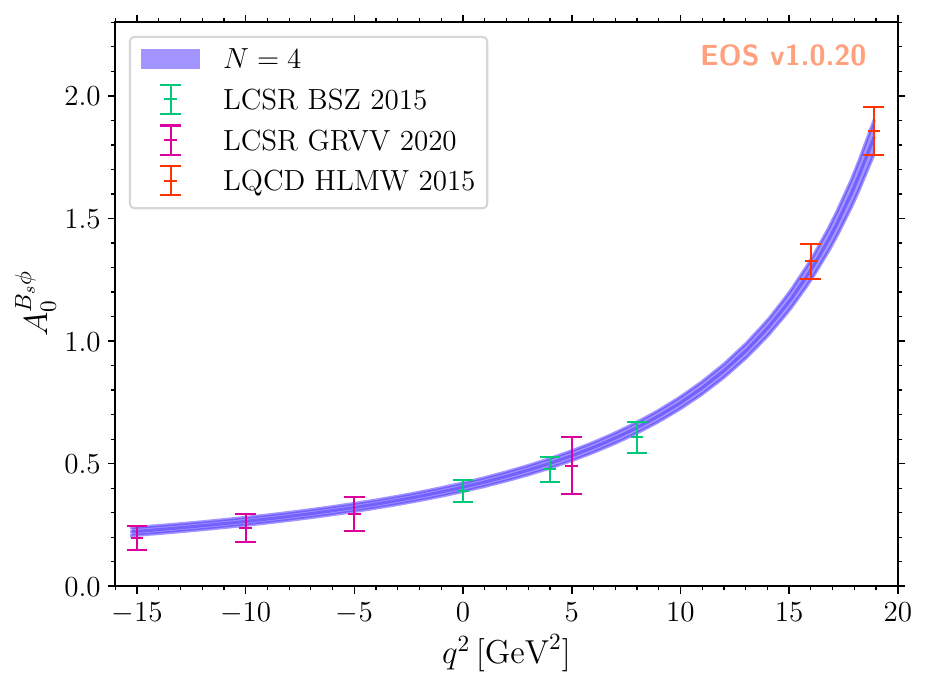}
    \end{subfigure}
    \\[0.7cm]
    \begin{subfigure}{0.48\textwidth}
        \centering
        \includegraphics[width=\textwidth]{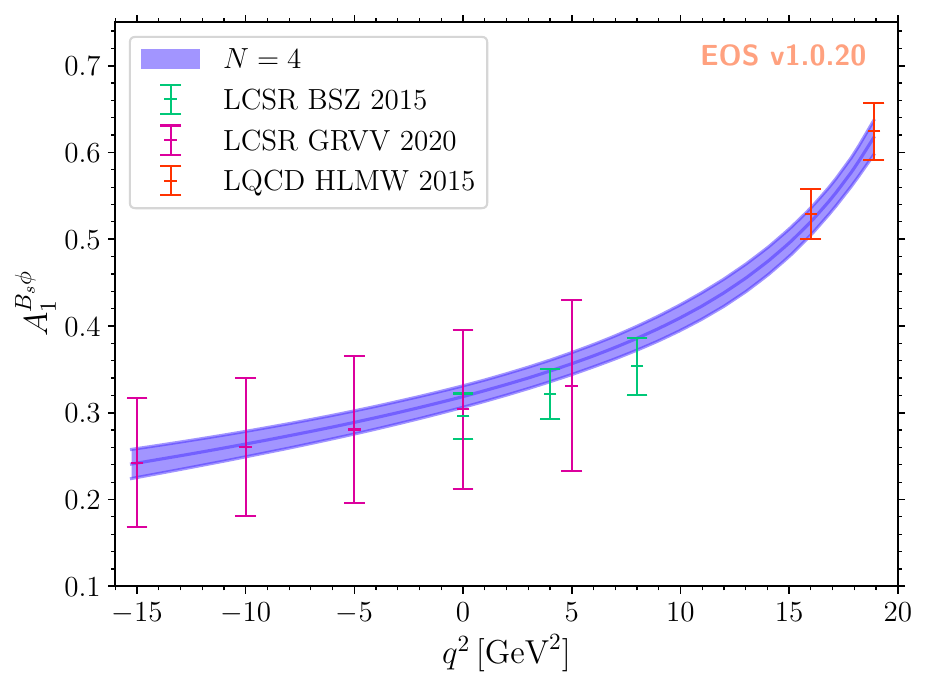}
    \end{subfigure}
    \hfill
    \begin{subfigure}{0.48\textwidth}
        \centering
        \includegraphics[width=\textwidth]{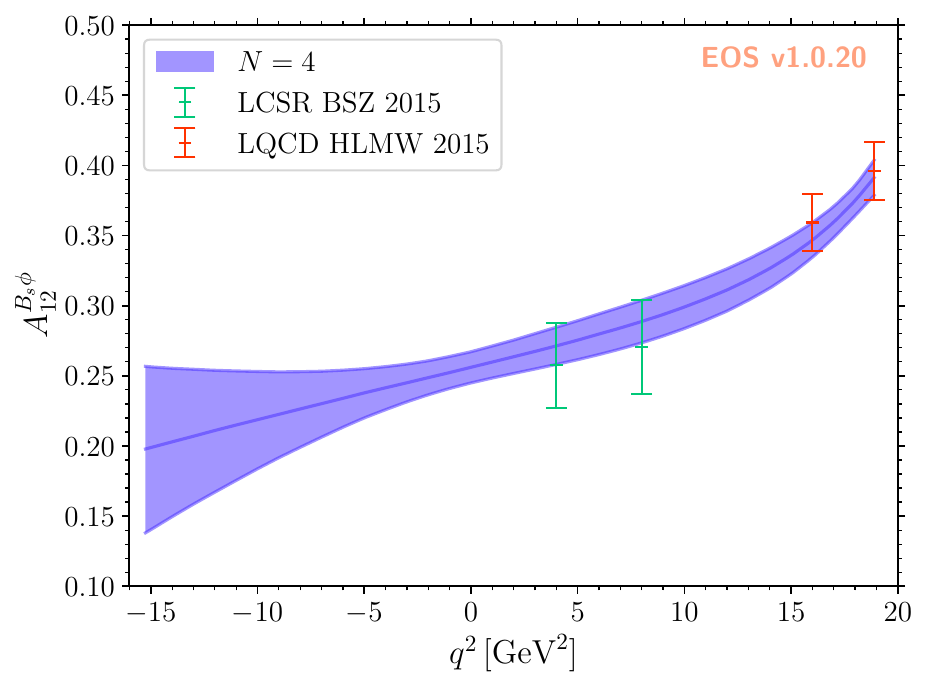}
    \end{subfigure}
    \captionsetup{format=plain} 
    \caption{
        Representative posterior FF results for the $b\to s$ analysis at truncation order $N=4$.
        The six panels show $f_+^{BK}$, $f_0^{BK}$, $V^{B_s\phi}$, $A_0^{B_s\phi}$, $A_1^{B_s\phi}$, and $A_{12}^{B_s\phi}$.
        In each panel, the blue band is the result of the combined GG fit, while the coloured points with error bars denote the lattice-QCD and LCSR inputs listed in the legend.
        The omitted FFs are provided in the supplementary material~\cite{suppl_unitb}.
        \label{fig:FF-bs}
    }
\end{figure}

\begin{figure}[p]
    \centering
    \begin{subfigure}{0.48\textwidth}
        \centering
        \includegraphics[width=\textwidth]{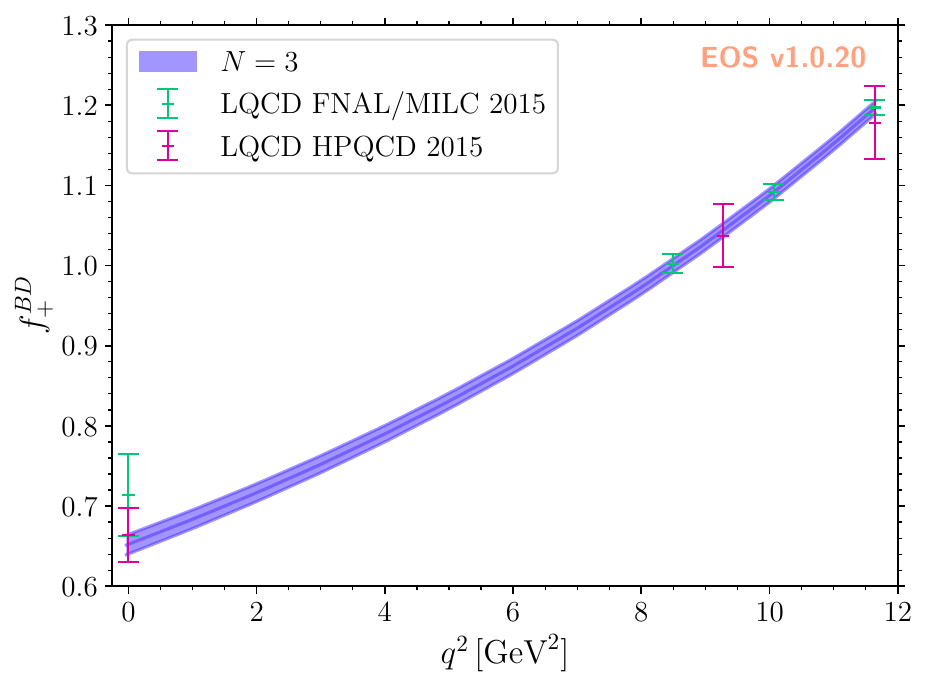}
    \end{subfigure}
    \hfill
    \begin{subfigure}{0.48\textwidth}
        \centering
        \includegraphics[width=\textwidth]{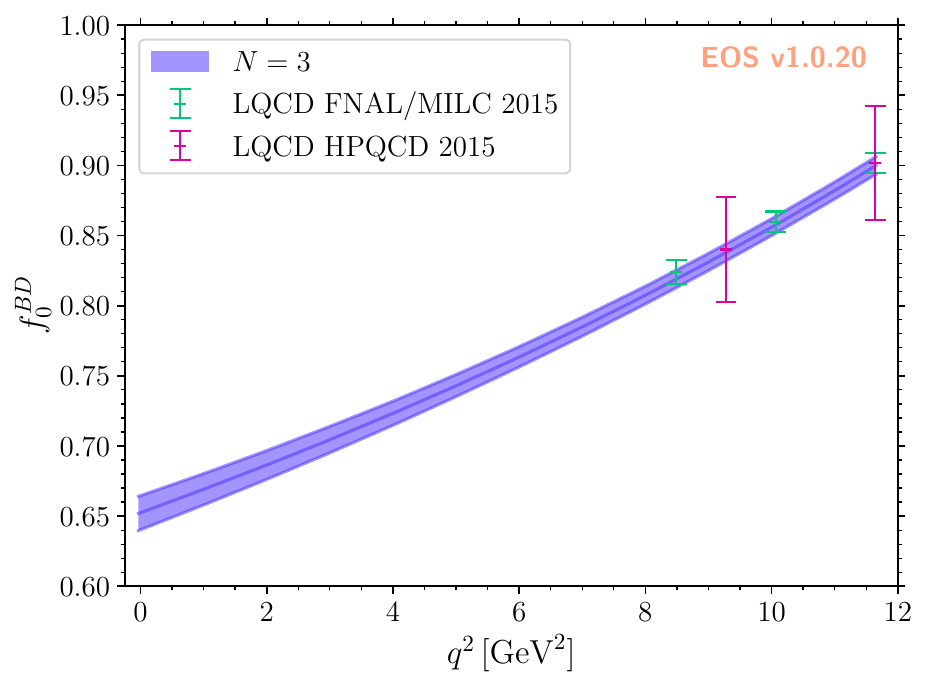}
    \end{subfigure}
    \\[0.7cm]
    \begin{subfigure}{0.48\textwidth}
        \centering
        \includegraphics[width=\textwidth]{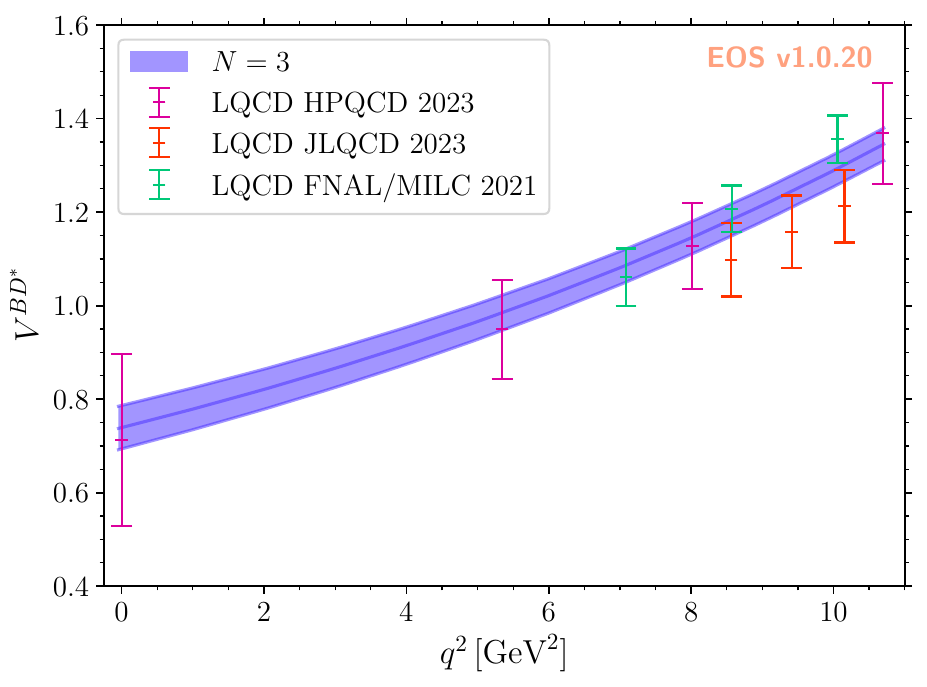}
    \end{subfigure}
    \hfill
    \begin{subfigure}{0.48\textwidth}
        \centering
        \includegraphics[width=\textwidth]{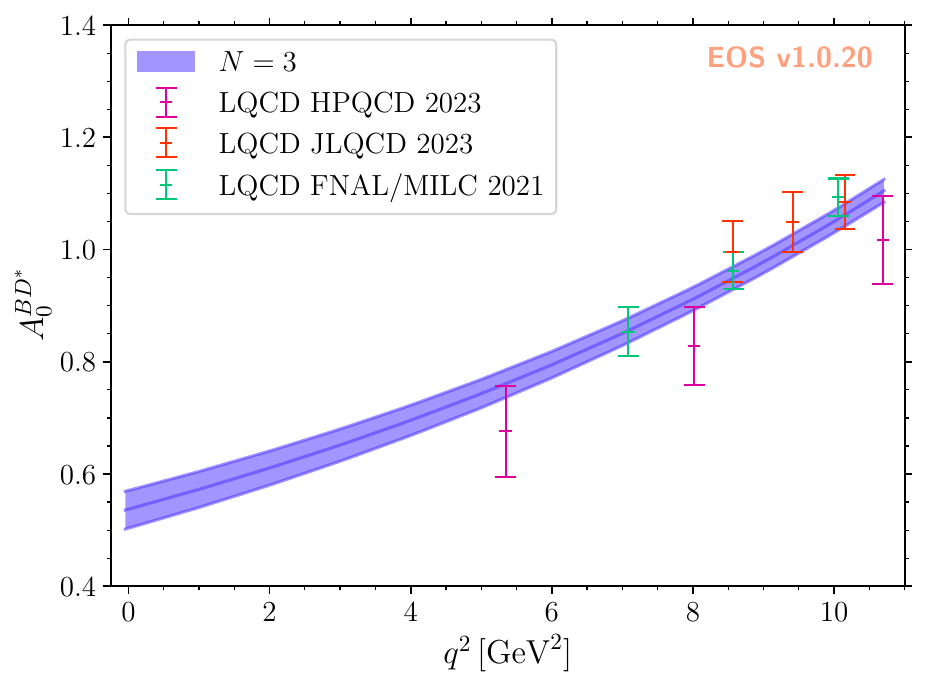}
    \end{subfigure}
    \\[0.7cm]
    \begin{subfigure}{0.48\textwidth}
        \centering
        \includegraphics[width=\textwidth]{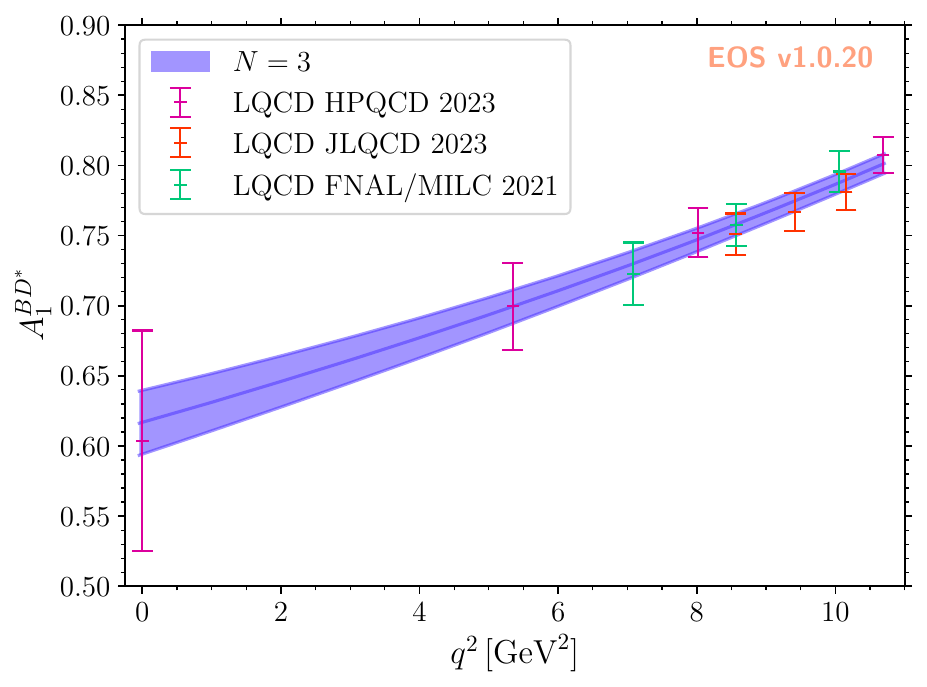}
    \end{subfigure}
    \hfill
    \begin{subfigure}{0.48\textwidth}
        \centering
        \includegraphics[width=\textwidth]{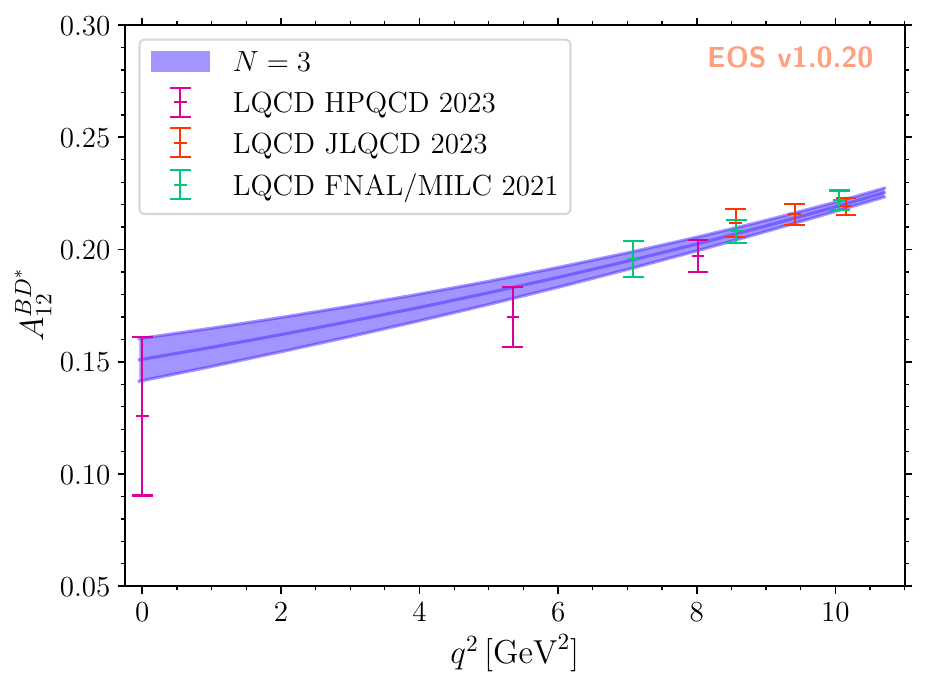}
    \end{subfigure}
    \captionsetup{format=plain} 
    \caption{
        Representative posterior FF results for the $b\to c$ analysis at truncation order $N=3$.
        The six panels show $f_+^{BD}$, $f_0^{BD}$, $V^{BD^*}$, $A_0^{BD^*}$, $A_1^{BD^*}$, and $A_{12}^{BD^*}$.
        In each panel, the blue band is the result of the combined GG fit, while the coloured points with error bars denote the lattice-QCD inputs listed in the legend.
        The omitted FFs are provided in the supplementary material~\cite{suppl_unitb}.
        \label{fig:FF-bc}
    }
\end{figure}

Representative FF results are displayed in Figs.~\ref{fig:FF-bu}--\ref{fig:FF-bc}.
For each quark-level transition, the figures display one pseudoscalar and one vector final state, with the posterior bands shown together with the input points.
The endpoint relations are imposed in the fits, as discussed in \refapp{endpointrel}; this is manifest, for instance, in the pseudoscalar panels.

In the $b\to u$ analysis, both the $B\to\pi$ and $B_s\to K^*$ bands smoothly connect the low-$q^2$ LCSR constraints with the high-$q^2$ lattice-QCD information.
Over the physical region, the relative uncertainties of the FFs displayed in \reffig{FF-bu} are typically at the $4\%$--$6\%$ level.
The main exception is $V^{B_sK^*}$, whose uncertainty increases to about $11\%$ near zero recoil, reflecting the lower precision of the available lattice-QCD information for this FF.

The $b\to s$ analysis exhibits a somewhat different pattern.
The $B\to K$ FFs are determined from lattice QCD alone and are already precise across the entire semileptonic region: the relative uncertainties of $f_+^{BK}$ and $f_0^{BK}$ decrease from about $3\%$ at $q^2=0$ to approximately $2\%$ at $q^2=s_-$.
For $B_s\to\phi$, the LCSR and lattice-QCD inputs are complementary, and their combination yields relative uncertainties of roughly $5\%$ for the four displayed FFs throughout the physical region.
The $B\to K^*$ results have a comparable precision.

Finally, in the $b\to c$ analysis the shorter kinematic range and the lattice-QCD inputs lead to narrow bands without recourse to LCSRs.
For $B\to D$, the relative uncertainty decreases from about $2\%$ at maximum recoil to $1\%$ at zero recoil, while for the four displayed $B\to D^*$ FFs it decreases from about $4\%$--$6\%$ to about $2\%$--$3\%$ over the same interval.
The corresponding non-tensor $B_s\to D_s^{(*)}$ predictions are generally determined at the few-per-cent level; among them, $V^{B_sD_s^*}(0)$ has the largest relative uncertainty, about $8\%$.

The sub-per-cent uncertainties reached for some FFs near zero recoil also signal that QED corrections can no longer be ignored in future precision applications.
Effects of nominal size $\O(\alpha_{\rm em})$, which are not treated systematically in the theory inputs used here, can be comparable to the quoted posterior uncertainties.
Their inclusion requires a consistent convention for the QED-corrected FFs and matching to infrared-safe observables, rather than simply adding an independent uncertainty to the pure-QCD result (see, e.g., Refs.~\cite{Giusti:2018guw,deBoer:2018ipi,Isidori:2020acz,Zwicky:2021olr,Beneke:2021jhp,Boer:2023vsg}).

The tensor FFs warrant a separate discussion.
Since no high-$q^2$ lattice-QCD input is included for $f_T^{B\pi}$ or $f_T^{B_sK}$ in the present analysis, their extrapolation towards zero recoil remains substantially less constrained than that of the corresponding vector and scalar FFs.
Likewise, the pseudoscalar $b\to c$ tensor FFs, which are constrained directly only through ratios at zero recoil, retain uncertainties of about $20\%$--$30\%$ at maximum recoil.
By contrast, the tensor FFs for vector final states benefit from direct lattice-QCD information and are typically determined at levels ranging from a few per cent to ten per cent over the semileptonic region.
The complete set of FF plots, including all tensor FFs and the additional channels not displayed here, is provided in the supplementary material~\cite{suppl_unitb}.
\\

\begin{figure}[t!]
    \centering
    \begin{subfigure}{0.48\textwidth}
        \centering
        \includegraphics[width=\textwidth]{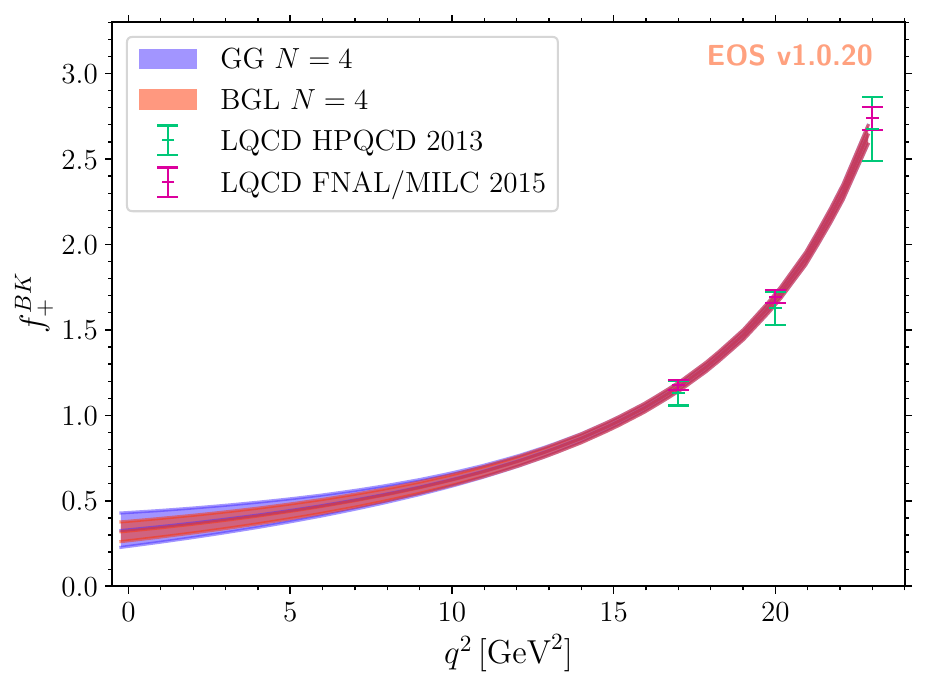}
    \end{subfigure}
    \hfill
    \begin{subfigure}{0.48\textwidth}
        \centering
        \includegraphics[width=\textwidth]{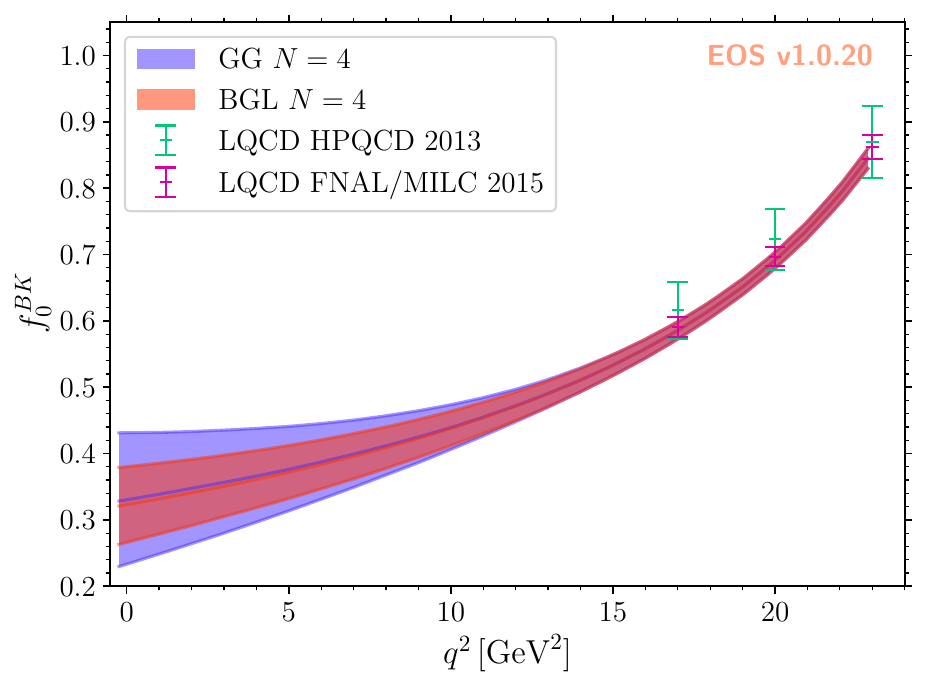}
    \end{subfigure}
    \caption{
    \label{fig:BK_BGL_GG_fit}
        Comparison between posterior $B\to K$ FF results obtained with the GG and plain BGL parametrisations for $f_+^{BK}$ (left) and $f_0^{BK}$ (right).
        Both fits are performed at truncation order $N=4$ and use only the lattice-QCD inputs of Refs.~\cite{Bouchard:2013eph,Bailey:2015dka}; the HPQCD 2022 dataset~\cite{Parrott:2022rgu} is omitted so that the difference between the two parametrisations remains visible.
        The two fits agree very well in the vicinity of the high-$q^2$ input points.
        However, towards low $q^2$ the GG parametrisation leads to a somewhat wider extrapolation band, which suggests that the plain BGL parametrisation may underestimate the extrapolation uncertainty in this case.
    }
    \vspace*{0.3cm}
    \begin{subfigure}{0.48\textwidth}
        \centering
        \includegraphics[width=\textwidth]{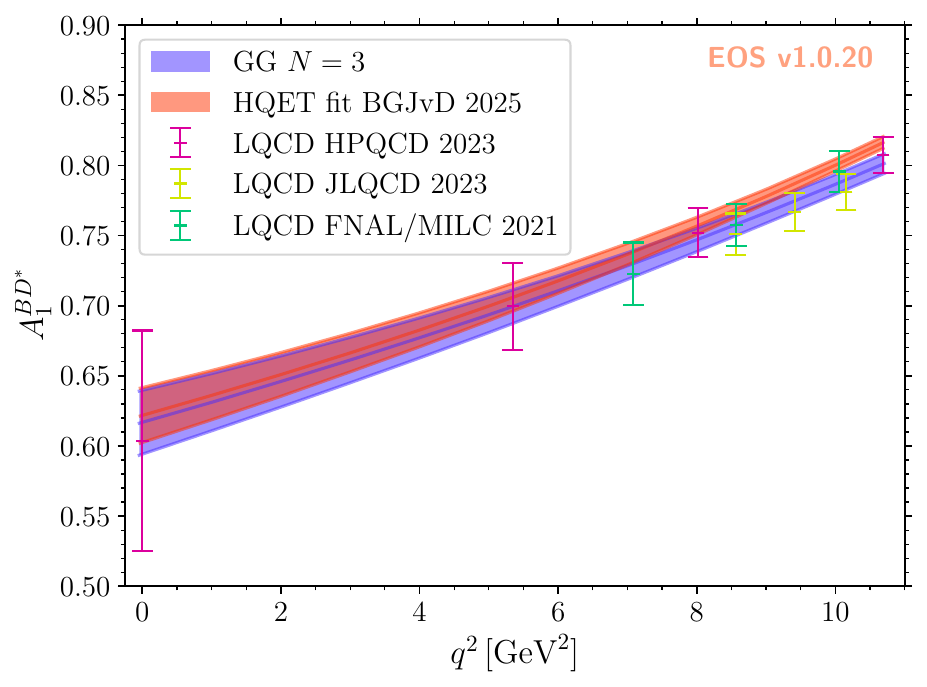}
    \end{subfigure}
    \hfill
    \begin{subfigure}{0.48\textwidth}
        \centering
        \includegraphics[width=\textwidth]{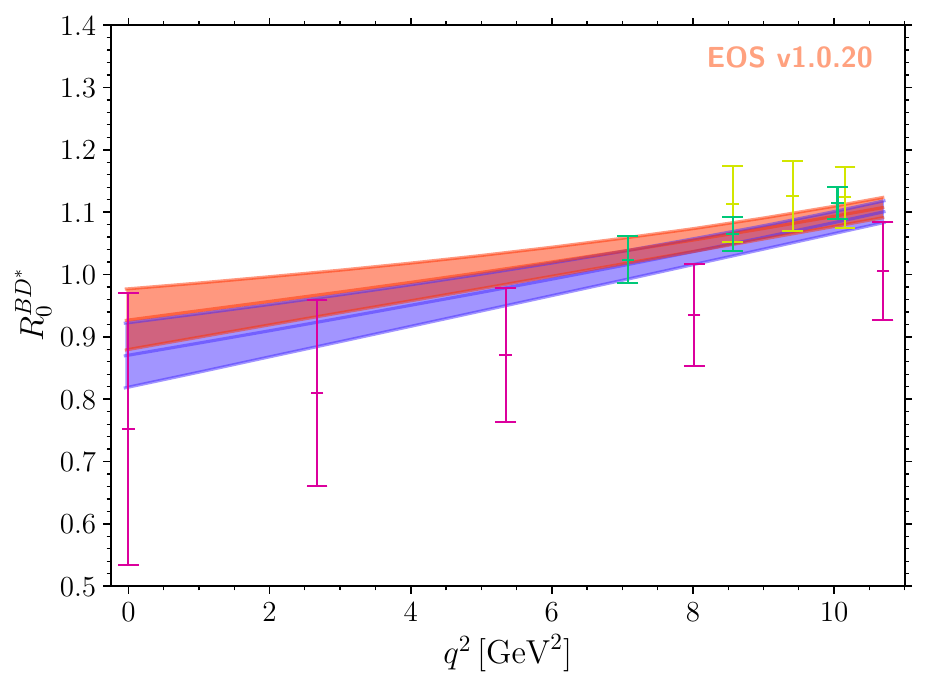}
    \end{subfigure}
    \begin{subfigure}{0.48\textwidth}
        \centering
        \includegraphics[width=\textwidth]{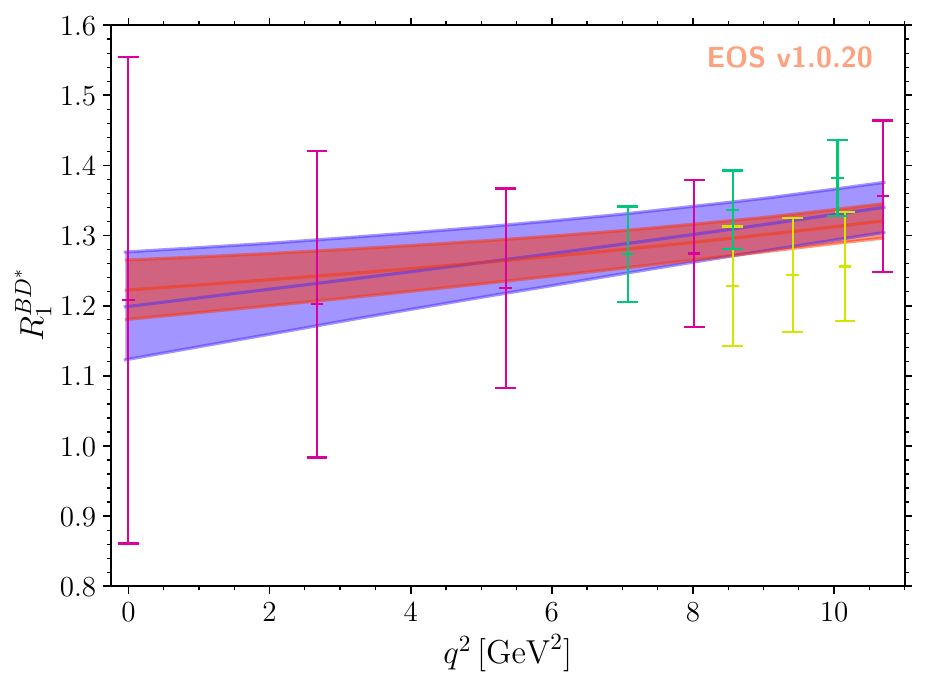}
    \end{subfigure}
    \hfill
    \begin{subfigure}{0.48\textwidth}
        \centering
        \includegraphics[width=\textwidth]{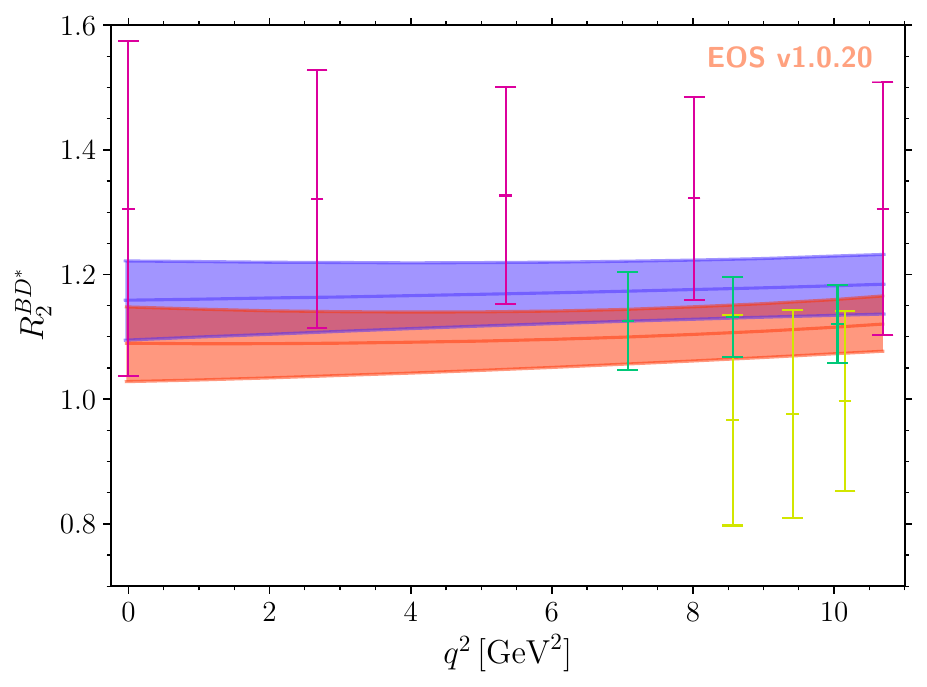}
    \end{subfigure}
    \captionsetup{format=plain} 
    \caption{
    \label{fig:BDstar_HQE_GG_fit}
        Comparison of the present GG fit with the nominal HQET fit of Ref.~\cite{Bordone:2025jur} for $B\to D^*$: top left $A_1^{BD^*}$, top right $R_0^{BD^*}$, bottom left $R_1^{BD^*}$, and bottom right $R_2^{BD^*}$.
    }
\end{figure}

The comparison with plain BGL has already been discussed in \refsubsec{GBGL} using fixed input.
Figure~\ref{fig:BK_BGL_GG_fit} shows the corresponding comparison in an actual fit to the $B\to K$ FFs.
To keep the effect visible on the scale of the plot, I omit the very precise HPQCD 2022 dataset~\cite{Parrott:2022rgu} and retain only the older lattice-QCD inputs of Refs.~\cite{Bouchard:2013eph,Bailey:2015dka}.
With this reduced input, the GG and plain BGL fits lead to very similar central values in the high-$q^2$ region where the data constrain the FFs most strongly.
The main visible difference is instead in the propagated uncertainty: the GG parametrisation yields somewhat broader bands away from the fitted points, especially towards low $q^2$.
This shows that, once the coefficients are fitted to data, the numerical impact of the subthreshold cut is less pronounced than in the fixed-input exercise, because the fit also incorporates additional sources of uncertainty beyond the pure extrapolation error.
Nevertheless, the conclusion is analogous to that reached in \refsubsec{GBGL}.
The choice between the BGL and GG parametrisations continues to have a noticeable effect on the extrapolation uncertainty, with plain BGL potentially underestimating it in the present case.

Although the GG parametrisation should clearly be preferred to the BGL parametrisation when rigorous unitarity bounds are required, this does not imply that every existing BGL analysis is numerically unreliable.
For many FFs, theory inputs are already sufficiently abundant and precise that the choice of parametrisation has little numerical impact at this level of precision.
In such cases, even an SPP can give numerically equivalent results over the region constrained by the inputs.
Thus, the difference between GG and BGL depends not only on the analytic structure of the FF, but also on the location and precision of the inputs, the extrapolation range, the truncation order, and the degree to which the unitarity bound is saturated.
For this reason, a universal estimate of this difference cannot be given; its numerical impact must instead be assessed case by case.
A systematic comparison would require repeating every FF analysis within both frameworks and lies beyond the scope of this work.
\\

I also compare the present $B\to D^*$ GG fit with the nominal HQET fit of Ref.~\cite{Bordone:2025jur}.
The comparison is shown in \reffig{BDstar_HQE_GG_fit} for $A_1^{BD^*}$ and for the ratios~\cite{Faller:2008tr}
\begin{align}
    R_0^{BD^*} &= \left(1 - \frac{q^2}{s_+}\right)\! \frac{A_0^{BD^*}}{A_1^{BD^*}}
    , &
    R_1^{BD^*} &= \left(1 - \frac{q^2}{s_+}\right)\! \frac{V^{BD^*}}{A_1^{BD^*}}
    , &
    R_2^{BD^*} &= \left(1 - \frac{q^2}{s_+}\right)\! \frac{A_2^{BD^*}}{A_1^{BD^*}}
    .
\end{align}
As expected, the HQET fit yields somewhat smaller uncertainties, even though it is based on the same lattice-QCD inputs.
This reduction is due to the inclusion of HQET constraints together with mild $SU(3)_F$ assumptions.
Overall, the two determinations are in very good agreement.
The agreement is particularly close for $A_1^{BD^*}$ and $R_1^{BD^*}$, while the largest visible differences occur for $R_0^{BD^*}$ and $R_2^{BD^*}$.
This pattern is qualitatively consistent with the discussion in Ref.~\cite{Bordone:2025jur}, where the largest visible shifts relative to other fits also occur in these two ratios.
At the same time, the uncertainty bands still overlap substantially over the semileptonic region.
This comparison therefore suggests a mild sensitivity to the additional HQET input rather than any pronounced tension between the two descriptions.
\\

I conclude this section by stressing that, \hi{once posterior samples for the FF coefficients have been obtained, \hbox{\EOS} can straightforwardly propagate them to any implemented observable, including branching ratios and angular observables.}
As an illustration, I use the samples from the present fits to predict the lepton-flavour-universality ratios $R\big(D_{(s)}^{(*)}\big)$ defined as
\begin{align}
    R\big(D_{(s)}^{(*)}\big) &=
    \frac{
        \mathcal{B}(\bar B_{(s)} \to D_{(s)}^{(*)} \tau^- \bar\nu_\tau)
    }{
        \mathcal{B}(\bar B_{(s)} \to D_{(s)}^{(*)} \ell^- \bar\nu_\ell)
    }\,,
    \qquad
    \text{with }\ell=e,\mu\,.
\end{align}
For these observables, I obtain
\begin{equation}
    \begin{aligned}
    R(D)     &= 0.3034 \pm 0.0036 \,, &\qquad\qquad
    R(D^*)   &= 0.2623 \pm 0.0043 \,, \\
    R(D_s)   &= 0.2987 \pm 0.0043 \,, &
    R(D_s^*) &= 0.2582 \pm 0.0039 \,,
    \end{aligned}
\end{equation}
with the correlation matrix
\begin{equation}
    \begin{pmatrix}
    1.0000 & 0.0002 & 0.0919 & 0.0616 \\
    0.0002 & 1.0000 & 0.0051 & 0.3403 \\
    0.0919 & 0.0051 & 1.0000 & 0.0199 \\
    0.0616 & 0.3403 & 0.0199 & 1.0000
    \end{pmatrix} 
    .
\end{equation}
Most of the correlations are small, with the largest one occurring between $R(D^*)$ and $R(D_s^*)$.
\\

Overall, the three combined analyses show that the available theory inputs are mutually compatible and yield precise FF predictions.
They also demonstrate that imposing unitarity jointly across channels can be numerically advantageous.
The released posterior samples make these results, including their correlations, directly reusable in phenomenological applications, as illustrated by the prediction of the lepton-flavour-universality ratios.

\clearpage

\clearpage

\section{Summary and conclusions}
\label{sec:concl}

The first part of the paper provides a pedagogical review of the analytic structure of the FFs, of the dispersion relations from which the unitarity bounds follow, and of the parametrisations that are most commonly used in FF analyses.
This fills a gap in the existing literature by providing the first unified and critical review of the BGL, BCL, CLN, and DM approaches.
Formulating the discussion in a common language allows their assumptions, first-principles content, and control of truncation uncertainties to be compared directly.

A central message is that the usefulness of a given parametrisation depends on the question one wants to answer.
If one merely needs a flexible description of the $q^2$ dependence, without aiming for a rigorous estimate of the truncation error or for the inclusion of additional theoretical input, simplified power-series parametrisations (SPPs) are the most suitable choice.
In this case, I find the optimised SPP of \refeq{OP} particularly natural, since it employs the correct branch point $s_\Gamma$ and performs the expansion around the origin of the optimal conformal map.
By contrast, the BSZ expansion is not optimal from the viewpoint of convergence, while the BCL parametrisation incorporates additional theoretical input and assumptions and therefore entails some degree of model dependence.
The CLN parametrisation entails an even higher degree of model dependence and is too restrictive for present-day precision analyses of $B\to D^{(*)}$ FFs.

When one does want a rigorous estimate of the truncation error, the number of available parametrisations is significantly reduced.
Neglecting subthreshold cuts for the moment, the only genuinely model-independent frameworks are the BGL parametrisation and the DM method.
I have argued in detail that these are not competing first-principles approaches, but two equivalent ways of implementing the same analyticity and unitarity information.
Any significant differences between the two should therefore be understood as implementation effects, such as the chosen truncation order, interpolation points, or statistical treatment.
\\

One of the main conceptual points of this paper is that the usual
BGL and DM constructions are strictly rigorous only in the absence
of subthreshold branch cuts.
For the local $B$-meson FFs considered here, this condition is essentially satisfied only for the $B\to\pi$ channel.
In all other cases, rescattering effects can move the first branch point from $s_+:=(m_B + m_M)^2$ to $s_\Gamma<s_+$, and hence the standard map $z(q^2;s_0,s_+)$ places part of the cut inside the unit disc.
Once this happens, the diagonal coefficient bound of BGL and the corresponding DM interpolation bounds no longer follow rigorously.
To overcome this problem, I have fully developed the BGL-like parametrisation introduced in Ref.~\cite{Gopal:2024mgb} --- which I refer to as GG --- and have shown how the same logic extends to a DM-like construction.
The resulting framework retains the main practical advantages of BGL.
It is straightforward to implement, preserves a diagonal coefficient bound, and therefore admits a direct estimate of the truncation error.
At the same time, it incorporates the correct analytic structure.
The only additional input required is an estimate, or simply an upper bound, for the contribution of the subthreshold cut.
This avoids having to model the detailed shape of that cut; if the estimate is conservative, the resulting bound is simply weaker and the final FF uncertainties can become somewhat larger.

I have also studied how the GG bounds depend on the practical choices entering the construction.
The numerical impact of varying the subtraction point $Q^2$ is small, and the standard choice $Q^2=0$ remains well justified.
Using additional subtractions in the dispersion relation is generally not beneficial.
The dependence on $\Chi_\Gamma^J$ is mild, which explains why conservative estimates of the subthreshold contribution do not drastically reduce the usefulness of the bounds.
By contrast, the choice of the conformal parameter matters at finite order, and the optimal choice $s_0=\sopt$ clearly improves convergence.
A comparison with plain BGL shows that, in some cases, the effect of neglecting subthreshold cuts is modest.
In other cases, however, such as $f_0^{BK}$, the effect is sizeable, and the use of BGL may lead to overly optimistic extrapolation bands.
\\

On the phenomenological side, I have implemented the GG parametrisation in \EOS under the label \texttt{G2026} and used it for three combined FF analyses, corresponding to $b\to u$, $b\to s$, and $b\to c$ transitions.
The $b\to u$ and $b\to c$ cases constitute the first analyses in which the corresponding FFs are studied simultaneously within a purely unitarity-based framework.
The three fits combine the available lattice-QCD information, as well as LCSRs in the $b\to u$ and $b\to s$ cases.
I impose the unitarity bounds directly in each statistical analysis and obtain FF predictions over the full semileptonic region.
All input datasets considered in this work are mutually compatible within the combined fits.
All numerical results, posterior samples, analysis files, and plots are provided in the supplementary material~\cite{suppl_unitb}.
These machine-readable and easy-to-use FF results are intended to serve as practical inputs for phenomenological studies by the community.
\\

Taken together, these results suggest a rather simple practical conclusion.
If unitarity is not to be imposed, an optimised SPP is a natural default choice.
Such SPPs are also useful for presenting lattice-QCD results, where one may wish to use as little additional theory input as possible in order not to distort the direct interpretation of the lattice calculation.
If rigorous unitarity bounds and a controlled truncation error are required, BGL or DM should be used only in cases where subthreshold cuts are absent.
%
%
For generic $B$-meson FFs, the appropriate replacement is the GG parametrisation, or equivalently the modified DM construction.

More broadly, the choice of parametrisation should not be regarded as a merely technical issue.
It provides the interface through which lattice-QCD and QCD sum-rule results are translated into phenomenological predictions.
Selecting the parametrisation that is best suited both to the analytic structure and to the intended application is therefore essential for making full use of the results produced by the theory community.
I expect the framework developed here to be useful not only for the channels analysed in this work, but also for future studies of other mesonic and baryonic FFs, and ultimately for extensions to non-local matrix elements.

\section*{Acknowledgements}

I am grateful to L. Vittorio and the Cambridge Pheno Working Group for helpful discussions.
I also thank I. Caprini for valuable comments on the manuscript.
I would like to express my gratitude to the \EOS maintainers, and especially to M.~Reboud and D.~van Dyk, for their comments and for their checks of the \EOS implementation.

This work has been funded by the Deutsche Forschungsgemeinschaft (DFG, German Research Foundation) -- Emmy-Noether Grant
No. 558599025.
\sloppy This work has also been partially supported by the STFC consolidated grants ST/T000694/1 and ST/X000664/1.

\noindent
\textcolor{white}{
I started this work motivated by my love for physics.\\
I finished it motivated by my love for someone special.\\
$\F + 18,1$.\\
Leave, love, live.
}
\vspace{-0.7cm}

\appendix
\addcontentsline{toc}{section}{Appendices}

\newcounter{APP}
\renewcommand{\theAPP}{\Alph{APP}}
\setcounter{APP}{0}
\renewcommand{\theequation}{\theAPP.\arabic{equation}}
\renewcommand{\theHequation}{\theAPP.\arabic{equation}}
\renewcommand{\thesubsection}{\theAPP.\arabic{subsection}}

\appsection{Helicity amplitudes, weight and outer functions}
\label{app:outer}

In this appendix, I define and compute the helicity amplitudes for $B$-meson decays.
Using these amplitudes, I then show how to determine the weight function introduced in \refsubsec{UB}.
Finally, I derive the outer functions from the weight functions.

\refstepcounter{subsection}
\subsection*{\thesubsection\quad Helicity amplitudes}

The derivation of the helicity amplitudes is most transparent in the $B$-meson rest frame. 
Following Ref.~\cite{Bharucha:2010im}, I choose the three-momentum of the final-state meson along the positive $z$ axis,
\begin{align}
    p^\mu &= (m_B,0,0,0)\,,
    &
    k^\mu &= (E_M,0,0,|\vec{q}|)\,,
    &
    q^\mu &= p^\mu-k^\mu=(q^0,0,0,-|\vec{q}|)\,,
\end{align}
with
\begin{align}
    q^0 = \frac{m_B^2-m_M^2+q^2}{2m_B}\,,
    \qquad
    E_M = \frac{m_B^2+m_M^2-q^2}{2m_B}\,,
    \qquad
    |\vec{q}| = \frac{\sqrt{\lamkin(q^2)}}{2m_B}\,.
\end{align}
For the virtual current with momentum $q^\mu$ I use the polarisation vectors
\begin{align}
    \eps_t^\mu(q) 
    &=
    \frac{q^\mu}{\sqrt{q^2}}\,,
    &
    \eps_\perp^\mu(q)
    &=
    (0,1,0,0)\,,
    &
    \eps_\para^\mu(q)
    &=
    (0,0,i,0)\,,
    &
    \eps_0^\mu(q) 
    &= 
    \frac{1}{\sqrt{q^2}} \big(|\vec{q}|,0,0,-q^0\big)\,.
    &
\end{align}
They satisfy
\begin{align}
    \eps_\lambda(q)\cdot \eps_{\lambda'}^*(q)
    &=
    g_{\lambda\lambda'}
    \,,
\end{align}
with $\lambda,\lambda'=t,\perp,\para,0$, and the completeness relation
\begin{align}
    g^{\mu\nu}
    =
    \eps_t^\mu \eps_t^{*\nu}
    -
    \eps_\perp^\mu \eps_\perp^{*\nu}
    -
    \eps_\para^\mu \eps_\para^{*\nu}
    -
    \eps_0^\mu \eps_0^{*\nu}
    \,.
\end{align}
Accordingly, the projectors introduced in \refeq{Pilambda} can be written, in four dimensions, as
\begin{align}
    \label{eq:projeps}
    \P_{\mu\nu}^{0}(q)
    &=
    \eps_{t,\mu}(q)\,\eps_{t,\nu}^*(q)
    \,,
    &
    \P_{\mu\nu}^{1}(q)
    &=
    \frac{1}{3}
    \sum_{\lambda=\perp,\para,0}
    \eps_{\lambda,\mu}(q)\,\eps_{\lambda,\nu}^*(q)
    \,.
\end{align}

For $B\to P$ transitions, I define the helicity amplitudes as
\begin{align}
    H_{\Gamma,\lambda}^{BP}
    &:=
    \, \eps_{\lambda,\mu}^*(q)\,
    \langle P(k)| J_\Gamma^\mu | \bar{B}(p)\rangle
    \,,
\end{align}
with $\lambda=t,\perp,\para,0$.
Using \refeqs{def-fpf0}{def-fT}, it follows that the only non-vanishing helicity amplitudes $H_{\Gamma,\lambda}^{BP}$ are given by
\begin{align} 
    \label{eq:HBP}
    H_{V,t}^{BP} &= \frac{m_B^2-m_P^2}{\sqrt{q^2}} \, f_0^{BP}\,,
    &
    H_{V,0}^{BP} &= \sqrt{\frac{\lamkin}{q^2}} \,f_+^{BP}\,,
    &
    H_{T,0}^{BP} &= i\frac{\sqrt{q^2\, \lamkin}}{m_B+m_P} \, f_T^{BP}\,.
\end{align}
The transverse projections vanish identically because the matrix elements involve only the four-vectors $p^\mu$, $k^\mu$, and $q^\mu$, all of which lie in the $t$-$z$ plane.

For $B\to V$ transitions, I define the helicity amplitudes as
\begin{align}
    H_{\Gamma,\lambda}^{BV}
    &:=
    \sum_{\eta\, {\rm pol.}}
    \eps_{\lambda,\mu}^*(q)\,
    \langle V(k, \eta)| J_\Gamma^\mu | \bar{B}(p)\rangle
    \,,
\end{align}
where I also use the polarisation vectors
\begin{align}
    \eta_\pm^\mu(k) &=  \frac{1}{\sqrt{2}} (0,\mp 1,-i,0)\,,
    \qquad
    \eta_0^\mu(k) = \frac{1}{m_M} \big(|\vec{q}|,0,0,E_M\big)\,.
\end{align}
%
%
Using \refeqs{def-V}{def-T2T3},\eqref{eq:def-A12},\eqref{eq:def-T23},  it follows that the only non-vanishing helicity amplitudes $H_{\Gamma,\lambda}^{BV}$ are given by
\begin{equation} 
\begin{aligned}
    \label{eq:HBV}
    H_{V,\perp}^{BV}
    &=
    i\frac{\sqrt{2 \lamkin}}{m_B+m_V} \, V^{BV}\,,
    &
    H_{A,t}^{BV}
    &=
    i\sqrt{\frac{\lamkin}{q^2}} \,A_0^{BV}\,,
    &
    H_{A,\para}^{BV}
    &=
    -i\sqrt{2}\,(m_B+m_V) \, A_1^{BV}\,,
    \\
    H_{A,0}^{BV}
    &=
    i\frac{8\,m_B m_V}{\sqrt{q^2}} \, A_{12}^{BV}\,,
    &
    H_{T,\perp}^{BV}
    &=
    -\sqrt{2\lamkin}\, T_1^{BV}\,,
    &
    H_{AT,\para}^{BV}
    &=
    -\sqrt{2}\,(m_B^2-m_V^2)\, T_2^{BV}\,,
    \\
    H_{AT,0}^{BV}
    &=
    \frac{4\,m_B m_V \sqrt{q^2}}{(m_B+m_V)} \, T_{23}^{BV}\,.  
\end{aligned}
\end{equation} 
The motivation for choosing the basis~\eqref{eq:BVFFbasis} for the $B\to V$ FFs should now be clear: each helicity amplitude depends on at most one FF.
This is why these FFs are commonly called helicity FFs.

\refstepcounter{subsection}
\subsection*{\thesubsection\quad Weight functions}

In \refeqa{defW}{WBGL}, I introduced the weight functions.
Using the helicity amplitudes computed in the previous subsection, it is then straightforward to obtain the coefficients reported in \reftab{outer}.
This is done by rewriting \refeq{ImPi} for a specific $BM$ contribution in the helicity basis.
For a two-particle state, the Lorentz-invariant phase space can be written as
\begin{align}
    \label{eq:LIPS}
    \sum \hspace*{-0.62cm} \int\limits_{BM} d\tau_{BM} (2\pi)^4 \delta^{(4)}(p_B+p_M-q)\Big|_{q^2=s} 
    =
    \kappa_I
    \int d\Omega\,
    \frac{\sqrt{\lamkin(s)}}{32 \pi^2 s}
    \,,
\end{align}
where $\kappa_I$ is the isospin factor defined in \refeq{kappaI}.
Using crossing symmetry~\eqref{eq:crossym} and decomposing the projectors $\P_{\mu\nu}^J$ in the polarisation basis introduced in \refeq{projeps}, the contraction in \refeq{ImPi} becomes
\begin{align}
    \P_{\mu\nu}^{0}(q)\,
    \braket{0 | J_\Gamma^{\mu} | BM}\,
    \braket{BM | J_\Gamma^{\dagger,\nu} | 0}
    &=
    \left|H_{\Gamma,t}^{BM}\right|^2
    \,,
    \\
    \P_{\mu\nu}^{1}(q)\,
    \braket{0 | J_\Gamma^{\mu} | BM}\,
    \braket{BM | J_\Gamma^{\dagger,\nu} | 0}
    &=
    \frac{1}{3}
    \sum_{\lambda=\perp,\para,0}
    \left|H_{\Gamma,\lambda}^{BM}\right|^2
    \,.
    \label{eq:HBMim}
\end{align}
By inserting \refeqs{LIPS}{HBMim} into \refeq{ImPi} and carrying out the solid-angle integration, $\int d\Omega = 4\pi$, for the contribution of a single FF one finds
\begin{align}
    \Im\,\Pi_\Gamma^{0}(s+i\eps)\Big|_{BM}
    &=
    \frac{\kappa_I}{16 \pi}
    \frac{\sqrt{\lamkin(s)}}{s}
    \left|H_{\Gamma,t}^{BM}(s)\right|^2
    \,,
    \\
    \Im\,\Pi_\Gamma^{1}(s+i\eps)\Big|_{BM}
    &=
    \frac{\kappa_I}{48 \pi}
    \frac{\sqrt{\lamkin(s)}}{s}
    \sum_{\lambda=\perp,\para,0}
    \left|H_{\Gamma,\lambda}^{BM}(s)\right|^2
    \,.
\end{align}
In the helicity basis chosen here, each non-vanishing amplitude depends on at most one FF.
Therefore, after inserting the expressions derived in the previous subsection, each contribution immediately takes the form of \refeq{defW}.
For example, using \refeq{HBP} reproduces \refeq{chiBpi}, while using \refeq{HBV} gives (for $k=2$, i.e. the minimal number of subtractions required)
\begin{align*}
    \chi_A^1(Q^2;2)\Big|_{BV}
    =
    \frac{\kappa_I}{24\pi^2}
    \int\limits_{s_+}^\infty ds\,
    \frac{\sqrt{\lamkin(s)}\,}{s^2\,(s+Q^2)^3}
    \left(
        s\,(m_B+m_V)^2
        \left|A_1^{BV}(s)\right|^2
        +
        32 \,m_B^2 m_V^2
        \left|A_{12}^{BV}(s)\right|^2
    \right)
    \,.
\end{align*}
Using $\lamkin(s)=(s-s_+)(s-s_-)$ and \refeq{spm}, all contributions can be cast into the canonical form~\eqref{eq:WBGL}, from which the parameters $K$, $a$, $b$, $c$, and $d$ are read off.

\begin{table}[t!]
    \newcommand{\pp}{\phantom{+}}
    \centering
    \renewcommand{\arraystretch}{1.4}
    \begin{tabular}{cccccccc}
        \toprule
        \multirow{2}{*}{Form factor}    &
        \multirow{2}{*}{$\chi_\Gamma^J$}&
        \multicolumn{5}{c}{Parameters}  \\
                                        & 
                                        & 
        $K$                             &
        $a$                             &
        $b$                             &
        $c$                             &
        $d$                             \\
        \toprule       
        $f_+^{BP}$                      & 
        $\chi_V^1$                      &
        $48$                            &
        3                               &
        3                               &
        2                               &
        3                               \\
        $f_0^{BP}$                      &
        $\chi_V^0$                      & 
        $16/(s_+ s_-)$                  &
        1                               &
        1                               &
        1                               &
        2                               \\
        $f_T^{BP}$                      & 
        $\chi_T^1$                      &
        $48 \, s_+$                     &
        3                               &
        3                               &
        1                               &
        4                               \\
        \midrule
        $V^{BV}$                        & 
        $\chi_V^1$                      &
        $24 \, s_+$                     &
        3                               &
        3                               &
        1                               &
        3                               \\
        $A_0^{BV}$                      & 
        $\chi_A^0$                      &
        $16$                            &
        3                               &
        3                               &
        1                               &
        2                               \\
        $A_1^{BV}$                      & 
        $\chi_A^1$                      &
        $24/s_+$                        &
        1                               &
        1                               &
        1                               &
        3                               \\
        $A_{12}^{BV}$                   & 
        $\chi_A^1$                      &
        $12/(s_+ - s_-)^2$              &
        1                               &
        1                               &
        2                               &
        3                               \\
        $T_1^{BV}$                      & 
        $\chi_T^1$                      &
        $24$                            &
        3                               &
        3                               &
        2                               &
        4                               \\
        $T_2^{BV}$                      & 
        $\chi_{AT}^1$                   &
        $24/(s_+ s_-)$                  &
        1                               &
        1                               &
        2                               &
        4                               \\
        $T_{23}^{BV}$                   & 
        $\chi_{AT}^1$                   &
        $48\,s_+/(s_+ - s_-)^2$         &
        1                               &
        1                               &
        1                               &
        4                               \\
        \bottomrule
    \end{tabular}
    \caption{
    \label{tab:outer} 
        Parameters of the weight and outer functions defined in Eqs.~\eqref{eq:WBGL},~\eqref{eq:outerBGL}, and~\eqref{eq:outerGG}.
        The values reported in this table correspond to the minimal number of subtractions for $\chi_\Gamma^J$, which is equal to $d-1$.
    }
\end{table}

\refstepcounter{subsection}
\subsection*{\thesubsection\quad Outer functions}

In the derivation of unitarity-bounded parametrisations, the weight functions $W_\F$ must be rewritten as the modulus squared of a function that is analytic in the open
unit disc (cf. \refsubsec{UBpar}).
These functions are called \emph{outer functions}, because once the conformal map has been performed they are determined by their modulus on the outer boundary of the domain, namely the unit circle.\footnote{
    The complementary factors are \emph{inner functions}, i.e. analytic functions with unit modulus for $|z|=1$. 
    Blaschke products~\eqref{eq:Bprod} are the standard inner factors: they encode zeros inside the disc without changing the boundary norm. 
}
Their role is to encode the known weight factor, thereby allowing the unitarity bound to be expressed in a simple form.

As explained in \refsubsec{UBpar}, the modulus squared of the outer functions must satisfy \refeq{phiBGLconstr}. Here I rewrite this condition for a generic threshold $s_\th$, which in practice is either $s_+$ or $s_\Gamma$ defined in \refeqa{spm}{sGamma}:
\begin{align}
    \label{eq:phigenconstr}
    \left|\phi_\F(s;s_\th)\right|^2
    =
    \frac{W_\F(s)}{\widetilde{\chi}_\Gamma^J(Q^2;d-1)}
    \left|\frac{dz(s;s_0,s_\th)}{ds}\right|^{-1}
    \qquad
    \text{for } s>s_\th
    \,.
\end{align}
Naively taking the square root of the above equation yields an outer function of the form
\begin{equation}
\begin{aligned}
    \phi_\F^{\rm naive} (s;s_\th)
    &:=
    \sqrt{\frac{\kappa_I}{K \pi \widetilde{\chi}_\Gamma^J(Q^2;d-1)}}\,  
    \left( \frac{s_\th - s}{s_\th - s_0} \right)^{\frac 1 4}
    \left( \sqrt{s_\th - s} + \sqrt{s_\th - s_0} \right)
    \\*
    & \times
    (s-s_+)^{\frac a4}
    (s-s_-)^{\frac b4}
    s^{-\frac{(c+3)}{2}}
    \left(\frac{s}{s+Q^2}\right)^{\frac{d}{2}}
    \,.
\end{aligned}
\end{equation}
Although this expression reproduces the correct modulus on the cut, it does not yet qualify as an acceptable outer function, because it exhibits zeros and singularities for $s<s_\th$ and hence has different analytic properties from the corresponding FFs.
The correct strategy is to replace each pole at $s=s_P$ with~\cite{Boyd:1997kz,Boyd:1997qw}
\begin{align} 
    \frac{1}{s-s_P} \mapsto -\frac{z(s;s_P,s_\th)}{s-s_P} 
    = 
    \frac{1}{\left( \sqrt{s_\th - s} + \sqrt{s_\th - s_P} \right)^2}
    \,.
\end{align}
This replacement simply introduces a Blaschke factor~\eqref{eq:Bprod}.
With this strategy one readily finds
\begin{equation}
\begin{aligned}
    \phi_\F (s;s_\th)
    & := 
    \sqrt{\frac{\kappa_I^{\phantom{n}}}{K \pi \widetilde{\chi}_\Gamma^J(Q^2;d-1)}} \,
    \left( \frac{s_\th - s}{s_\th - s_0} \right)^{\frac 1 4}
    \left( \sqrt{s_\th - s} + \sqrt{s_\th - s_0} \right)
    \\* 
    & \times 
    \left( s_+ - s \right)^{\frac a 4} 
    \left( \sqrt{s_\th - s} + \sqrt{s_\th - s_-} \right)^{\frac b 2}
    \left( \sqrt{s_\th - s} + \sqrt{s_\th} \right)^{-(c+3)}
    \\* 
    & \times
    \left(
    \frac{
        \sqrt{s_\th - s} + \sqrt{s_\th}
    }{
         \sqrt{s_\th - s} + \sqrt{s_\th + Q^2}
    }
    \right)^d
    \,.
\end{aligned}
\end{equation}
For the BGL parametrisation, one has $s_\th=s_+$, and \refeq{outerBGL} is recovered. For the GG parametrisation, one instead has $s_\th=s_\Gamma$, which reproduces part of \refeq{outerGG}.
By construction, after applying the conformal map~\eqref{eq:zmap}, $\phi_\F$ is analytic and non-vanishing in the open unit disc, and hence it is an outer function.
As noted in Ref.~\cite{Simula:2025fft}, in the case $s_\th=s_\Gamma$ one may still choose to impose \refeq{phigenconstr} only for $s>s_+$.
This can lead to different, inequivalent choices of outer functions and, consequently, to different values of $\Delta\chi_\Gamma^J$ (see \refeq{defDeltaChiGG}).
I do not explore this possibility in the present work, although it is certainly an interesting direction for future investigation.

\appsection{Note on form-factor predictions}
\label{app:FFcal}

Lattice QCD and LCSRs are two different methods for predicting FFs.
One of the main advantages of lattice QCD is that its uncertainties can, in principle, be systematically reduced, whereas the uncertainties of LCSR calculations cannot be lowered below a certain threshold.
However, LCSRs are currently able to provide predictions for matrix elements in processes for which lattice QCD results are not available, or are available only to a limited extent.
For instance, lattice-QCD calculations of FFs involving QCD-unstable hadrons --- such as the $B\to K^*$ FFs --- are still under development~\cite{Leskovec:2025gsw,Erben:2026ylp}.
It is also important to emphasise that, even though some systematic uncertainties in LCSRs are irreducible, they can nevertheless be estimated.
This is typically done by varying the Borel parameter in the LCSR calculation, in analogy with scale variation in a perturbative QCD calculation, and by assessing the relative contribution of excited states with respect to the ground state.
The key point is that, even after including progressively higher-order corrections in $\alpha_s$ and higher-order power corrections, the dependence on the Borel parameter and, in particular, on the quark--hadron duality approximation does not vanish.
This is, however, not a cause for concern, since relative uncertainties of about $15\%$--$20\%$, which are typical for LCSRs, are often sufficient to obtain reasonably precise predictions, especially when LCSR results are combined with lattice-QCD computations and unitarity bounds.
In this respect, an important and fortunate feature is the complementarity between the two methods: LCSR calculations are applicable only at low $q^2$, in order to ensure a sufficiently rapid convergence of the light-cone OPE, whereas lattice-QCD computations achieve their highest precision in the high-$q^2$ region, near the zero-recoil point.
As a result, the combination of LCSR and lattice-QCD inputs allows for a controlled determination of the FFs across the full physical kinematic range.

That uncertainties cannot be reduced below a certain threshold is not an unusual feature in quantum field theory.
In perturbatively dominated QCD calculations, a truncation at any finite order necessarily leaves a residual scale dependence because higher-order terms are omitted.
Since the perturbative expansion in $\alpha_s$ is generally only asymptotic rather than convergent, one should not expect this uncertainty to vanish in any strict sense by going to arbitrarily high order.
\\

There are two main variants of LCSRs, which differ in the choice of the distribution amplitudes (DAs) entering the light-cone expansion; see Refs.~\cite{Colangelo:2000dp,Khodjamirian:2020btr,Khodjamirian:2023wol} for reviews.
The DAs are universal non-perturbative functions that describe the momentum distribution of the partons inside a given hadron.
In light-meson LCSRs, the sum rule is formulated in terms of the DAs of the light meson in the final state, such as the pion, kaon, or a light vector meson.  
For representative calculations, see e.g. Refs.~\cite{Ball:2004ye,Ball:2004rg,Duplancic:2008ix,Duplancic:2008tk,Kang:2013jaa,Bharucha:2015bzk,Khodjamirian:2017fxg,Cheng:2017sfk,Leljak:2021vte,Bolognani:2023mcf,Cheng:2025hxe}.
By contrast, in $B$-meson LCSRs, the sum rule is formulated in terms of the DAs of the initial $B$ meson. 
For representative calculations, see e.g. Refs.~\cite{Faller:2008tr,Cheng:2017smj,Gubernari:2018wyi,Descotes-Genon:2019bud,Gubernari:2020eft,Biswas:2025drz}.
The two approaches rely on different non-perturbative inputs and are affected predominantly by distinct sources of theoretical uncertainty.
Therefore, light-meson and $B$-meson LCSR calculations can be regarded as independent and can thus be combined in a FF analysis.

\appsection{Meson masses and thresholds}
\label{app:meson-masses}

\begin{table}[t!]
    \newcommand{\pp}{{\phantom{*}}}
    \centering
    \renewcommand{\arraystretch}{1.4}
    \begin{tabular}{
        c@{\hspace{0.4cm}}
        c@{\hspace{0.4cm}}
        c@{\hspace{0.4cm}}
        c@{\hspace{0.4cm}}
        c@{\hspace{0.4cm}}
        c
    }
        \toprule
        Transition                      &
        Thresholds                      &
        PDG name                        &
        $m_{\Gamma,i}^J$                &
        Mass $[\GeV]$                   &
        Ref.  \\[0.1cm]
        \toprule
        \multirow{4}{*}{\makecell{
            $b\to u$  \\[0.1cm]
            $b\to d$
        }}                              &
        \multirow{4}{*}{\makecell{
          $\begin{aligned}
            \sqrt{s_V} &= 5.414 \,\GeV \\[0.1cm]
            \sqrt{s_A} &= 5.460 \,\GeV \\[0.0cm]
            \sqrt{s_+^{B\pi}} &= m_B + m_{\pi} = \sqrt{s_V}
          \end{aligned}$
        }}                              & 
        $B^{\pm},\,B^0$                 &
        $m_{A,1}^0$                     & 
        $5.279^\pp$                     &
        \cite{PDG:2024cfk}              \\

                                        &
                                        &
        ---                             &
        $m_{V,1}^0$                     & 
        $5.642^*$                       & 
        FP      \\ 
                                        & 
                                        & 
        $B^*$                           &
        $m_{V,1}^1$                     & 
        $5.325^\pp$                     &
        \cite{PDG:2024cfk}              \\
                                        & 
                                        & 
        $B_1(5721)$                     &
        $m_{A,1}^1$                     & 
        $5.726^*$                       &
        \cite{PDG:2024cfk}              \\
        \midrule
        \multirow{4}{*}{$b\to s$}       &
        \multirow{4}{*}{\makecell{
          $\begin{aligned}
            \sqrt{s_V} &= 5.502 \,\GeV \\[0.1cm]
            \sqrt{s_A} &= 5.550 \,\GeV \\[0.1cm]
            \sqrt{s_+^{BK}} &= m_B + m_K  = 5.773 \,\GeV
          \end{aligned}$
        }}                              & 
        $B_s^0$                         &
        $m_{A,1}^0$                     & 
        $5.367^\pp$                     &
        \cite{PDG:2024cfk}              \\
                                        & 
                                        & 
        ---                             &
        $m_{V,1}^0$                     & 
        $5.711^*$                       &
        \cite{Lang:2015hza}             \\ 
                                        & 
                                        & 
        $B_s^*$                         &
        $m_{V,1}^1$                     & 
        $5.415^\pp$                     &
        \cite{PDG:2024cfk}              \\
                                        &  
                                        & 
        $B_{s1}(5830)^0$                &
        $m_{A,1}^1$                     & 
        $5.829^*$                       &
        \cite{PDG:2024cfk}              \\
        \midrule
        \multirow{6}{*}{$b\to c$}       &
        \multirow{6}{*}{\makecell{
          $\begin{aligned}
            \sqrt{s_V} &= 6.409 \,\GeV \\[0.1cm]
            \sqrt{s_A} &= 6.463 \,\GeV \\[0.1cm]
            \sqrt{s_+^{BD}} &= m_B + m_D  = 7.144 \,\GeV
          \end{aligned}$
        }}                              &
        $B_c^\pm$                       &
        $m_{A,1}^0$                     & 
        $6.274^\pp$                     &
        \cite{PDG:2024cfk}\\ 
                                        & 
                                        & 
        $B_c(2S)^\pm$                   &
        $m_{A,2}^0$                     & 
        $6.871^*$                       &
        \cite{CMS:2019uhm,LHCb:2019bem} \\
                                        & 
                                        & 
        ---                             &
        $m_{V,1}^0$                     & 
        $6.707^*$                       &
        \cite{Dowdall:2012ab}           \\ 
                                        & 
                                        & 
        ---                             &
        $m_{V,1}^1$                     & 
        $6.328^\pp$                     &
        \cite{Dowdall:2012ab}           \\  
                                        & 
                                        & 
        ---                             &
        $m_{V,2}^1$                     & 
        $6.922^*$                       &
        \cite{Dowdall:2012ab}           \\  
                                        & 
                                        & 
        ---                             &
        $m_{A,1}^1$                     & 
        $6.739^*$                       &
        \cite{Dowdall:2012ab}           \\ 
        \bottomrule
    \end{tabular}
    \caption{ 
    \label{tab:states}
    List of the relevant mesons and their masses organised according to their quark content and quantum numbers.
    The index $\Gamma$ denotes one of the currents defined in \refeq{JGamma}, $J$ is the spin of the state, and ``$i$'' is simply an enumerative index.
    The masses for $\Gamma= T, \,AT$ are not explicitly listed since $m_{V,i}^J \equiv m_{T,i}^J$ and $m_{A,i}^J \equiv m_{AT,i}^J$.
    I only include states that have either been observed experimentally or been determined in lattice QCD, together with the special case of the fictitious scalar pole in $b\to u,d$ transitions labelled by FP.
    The role of fictitious poles is explained in \refsubsec{X}.
    Masses exceeding $\sqrt{s_\Gamma}$ are indicated with an asterisk, as they require distinct treatment in the GG parametrisation.
    The thresholds $s_\pm$ and $s_\Gamma$ are defined in \refeqa{spm}{sGamma}, respectively.
    }
\end{table}

For the reader's convenience, I collect in \reftab{states} the mass inputs for the $b\to u,d$, $b\to s$, and $b\to c$ transitions considered in this work.
I do not distinguish between $b \to u$ and $b \to d$ transitions, since the corresponding mass difference is negligible compared with the FF uncertainties.

In the numerical implementation of the unitarity bounds, these masses enter in several ways.
They determine the thresholds $s_\pm$ and $s_\Gamma$ introduced in \refeqa{spm}{sGamma}.
They also determine which states satisfy $(m_{\Gamma,i}^J)^2 < s_\Gamma$ and therefore contribute explicitly to the one-particle part of the correlator, as discussed in \refsubsec{UB}.
In channels with resonances satisfying $s_\Gamma < (m_{\Gamma,i}^J)^2 < s_+$, the same masses also enter the hatted quantities of the GG parametrisation through the modified weights and outer functions; see Eqs.~\eqref{eq:outerGG},~\eqref{eq:hDchi}, and~\eqref{eq:Chi-def}.

\appsection{Endpoint relations for GG coefficients}
\label{app:endpointrel}

For practical applications, the series in \refeq{GG} must be truncated at a finite order $N$, because only a finite number of parameters can be determined from a statistical analysis:
\begin{align}
    \label{eq:GGtruncapp}
    \F(q^2)
    =
    \frac{1}{\B_\Gamma^J(q^2;s_\Gamma)\,\widehat\phi_\F(q^2)}
    \sum_{n=0}^{N}
    \alpha_{\F,n}^{\rm GG}\,
    z(q^2;\sopt,s_\Gamma)^n
    \,.
\end{align}
Here $\sopt$ is the optimal choice of $s_0$ given in \refeq{s0opt}.
Since $\sopt$ depends on $s_\Gamma$, its numerical value is FF-dependent.
The values of $s_\Gamma$ follow from combining the information in \reftab{outer} and \reftab{states}.
For $B\to P$ FFs one has $s_\Gamma=s_V$.
For $B\to V$ FFs, one finds $s_\Gamma=s_V$ for $V^{BV}$ and $T_1^{BV}$, whereas $s_\Gamma=s_A$ for $A_0^{BV}$, $A_1^{BV}$, $A_{12}^{BV}$, $T_2^{BV}$, and $T_{23}^{BV}$.

In the truncated GG parametrisation, the FFs must still satisfy the endpoint relations given in \refeqa{endp0}{endpsm}.
In practice, these relations can be used to eliminate one GG coefficient for each constraint.
In the implementation used here, the dependent coefficients are chosen as $\alpha_{f_0,0}^{\rm GG}$, $\alpha_{A_{12},0}^{\rm GG}$, $\alpha_{A_1,0}^{\rm GG}$, $\alpha_{T_2,0}^{\rm GG}$, and $\alpha_{T_{23},0}^{\rm GG}$, while all remaining GG coefficients are treated as independent fit parameters.
One then obtains
\begin{align}
    \alpha_{f_0,0}^{\rm GG}
    &=
    \frac{
        \B_V^0(0;s_V)\,\widehat\phi_{f_0}(0)
    }{
        \B_V^1(0;s_V)\,\widehat\phi_{f_+}(0)
    }
    \sum_{n=0}^{N}
    \alpha_{f_+,n}^{\rm GG}\,
    z(0;\sopt,s_V)^n
    -
    \sum_{n=1}^{N}
    \alpha_{f_0,n}^{\rm GG}\,
    z(0;\sopt,s_V)^n
    \,,
    \\
    \alpha_{A_{12},0}^{\rm GG}
    &=
    \frac{m_B^2-m_V^2}{8\,m_B m_V}\,
    \frac{
        \B_A^1(0;s_A)\,\widehat\phi_{A_{12}}(0)
    }{
        \B_A^0(0;s_A)\,\widehat\phi_{A_0}(0)
    }
    \sum_{n=0}^{N}
    \alpha_{A_0,n}^{\rm GG}\,
    z(0;\sopt,s_A)^n
    -
    \sum_{n=1}^{N}
    \alpha_{A_{12},n}^{\rm GG}\,
    z(0;\sopt,s_A)^n
    \,,
    \\
    \alpha_{A_1,0}^{\rm GG}
    &=
    \frac{8\,m_B m_V}{m_B^2-m_V^2}\,
    \frac{
        \widehat\phi_{A_1}(s_-)
    }{
        \widehat\phi_{A_{12}}(s_-)
    }
    \sum_{n=0}^{N}
    \alpha_{A_{12},n}^{\rm GG}\,
    z(s_-;\sopt,s_A)^n
    -
    \sum_{n=1}^{N}
    \alpha_{A_1,n}^{\rm GG}\,
    z(s_-;\sopt,s_A)^n
    \,,
    \\
    \alpha_{T_2,0}^{\rm GG}
    &=
    \frac{
        \B_A^1(0;s_A)\,\widehat\phi_{T_2}(0)
    }{
        \B_V^1(0;s_V)\,\widehat\phi_{T_1}(0)
    }
    \sum_{n=0}^{N}
    \alpha_{T_1,n}^{\rm GG}\,
    z(0;\sopt,s_V)^n
    -
    \sum_{n=1}^{N}
    \alpha_{T_2,n}^{\rm GG}\,
    z(0;\sopt,s_A)^n
    \,,
    \\
    \alpha_{T_{23},0}^{\rm GG}
    &=
    \frac{(m_B+m_V)^2}{4\,m_B m_V}\,
    \frac{
        \widehat\phi_{T_{23}}(s_-)
    }{
        \widehat\phi_{T_2}(s_-)
    }
    \sum_{n=0}^{N}
    \alpha_{T_2,n}^{\rm GG}\,
    z(s_-;\sopt,s_A)^n
    -
    \sum_{n=1}^{N}
    \alpha_{T_{23},n}^{\rm GG}\,
    z(s_-;\sopt,s_A)^n
    \,.
\end{align}
Here, I used the fact that the Blaschke products defined in \refeq{Bprod} depend only on the corresponding pole masses.
With the identifications $m_{T,i}^J\equiv m_{V,i}^J$ and $m_{AT,i}^J\equiv m_{A,i}^J$ given in \reftab{states}, it then follows that $\B_T^1\equiv\B_V^1$ and $\B_{AT}^1\equiv\B_A^1$.
The constraints involving $A_1^{BV}$ and $T_{23}^{BV}$ are imposed sequentially:
first $\alpha_{A_{12},0}^{\rm GG}$ and $\alpha_{T_2,0}^{\rm GG}$ are determined, and only afterwards $\alpha_{A_1,0}^{\rm GG}$ and $\alpha_{T_{23},0}^{\rm GG}$ are evaluated.
Endpoint relations can also be imposed in the DM framework, as explained in Ref.~\cite{Martinelli:2023fwm}.

\bibliographystyle{myJHEP}
\bibliography{references}

\end{document}